\documentclass[
				a4paper,fleqn,usenatbib,useAMS]{mnras}

\usepackage{newtxtext,newtxmath}

\usepackage[T1]{fontenc}
\usepackage{ae,aecompl}

\usepackage{graphicx}	
\usepackage{amsmath}	
\usepackage{amssymb}	

\usepackage[version=3]{mhchem}
\usepackage{datetime}
\usepackage{booktabs}
\usepackage{float}
\usepackage{gensymb}
\usepackage{wasysym}
\usepackage{amsfonts}
\usepackage{courier}
\usepackage{caption}
\usepackage{subcaption}
\usepackage{multicol}   
\usepackage{multirow}
\usepackage{pdflscape}	
\usepackage[para]{threeparttable}
\usepackage{mathrsfs}
\usepackage{longtable}
\usepackage{xtab,afterpage}

\newdateformat{mydate}{\twodigit{\THEDAY}{ }\monthname[\THEMONTH], \THEYEAR}
\mydate
\newdate{reportdate}{06}{6}{2017}
\newdate{datadate}{02}{2}{2017}

\newcommand{\rom}[1]{%
  \textup{\uppercase\expandafter{\romannumeral#1}}%
}

\title[Prospecting for exo-Earths]{Prospecting for exo-Earths in multiple planet systems with a gas giant}
       
\author[M. T. Agnew et al.]{
Matthew T. Agnew,$^{1}$ 
Sarah T. Maddison,$^{1}$
Jonathan Horner$^{2}$
\\
$^{1}$Centre for Astrophysics and Supercomputing, Swinburne University of Technology, Hawthorn, Victoria 3122, Australia\\
$^{2}$University of Southern Queensland, Centre for Astrophysics, Toowoomba, Queensland 4350, Australia\\
}

\date{Accepted 2018 September 7. Received 2018 September 6; in original form 2018 August 13}

\pubyear{2017}

\begin{document}
\label{firstpage}
\pagerange{\pageref{firstpage}--\pageref{lastpage}}
\maketitle

\begin{abstract}
In this work, we hunt for the best places to find exo-Earths in the currently known exoplanet population. While it is still unclear whether Jupiter had a beneficial or detrimental effect on the creation of the right environment for a habitable Earth to develop,  we focus on the 51 multiple planet systems that have at least one Jupiter-like planet and aim to identify which would be good candidates to host an exo-Earth. We conduct a series of numerical simulations to identify dynamically stable regions of the habitable zone of the multiple exoplanet systems capable of hosting an Earth-mass planet. We produce a candidate list of 16 systems that could host such a stable exo-Earth in their habitable zone, and for which the induced radial velocity signal of a hypothetical one, two or four Earth-mass planet on the host star would be detectable with the ESPRESSO spectrograph. We find that whilst the gravitational interactions with the massive planet nearest the habitable zone are critical in determining stability, the secular resonant interactions between multiple planets can also have a dramatic influence on the overall stability of the habitable zone. 
\end{abstract}



\section{Introduction}
\label{sec:introduction}
The high precision spectroscopy instruments that will be available on the next generation of ground-based and space-based telescopes will usher in a new era in the search for life on potentially Earth-like worlds. The sensitivity of such instruments will enable us to detect small, rocky planets in the habitable zone (HZ) of exoplanetary systems \citep{Pasquini2010,Pepe2014,Hernandez2017}. However, with over 3700\footnote{As of 26 June 2018 (NASA Exoplanet Archive, exoplanetarchive.ipac.caltech.edu).} confirmed exoplanets to date, and that number expected to more than double with NASA's Transiting Exoplanet Survey Satellite \citep[TESS;][]{Ricker2014,Sullivan2015,Barclay2018}, it is important to provide direction to help planet hunters to identify the most promising candidates in their search for potentially habitable exoplanets.

One of the key goals of the next generation of exoplanet surveys will be the discovery of a planet that could be considered to be a twin to the Earth, in order to facilitate the search for evidence of life elsewhere. We currently know of only one location where life has established and has thrived, so to maximise our chances of success in the search for life beyond the Solar system, it makes sense to first look at those systems that most closely resemble our own. A large body of research has considered the impact that Jupiter may have had in establishing the right conditions for life on Earth -- from its role in facilitating Earth's composition and hydration \citep[e.g.][]{Bond2010,Carter-Bond2012,Bond2010b,Martin2013a,Quintana2014,OBrien2014}, to suggestions that it might have served to shield our planet from an impact regime that would otherwise have proven inimical to life \citep{Wetherill1994,Ward2000}. In recent years, a number of studies have suggested that Jupiter's role as an impact shield may have, at the very least, been significantly overstated \citep[e.g.][]{Horner2008,Horner1908,Horner2010b,Horner2012,Horner2013,Grazier2016}, the existence of Jupiter in our own Solar system nevertheless provides a point from which to begin; an Earth-like planet coexisting with a Jupiter-like planet. Regardless of whether Jupiter fostered or hindered the development of life on Earth, it is clear that the dynamics of a massive body will have an impact on the dynamics of nearby planets.

A wide range of tools exist that facilitate the computational study of the dynamical interaction of planetary systems. Such tools have been used to perform a variety of studies that investigate the stability of planetary systems. Some authors have used those tools to examine the effects of outer, giant planets on inner, rocky planets \citep[e.g.][]{Carrera2016,Kaib2016a,Mustill2016}, whilst others have used those tools to investigate the dynamical feasibility of proposed multi-planet systems \citep[e.g.][]{Horner2011,Wittenmyer2012,Wittenmyer2014b}.

Other studies have used such simulations to predict the stability of hypothetical additional bodies in existing systems \citep{Raymond2005,Raymond2006a,Wittenmyer2013} and, with improved computing power, simulation suites continue to be used assess the dynamical stability of exoplanetary systems and predict the feasibility of additional, as yet unseen companions \citep{Kane2015,Thilliez2016,Agnew2017,Agnew2018}. While there are several bodies of work that have advanced analytical, semi-analytical and qualitative classifications \citep{Giuppone2013,Laskar2017,Agnew2018} in order to rapidly and robustly assess system stability, the potential chaos of multi-body systems means that numerical simulations of planetary systems remain an important tool in studying dynamical stability.

In our previous work, we developed a framework to predict which single Jovian planet systems are capable of hosting a dynamically stable and potentially habitable rocky planet \citep{Agnew2017,Agnew2018}. It was found that the proximity of a massive Jupiter-like planet to the HZ is critical in determining the overall stability of the HZ. This can take the form of completely dynamically stable HZs when the Jovian planet is well separated from the HZ (e.g. hot Jupiters or Jupiters far beyond the outer boundary of the HZ); or in the form of stable islands of mean motion resonance with the Jovian planet close to the HZ (e.g. with an orbit embedded within, or traversing, the HZ). In this work, we study the effects of the gravitational interactions of multiple planets on potential exo-Earths in the HZ of multi-planet systems, in order to understand what sort of planetary architectures can maintain stable HZs.

In section \ref{sec:method}, we describe the method used to determine our source list of multiple Jovian planet systems to model and the numerical technique used to predict which systems are capable of hosting dynamically stable exo-Earths in their HZs.  We present and discuss our results in section \ref{sec:results}, highlighting the dynamical stability analysis and the candidate list of observable exo-Earths, and summarise our findings in section \ref{sec:summary}.

\section{Method}
\label{sec:method}
In this work, we numerically search the catalogue of known multiple planet systems to determine which, if any, would be capable of hosting dynamically stable and potentially habitable exo-Earths. Before introducing additional bodies into the HZ of each system, we first need to confirm the dynamical stability of the known planets using existing best-fit planetary and stellar parameters. If we find a multi-planet system to be dynamically unstable, then it is a candidate for further observations or numerical analysis \citep[e.g.][]{Robertson2012,Wittenmyer2012}. Such studies are beyond the scope of this work, but we will report on those systems which we find to be dynamically unstable. If a system is found to be stable, we can then begin to assess the stability of its HZ, first with massless test particles, and then with massive bodies for those systems which experience gravitational stirring in their HZ due to the orbits of the known planets. 

\subsection{System Selection}
\label{subsec:predicting_stable_regions}
\begin{table}
    \caption{The radius and mass limits we use to classify exoplanets. Mass is used only in the case of missing radius data.}
    \label{tab:planet_classification}
    \centering
    \begin{tabular}{l c c c c}
        \toprule                     
        				& $r_{\mathrm{min}}$ 		& $r_{\mathrm{max}}$  		& $m_{\mathrm{min}}$  		& $m_{\mathrm{max}}$ 	\\                     
        				& ($\textrm{r}_{\oplus}$) 	& ($\textrm{r}_{\oplus}$) 	& ($\textrm{M}_{\oplus}$) 	& ($\textrm{M}_{\oplus}$)	\\				
        \midrule
        Terrestrials	& 0							& $<1.5$						& 0							& $<1.5$					\\
        Super-Earths	& 1.5						& $<2.5$						& 1.5						& $<10$					\\
        Neptunians	& 2.5						& $<6$						& 10							& $<50$					\\
        Jovians		& 6							& $>6$						& 50							& $>50$			       	\\
        \bottomrule
    \end{tabular}
\end{table}

We classify an exoplanet as either a terrestrial, a super-Earth, a Neptunian or a Jovian using a radius classification scheme (or mass classification scheme in lieu of available radius data). The radius and mass cuts we use are shown in Table~\ref{tab:planet_classification}. By applying this classification scheme to the systems in the NASA Exoplanet Archive\footnote{ exoplanetarchive.ipac.caltech.edu}, we find that there are 135 multiple planet systems \textit{with at least one Jovian planet}\footnote{As of 27 April 2018}. We immediately eliminate 77 systems for which the necessary stellar or planetary properties required to carry out numerical simulations are unavailable or unknown. As there are so few exoplanet systems for which the mutual inclinations of planets have been measured, we accept all systems with missing inclination and longitude of ascending node values, and make the simplifying assumption that all systems considered are co-planar. Whilst this is an idealised scenario, there is research to support shallow, near co-planar mutual inclinations for multiple planet systems \citep{Lissauer2011,Lissauer2011a,Fang2012,Figueira2012,Fabrycky2014}. We consider the implications of non-zero mutual inclinations on the HZ stability of a system in lesser detail in Section~\ref{subsubsec:incl}. There are 58 systems remaining for which we have all the necessary stellar and planetary properties, noting that none of these have inclination data available. 

We calculate the HZ boundaries for each star following the approach of \cite{Kopparapu2014}. They provide a method for calculating the HZ boundaries of F, G, K and M spectral type main-sequence stars that is only valid for stars with $2600$~K $\leq T_\mathrm{eff} \leq 7200$~K. The distance from the star for the edges of the HZ are:
\begin{align}
	d_{\textrm{HZ}} = \sqrt{\frac{L/L_\odot}{S_{\textrm{eff}}}} \,\,\, \rm{au},
\end{align}
where $L$ is the luminosity of the star, and $S_{\textrm{eff}}$ is calculated as
\begin{align}
	S_{\textrm{eff}} = S_{\textrm{eff}\odot} + aT_\star + bT_\star^2+cT_\star^3+dT_\star^4 \, ,
\end{align}
where $T_\star = T_{\textrm{eff}} - 5780$ K, and $a$, $b$, $c$, $d$ and $S_{\textrm{eff}\odot}$ are constants depending on the planetary mass considered within the HZ, $M_\textrm{pl}$, and the HZ boundary regime being used. We use the the conservative HZ boundaries \citep[Runaway Greenhouse and Maximum Greenhouse from][]{Kopparapu2014}, and a $1\ \textrm{M}_{\oplus}$ planet. This gives us the constants as shown in Table~\ref{tab:HZ_params}. Whilst we will ultimately consider $2\ \textrm{M}_{\oplus}$ and $4\ \textrm{M}_{\oplus}$ planets in our results when we calculate the induced Doppler wobble on the star, the resulting small variations in the HZ boundaries are trivial (in the order of a few per cent for a star like the Sun). 

\begin{table}
	\caption{The constants provided by \protect\cite{Kopparapu2014} that we use to calculate the edges of the HZ for our simulations.}
   	\label{tab:HZ_params}
   	\centering
	\begin{tabular}{c c c c c}
    	 \cmidrule(r){3-3}  \cmidrule(r){5-5}  
    						& 	&{Runaway Greenhouse}  & & {Maximum Greenhouse }\\

    	\cmidrule(r){1-1} \cmidrule(r){3-3}  \cmidrule(r){5-5} 	 
        $a$							&	& 	$1.332\times10^{-4}$& 	&	$6.171\times10^{-5}$\\
        $b$							&	& 	$1.58\times10^{-8}$& 	&	$1.698\times10^{-9}$\\
        $c$							&	& 	$-8.308\times10^{-12}$& 	&	$-3.198\times10^{-12}$\\
        $d$							&	& 	$-1.931\times10^{-15}$& 	&	$-5.575\times10^{-16}$\\
        $S_{\textrm{eff}\odot}$		&	& 	$1.107$& 				&	$0.356$\\
    	\cmidrule(r){1-1} \cmidrule(r){3-3}  \cmidrule(r){5-5} 
   	\end{tabular}
\end{table}

Given that the HZ calculation of \cite{Kopparapu2014} is only valid for temperatures in the range $2600$~K $\leq T_\mathrm{eff} \leq 7200$~K, we eliminate a further seven systems where the host star is either too hot or too cold for such analysis to be valid. This yields the final sample of 51 multiple planet systems with at least one Jovian planet.

\subsection{Dynamical Simulations}
\label{subsec:dynamical_analysis}
We conduct a series of numerical simulations to help determine which of the 51 system could potentially host a habitable exo-Earth. We use the N-body package \textsc{SWIFT} \citep{Levison1994}, using the regularised mean variable symplectic (RMVS) integrator \citep{Levison2000}, with a combination of massive planets and massless test particles (TPs). Our three sets of simulations consist of: 1) a planetary stability test, using the existing known planets of each system, 2) a HZ stability test, using massless TPs in the HZ of each system, along with the known planets, and 3) a gravitationally stirred HZ test, whereby we explore the dynamical stability of a $1\ \textrm{M}_{\oplus}$ planet in system for which the existing planets gravitationally perturb the HZ. 

For the planetary dynamics test, we use the best-fit stellar and planetary parameters for each system from the NASA Exoplanet Archive. This includes the stellar mass and effective temperature of the host star, and the semi-major axis, eccentricity, longitude of periastron and time of periastron passage (from which the mean anomaly is computed) for all the known planets in each system  (the initial conditions for all planets in a given system can be found in Table~\ref{tab:xtab}). These simulations are run for $10^8$ years, using an integration time step of $1/50$ of the orbital period of the innermost planet. The simulations terminate if two of the planets experience a mutual close encounter, considered to be approaching to within one Hill radius of the other, or if a planet is ejected from the system, defined in this work as reaching an astrocentric distance of $250$~au. 

The second set of simulations investigates the stability of the HZ for those systems found to pass the planetary stability test. 1000 massless TPs are randomly distributed throughout the HZ with orbital parameters within the ranges shown in Table~\ref{tab:tp_params}. Simulations are run for $10^7$ years, using an integration time step of $1/50$ of the orbital period of the innermost object (TP or planet). Simulations are terminated if all the TPs are removed by ejection from the system (defined as $r > 250$~au). 

In systems which experience substantial gravitational stirring of the HZ by another planet, a surviving swarm of stable TPs does not necessarily indicate that a $1\ \textrm{M}_{\oplus}$ body would also be stable. It is possible for the orbit of a putative exo-Earth to be stabilised, or destabilised, by the mutual gravitational interactions with the known exoplanets.  This motivates the third set of simulations which investigates the stability of a $1\ \textrm{M}_{\oplus}$ planet in the HZ of gravitationally stirred systems. We focus on those systems from the second set of simulations for which some massless TPs survived on stable orbits throughout the entire simulation, but we ignore those systems that had large unperturbed regions within their HZ. We assume that such systems are inefficiently stirred and that the $1\ \textrm{M}_{\oplus}$ body will remain stable, and hence massive body simulations would be a waste of computational resources. 

We follow the dynamical evolution of a $1\ \textrm{M}_{\oplus}$ body with orbital parameters given in Table~\ref{tab:tp_params}. In each simulation, the initial conditions of the known massive planets were set to their observed best-fit values, whilst the orbital parameters of the $1\ \textrm{M}_{\oplus}$ body were changed systematically from run to run. A total of $28,560$ ($51\times16\times5\times7$) simulations were performed for each system studied. These simulations were also run for $10^7$ years with an integration time step of of $1/50$ of the orbital period of the innermost planet, and simulations were terminated if one of the planets experienced a close encounter (within 1 Hill radius) with another or any of the planets were ejected from the system.

\begin{table}
	\caption{The orbital parameters of the TPs and $1\ \textrm{M}_{\oplus}$ body for the simulations. The TPs were randomly distributed between the minimum and maximum values given. Each $1\ \textrm{M}_{\oplus}$ simulation used a unique set of orbital parameters, where the parameters are incremented over the given range for the number of values shown.}
   	\label{tab:tp_params}
   	\centering
	\begin{tabular}{c c c c c c c c}
    		\cmidrule(r){3-4}  \cmidrule(r){5-8}  
    	& & \multicolumn{2}{c}{TPs}  & & \multicolumn{3}{c}{$1\ \textrm{M}_{\oplus}$ }\\
        \cmidrule(r){3-4}  \cmidrule(r){5-8}                     
        						&	& Min					& Max & 			& Min					& Max  		& $\#$ of Steps\\
        \cmidrule(r){1-2} \cmidrule(r){3-4}  \cmidrule(r){5-8}  
        $a$ (AU)				&	& HZ$_{\mathrm{min}}$	& HZ$_{\mathrm{max}}$	&	& HZ$_{\mathrm{min}}$	& HZ$_{\mathrm{max}}$	& 51\\
        $e$					&	& 0.0					& 0.3 &			& 0.0					& 0.3 					& 16\\
        $i$ ($\degree$)		&	& 0.0					& 0.0 &			& 0.0					& 0.0 					& 1 \\
        $\Omega$ ($\degree$)	&	& 0.0					& 0.0 &	 		& 0.0					& 0.0 					& 1 \\
        $\omega$ ($\degree$)	&	& 0.0					& 360.0 & 		& 0.0					& 288.0 					& 5\\
        $M$ ($\degree$)		&	& 0.0					& 360.0 	&		& 0.0					& 288.0 					& 7\\
        \cmidrule(r){1-2} \cmidrule(r){3-4}  \cmidrule(r){5-8}  
   	\end{tabular}
\end{table}

\section{Results and Discussion}
\label{sec:results}
Here, we present the results of our three sets of simulations, present an in-depth look at the complex stabilising behaviour found in some of the multiple planet systems, and examine the implications of mutual inclinations between the orbits of the TPs and the planets. Finally, we present a candidate list estimating the radial velocity signal that an exo-Earth would induce on its host star. for those systems where we determined that a stable exo-Earth could exist within the HZ. This candidate list is intended to help direct future searches for Earth-like planets in these systems.

\subsection{Planetary Stability}
For the 51 planetary systems that satisfied the selection process outlined in Section~\ref{subsec:predicting_stable_regions}, we conduct a dynamical stability analysis as described in Section~\ref{subsec:dynamical_analysis}. Table~\ref{tab:unst_syst} shows the nine systems we found to be unstable using their observed best-fit orbital parameters. Five of these systems have had further numerical analyses performed to identify more appropriate parameters. A further two are currently being investigated (HD 67087 and HD 181433), and two require similar dynamical investigation. Whilst numerical simulations are used to suggest more dynamically feasible orbital parameters, ultimately further observations of these systems are required to better constrain each planet's orbit.

\begin{table}
\caption{Multiple exoplanet systems found to be dynamically unstable with the currently accepted best-fit data. Some systems have been subject to numerical re-analysis, while others are candidates for a similar numerical analysis must still be carried out.}
	\label{tab:unst_syst}	
    \centering
    \begin{tabular}{l c r}
        \toprule                     
        System				& Destabilisation 	& Re-analysis\\
        					& Time (yr) 		& Reference\\
        \midrule
        HD 5319		& $6.2\times10$  &	\cite{Kane2016a}\\
        24 Sex		& $3.6\times10^{4}$  &	\cite{Wittenmyer2012} \\
        HD 200964	& $1.4\times10^{2}$  &	\cite{Wittenmyer2012} \\
        HD 33844	& $7.0 $  &	\cite{Wittenmyer2016a}\\
        BD+20 2457	& $4.9\times10$  &	\cite{Horner2014}\\
        HD 67087	& $3.1\times10$  &	Marshall et al. (in prep.)\\
        HD 181433	& $2.2\times10$  &	Horner et al. (in prep.)\\
        HD 133131 A	& $1.0\times10^{5}$  &	To be investigated\\
        HD 160691 	& $5.4\times10^{7}$  &	To be investigated\\
        \bottomrule
    \end{tabular}
\end{table} 

It should be emphasised the planetary masses provided are minimum masses (i.e. $m_\mathrm{pl}=m\sin{I}$, where $I$ is the inclination of the planet's orbit with respect to our line of sight). As such, the actual masses of each planet will also vary somewhat depending on the inclination of the orbit of each planet relative to us, and that may impact on the overall stability of each system.

\subsubsection{Angular Momentum Deficit Comparison}
While numerical simulations provide a thorough and robust assessment of the dynamical stability of a planetary system, they can be computationally expensive and time consuming, especially as the number of planets increases or for systems where planets are spread over a large range of orbital periods. Numerical and theoretical predictions complement one another and can be applied to different types of system architectures. In the case of tightly orbiting hot Jupiter systems, the small integration timestep needed to accurately resolve the short orbital period is a very inefficient use of computational resources. In such a scenario, theoretical predictions potentially offer a more appropriate method to assess the stability of the system. Alternatively, multiple planet systems with complex stabilising resonant mechanisms require the robustness of numerical methods to ensure the stability is identified.

To address this issue, \cite{Laskar2017} present the angular momentum deficit (AMD) stability criterion. The AMD is a conserved quantity that indicates the variability of averaged planetary systems, where zero corresponds with co-planar, circular motion, and higher values indicate chaotic behaviour. The AMD stability criterion can be used to predict the potential stability (or otherwise) of a given planetary system given the masses, $m$, semi-major axes, $a$, and eccentricities $e$, of the bodies in the system. For each pair of adjacent planets, we can calculate the AMD stability coefficient $\beta$ given by equation (58) of \cite{Laskar2017}. For any pair of planets, the AMD coefficient $\beta < 1$, this means collisions are not possible and hence the pair is considered AMD stable. A multiple planet system is considered AMD stable if all adjacent pairs of planets are AMD stable. 

\begin{figure}
	\centering
	\includegraphics[width=\linewidth]{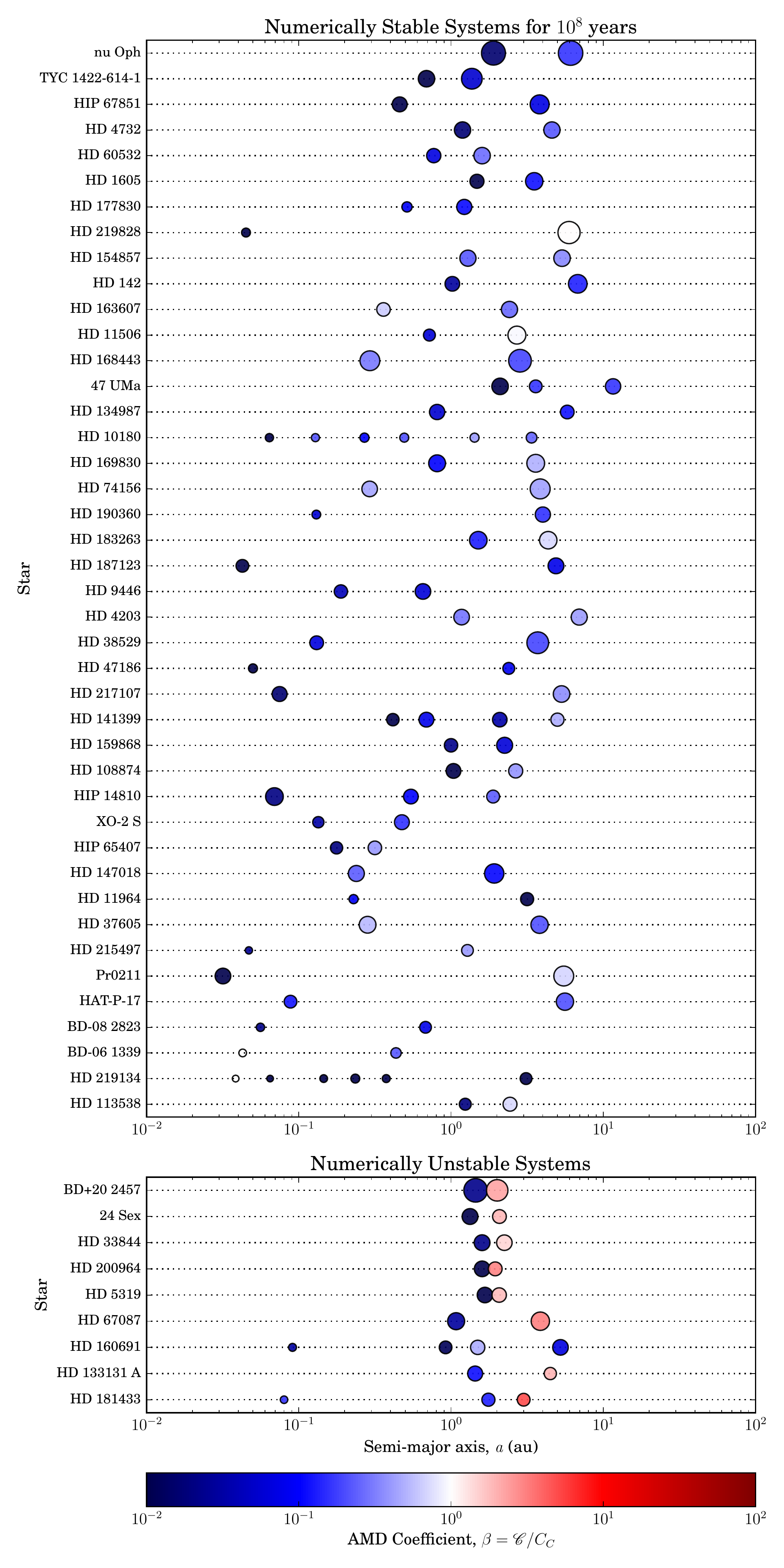}
	\caption{An AMD stability plot adapted from \protect\cite{Laskar2017} showing all the multiple planet systems we consider in our analysis. Planets in each system are represented by circles, with size  proportional to the planet mass to the third power, $m^{1/3}$, and colour representing the AMD stability coefficient, $\beta$, of each planet with its inner neighbour (and the innermost planet with the star). When $\beta > 1$ (shown here in red) the planet, and consequently the system, is AMD unstable.}
	\label{fig:amd}
\end{figure}

Using this approach, we can compute the AMD stability coefficient for each of our systems to predict their stability or instability, and compare that with the results of our numerical simulations. In Figure~\ref{fig:amd} we show those systems which we found to be numerically stable for $10^8$ years in the upper plot, and numerically unstable systems are shown in the lower plot. Following \cite{Laskar2017}, a blue planet indicates AMD stability, whilst a red planet indicates AMD instability. CIt can quickly be seen that all those systems which we found to be numerically stable are also AMD stable. Conversely, all of our numerically unstable systems are also found to be AMD unstable, with the exception of HD 160691.

\subsubsection{HD 160691}
\begin{figure}
	\begin{subfigure}{\linewidth}
    	\centering
		\includegraphics[width=\linewidth]{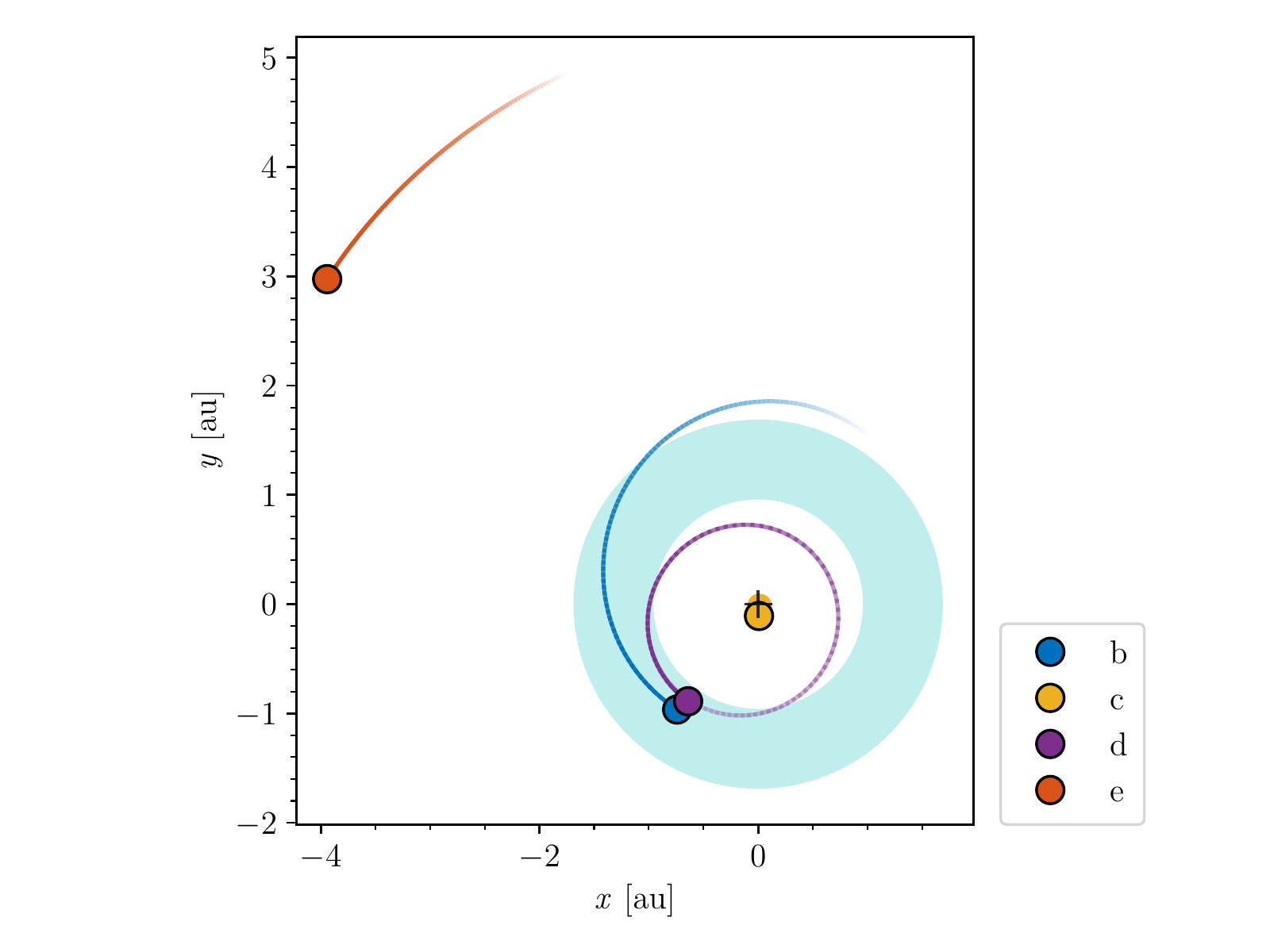}
		\caption{}\label{fig:hd_160691_cart}
	\end{subfigure}
	
	\begin{subfigure}{\linewidth}
    	\centering
		\includegraphics[width=\linewidth]{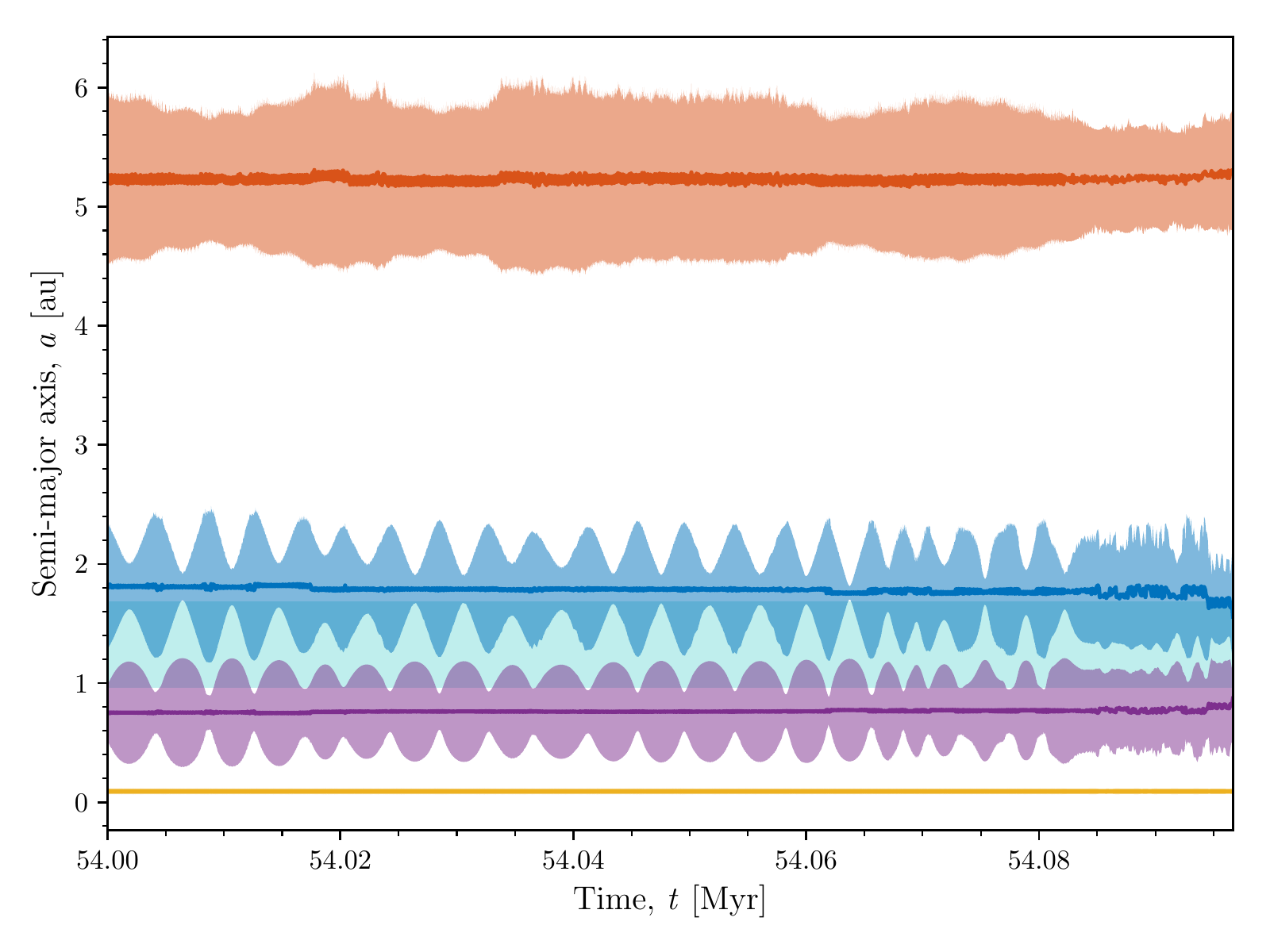}
		\caption{}\label{fig:hd_160691_sma}
	\end{subfigure}
	
	\begin{subfigure}{\linewidth}
    	\centering
		\includegraphics[width=\linewidth]{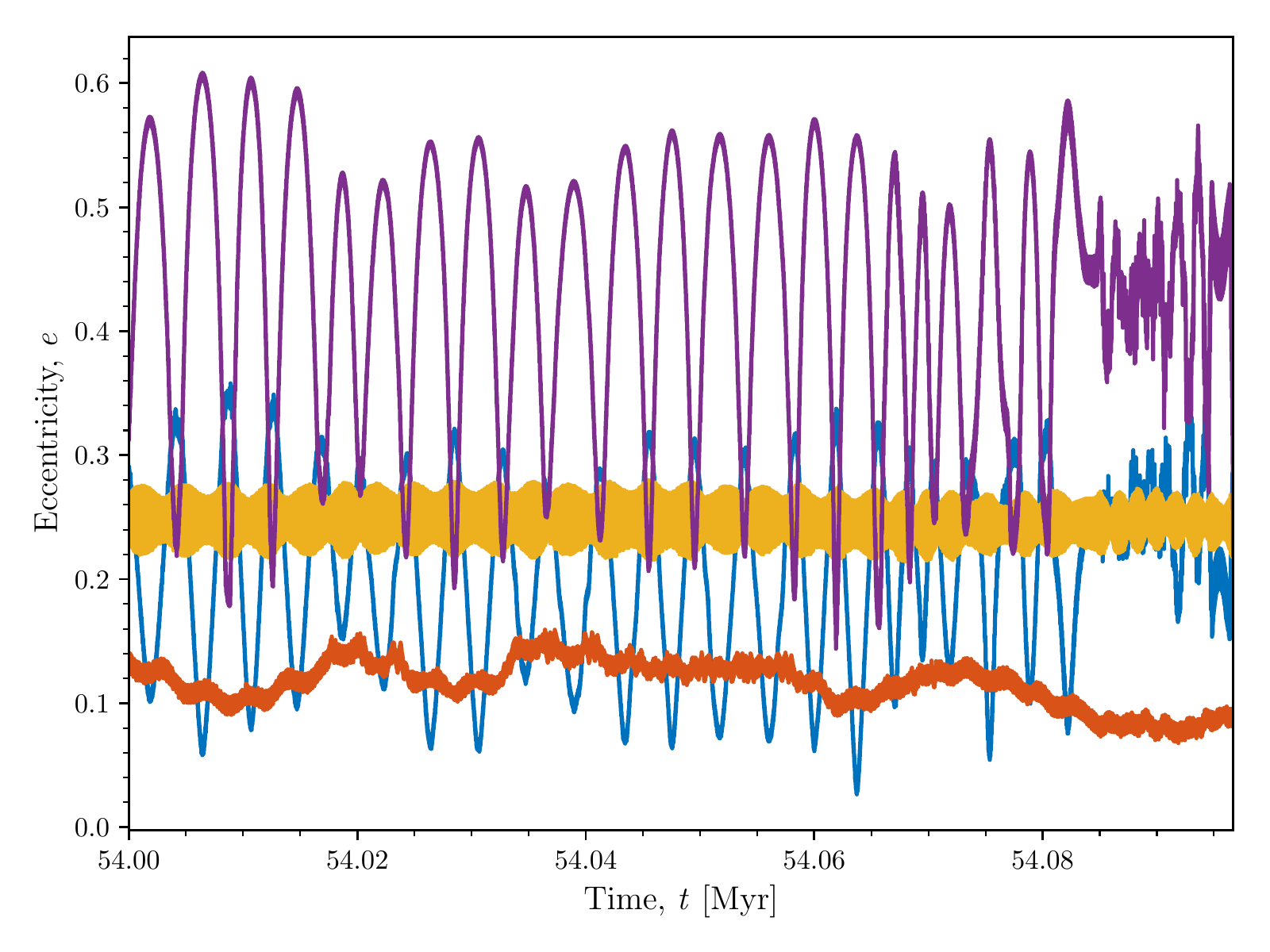}
		\caption{}\label{fig:hd_160691_ecc}
	\end{subfigure}
\caption{The Cartesian plot, and the evolution of the semi-major axes and eccentricities of the four planets in the HD~160691 system. The Cartesian plot shows the orbits for the last year leading up to the close encounter. The fainter, coloured regions in the semi-major axis plot shows the apsides of the planets orbits.}\label{fig:hd_160691}
\end{figure}

HD 160691 contains 4 planets that have large separations and relatively circular orbits \citep{Pepe2007}. Intuitively, and quantitatively considering its AMD stability, it appears stable. Our numerical analysis shows the system to be stable for more than $50~\mathrm{Myr}$. However, it becomes unstable shortly after that point (see Figure~\ref{fig:hd_160691}). This is due excitation between several of the planets that ultimately leads to HD~160691~b and HD~160691~d experiencing a close encounter. Figure~\ref{fig:hd_160691_cart} shows the final year leading up to the close encounter, while Figures~\ref{fig:hd_160691_sma} and \ref{fig:hd_160691_ecc} show the evolution of the planetary semi-major axes and eccentricities. 

While we have one discrepancy between numerical stability and AMD stability, overall the analytical model accurately predicts those systems that are unstable. AMD stability is a demonstrably powerful tool for determining system stability of multiple planet systems, and is a time efficient approach compared with our numerical analysis.

\subsection{HZ Stability}
\label{subsec:tp_results}
\begin{figure*}
	\centering
	\includegraphics[width=\linewidth]{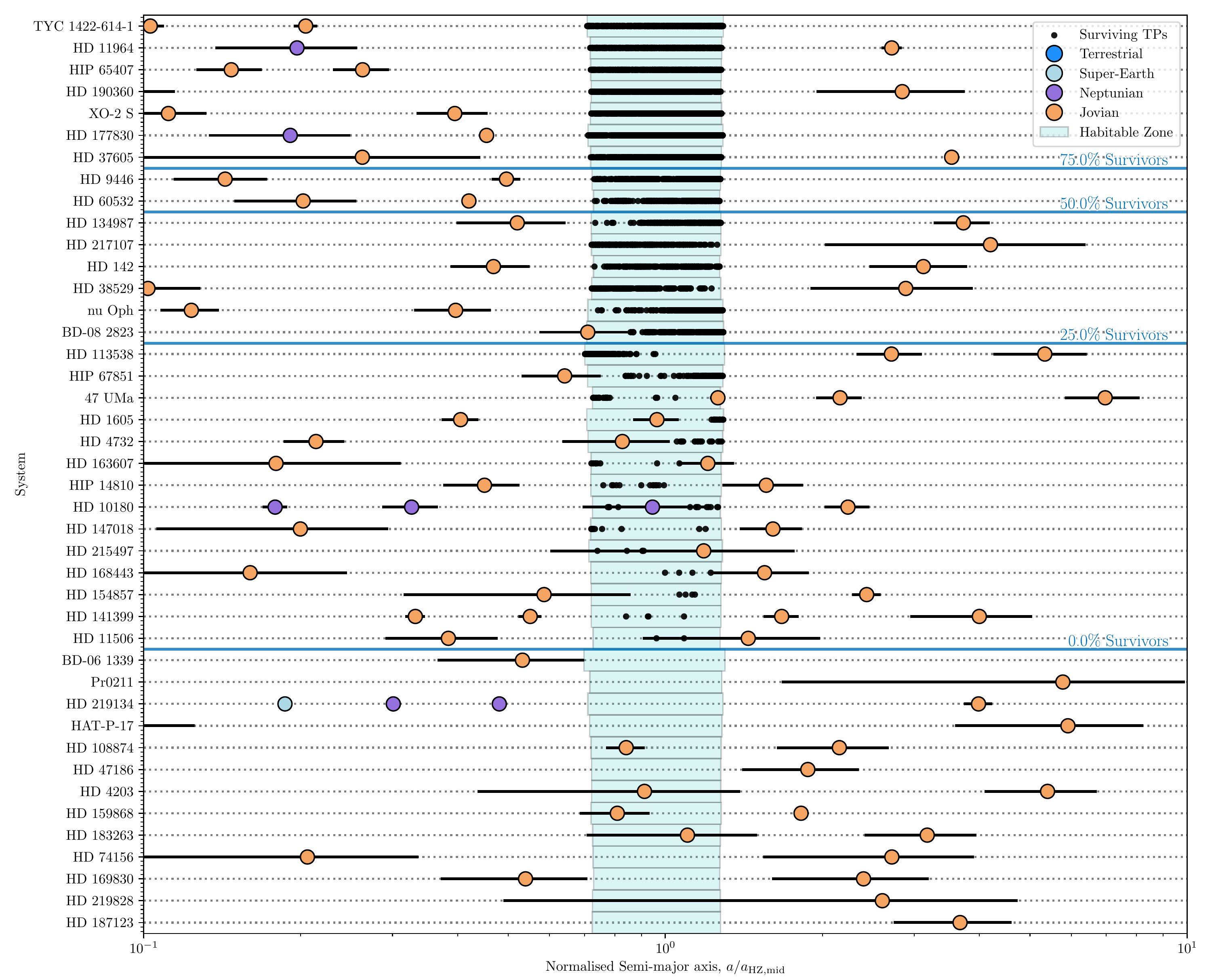}
	\caption{All 42 currently known multiple planet systems with at least one Jovian planet orbiting stars with $2600$ K $\leq T_\mathrm{eff} \leq 7200$~K. The $x$-axis is normalised semi-major axis, defined as $a_\mathrm{norm} = a_\mathrm{body}/a_\mathrm{HZ,mid}$. The planets are coloured according to planetary classifications given in Table~\ref{tab:planet_classification}, while the error bars represent the apsides of their orbits. The green region represents the HZ of each system. The systems are sorted by the fraction of TPs that survived the $10^7$ year duration of our simulations, from most stable (top) to least stable (bottom). The locations of surviving TPs are marked by black dots}.
	\label{fig:massive_selection}
\end{figure*}

For those 42 systems which passed the planetary stability test in the first set of simulations, we conduct simulations of massless TPs spread throughout the HZ as defined by \cite{Kopparapu2014}. The aim of this second set of simulations is to test the stability of the HZ and determine which systems require further numerical investigation. System which are found to have completely unstable HZs, as well as those with completely, or nearly completely, stable HZs, require no further analysis. For the latter group, an extensive suite of massive body simulations would yield little value at high computational cost. 

Figure~\ref{fig:massive_selection} shows the HZ of each system, all the planetary bodies, and any surviving TPs after $10^7$ years of the simulation (called \textit{survivors}). Note that planets located significantly interior or exterior to the HZ  are not shown. We plot the various bodies and HZ boundaries against a normalised semi-major axis, where unity corresponds with the semi-major axis at the mid-point of the HZ, such that $a_\mathrm{norm} = a_\mathrm{body}/a_\mathrm{HZ,mid}$. The normalised semi-major axis allows us to align the HZs of all systems for easy comparison across systems. The planets are plotted with colours corresponding to the classification scheme of Table~\ref{tab:planet_classification}, and the error bars representing the apsides of the planet's orbit.

The systems are sorted vertically in order of the fraction of TPs that survive the full duration of the integration. The horizontal blue lines show survival threshold percentages of $0\%$, $25\%$, $50\%$, and $75\%$. The systems below the $0\%$ line are those for with no survivors, and we do not investigate this group of systems further. Conversely, those above the $25\%$ line show large regions of relatively unperturbed TPs. This suggests gravitational stirring by existing planets is not sufficiently large to disturb the bulk of the HZ. We argue that a suite of simulations with a massive body for all such systems ($>25\%$ survivors) would be a waste of computational resources, as the majority of the bodies would be stable and therefore such simulations would provide little scientific value. 

Of all 42 systems simulated, there are 15 systems which have mostly stable HZs, 13 have unstable HZs and 14 have gravitationally stirred HZs. These latter 14 systems are those with between $0\%$ and $25\%$ survivors, whose HZs contain narrow regions and points of stability. Similar behaviour was observed for a number of single Jovian planet systems investigated by \cite{Agnew2017}. These smaller regions may be the result of unperturbed islands of stability, whilst points often correspond to the location of stabilising mean motion resonances (MMRs) with one of the existing planets. As MMRs are the result of mutual gravitational interactions, it is critical to run massive body simulations to understand whether this is indeed the mechanism by which stability is achieved in these systems.

\subsubsection{Effect of Inclination}
\label{subsubsec:incl}
\begin{figure*}
	\centering
	\includegraphics[width=0.75\linewidth]{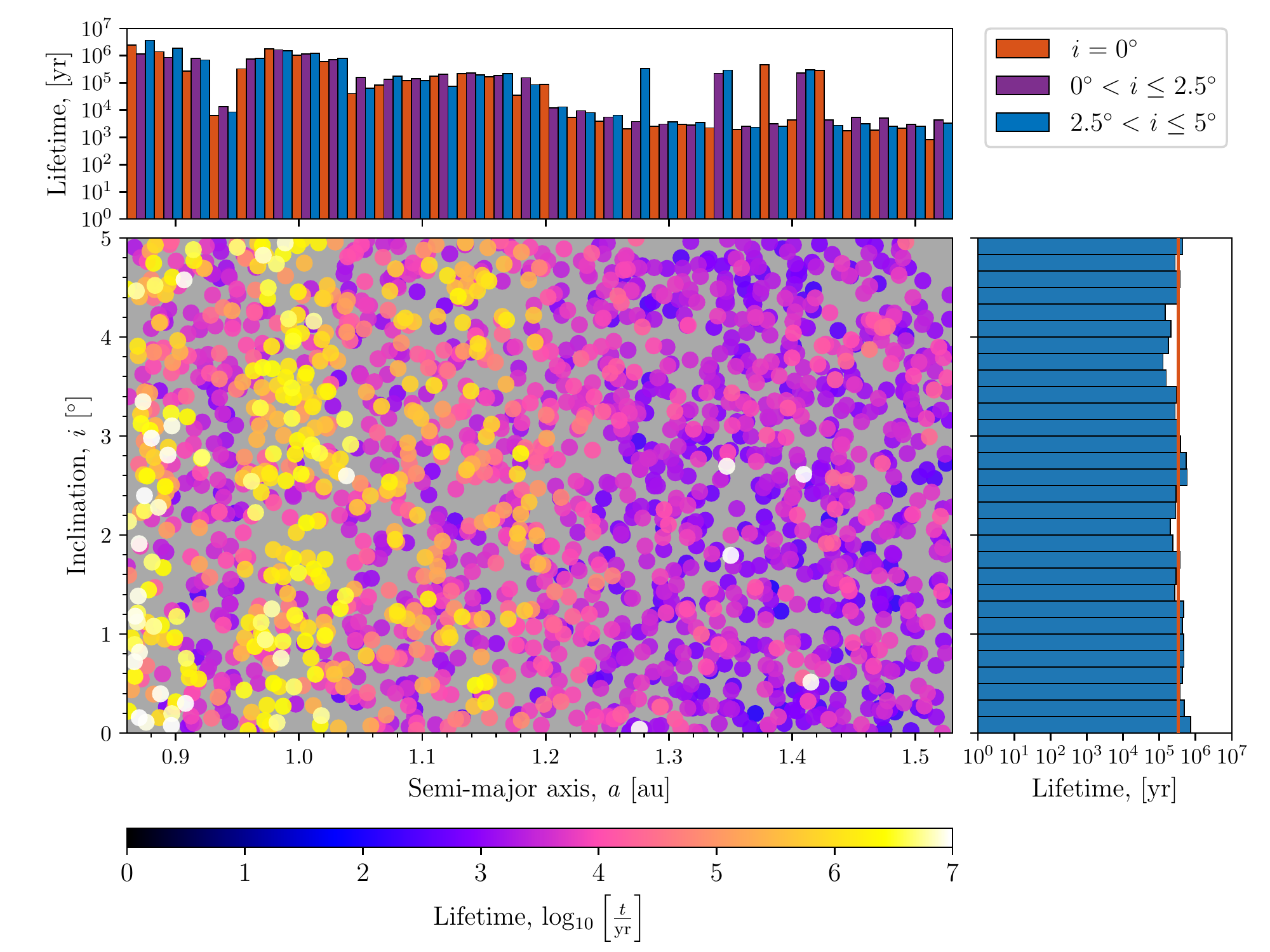}
	\caption{The ($a,i$) stability map for the HD~147018 system. The colour scale for the TP lifetimes is logarithmic. The top histogram shows the binned mean lifetimes for the original 1000 co-planar TPs (orange), and the $0\degr<i\leq2.5\degr$ TPs (purple), and $2.5\degr<i\leq5\degr$ TPs (blue) shallow inclination TPs. The bins are only $1/3$ the width of their actual size for readability. The histogram on the right shows the binned mean lifetimes for the $0\degr<i\leq5\degr$ TPs, with the mean lifetime of all co-planar TPs overlaid in orange.}.
	\label{fig:incl}
\end{figure*}

Whilst we have assumed that all systems are co-planar, a number of studies have shown that multiple body systems are typically not perfectly planar, and generally feature shallow TP mutual inclinations \citep{Lissauer2011,Lissauer2011a,Fang2012,Figueira2012,Fabrycky2014}. We therefore investigate the impact of shallow inclinations in the 14 systems of interest (i.e. those with between $0\%$ and $25\%$ survivors shown in Figure~\ref{fig:massive_selection}).

We repeat the massless TP simulations as outlined in Section~\ref{subsec:dynamical_analysis} for these 14 systems, placing 2000 TP in the HZ with random  inclinations, $i$, between $0\degr$ and $5\degr$, and longitudes of ascending node, $\Omega$, between $0\degr$ and $360\degr$. For two of the systems (HD~10180 and HD~215497), we attempted a suite of such simulations. 

An example of our inclination investigation is shown in Figure~\ref{fig:incl} for HD~147018. The ($a$, $i$) stability map shows the lifetime of each of the 2000 massless TP coloured logarithmically, and indicates that orbital inclination has very little to no effect on the stability of TPs \textit{with shallow inclinations}. The upper histogram bins the stability of the TPs by semi-major axis and shows the stable and unstable regions for both the original 1000 TP co-planar case, and the slightly inclined 2000 TP case are the same (outside of stochastic variations i.e. single stable TPs). It should be noted that the points in Figure~\ref{fig:massive_selection} correspond with those TPs that are stable for the entire simulation, whereas the histogram shows the mean lifetime of all TPs that fall within the $a$ and $i$ bins.

The histogram on the right of Fig.~\ref{fig:incl} bins the stability of TPs by inclination. We see very little variation, and no trend between TP stability and inclination. We overlay the mean survival time of the co-planar TPs (in orange), again showing little variance.  This result suggests that our assumption of co-planarity has little impact on our results. The results for the other systems are shown in Figure~\ref{fig:inc_all}.

In the case of HD~10180, the massive bodies do not as efficiently clear the TPs from the system due to their inclination. As a result, the system take significantly longer to simulate, as not many of the TPs have been ejected to an astrocentric distance of $250$~au, where they would be removed from the simulation. The result of this is that the simulation takes an inordinate length of time to complete. For this system we instead compare the co-planar and shallow inclined cases for the first million years to see if the systems begin to diverge. It was found that the survival time of the TPs in the shallow inclined cases are longer than the co-planar case in just the first million years (see Figure~\ref{fig:inc_hd10180}).

Looking at the results of all systems in Figure~\ref{fig:inc_all}, we can see that the shallow inclinations either extend the TP lifetimes or have no effect. In either case, the TPs are generally removed within the $10^7$~years of the simulation, and hence the assumption of co-planarity  remains a reasonable starting point for our next series of simulations with an Earth-mass body in gravitationally stirred HZs.

It should be emphasised that this investigation considers only the inclinations of the TPs, and not the massive bodies (i.e. $i_\textrm{tp}<5\degree$ but $i_\textrm{pl}=0\degree$). Without further constraints on the planetary orbital inclinations, the parameter space for each system grows significantly such that a systematic analysis of all systems extends beyond the scope of this work.

\subsection{Gravitationally stirred HZs}
\label{subsec:mass_results}
\begin{figure}
	\centering
	\includegraphics[width=\linewidth]{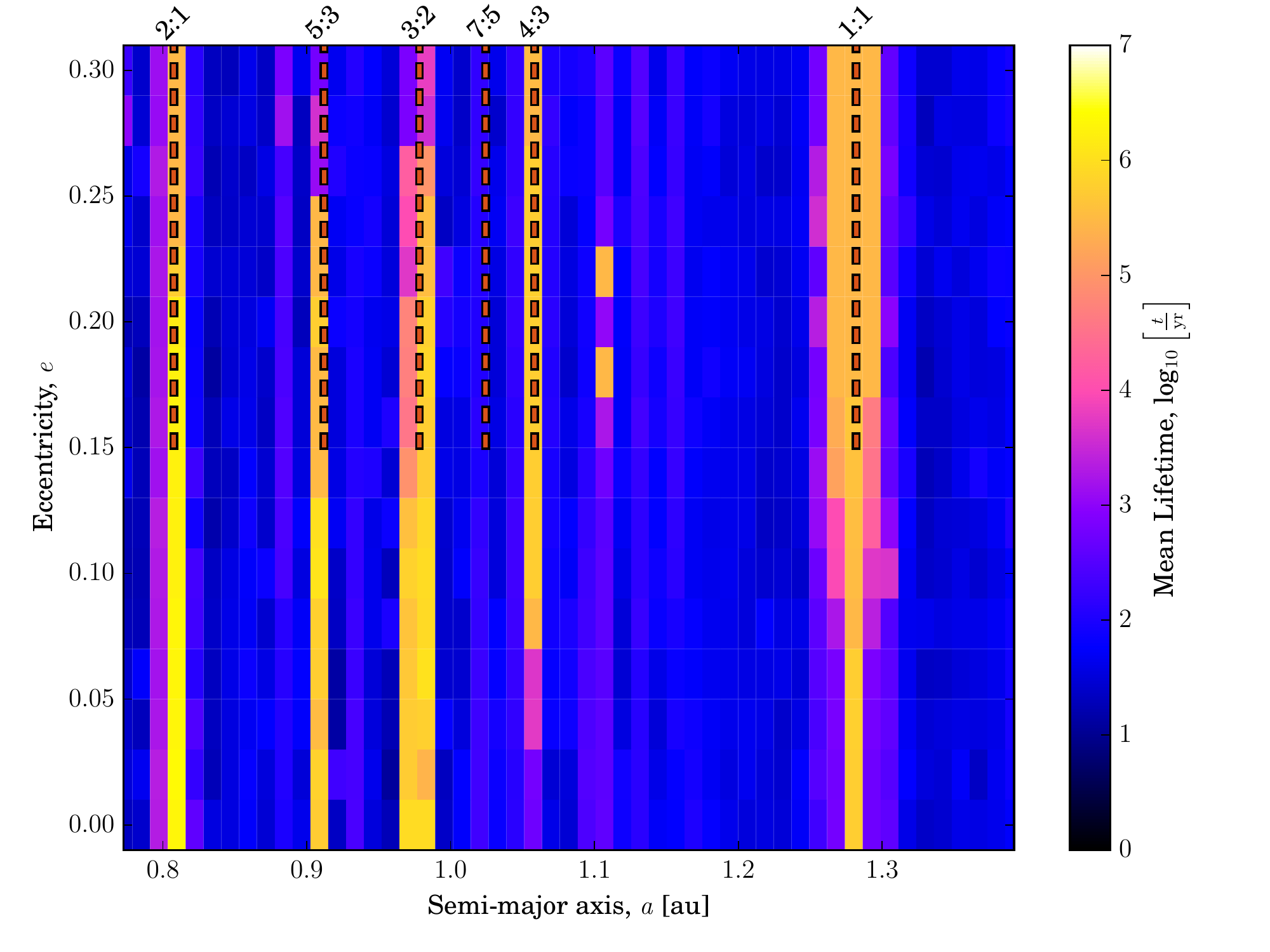}
    \caption{The ($a,e$) stability map for the HD~215497 system. The colour scale for the lifetimes is logarithmic, and each bin represents the mean lifetime of the 35 massive bodies that began the simulation at that particular ($a,e$) value, with different $\omega$ and $M$ values. The semi-major axes that align with the MMRs of each planet are overlaid, starting with the innermost planet at the bottom and progressively moving upwards for each planet further out.}
	\label{fig:hd_215497_map}
\end{figure}

\begin{figure}
	\centering
	\includegraphics[width=\linewidth]{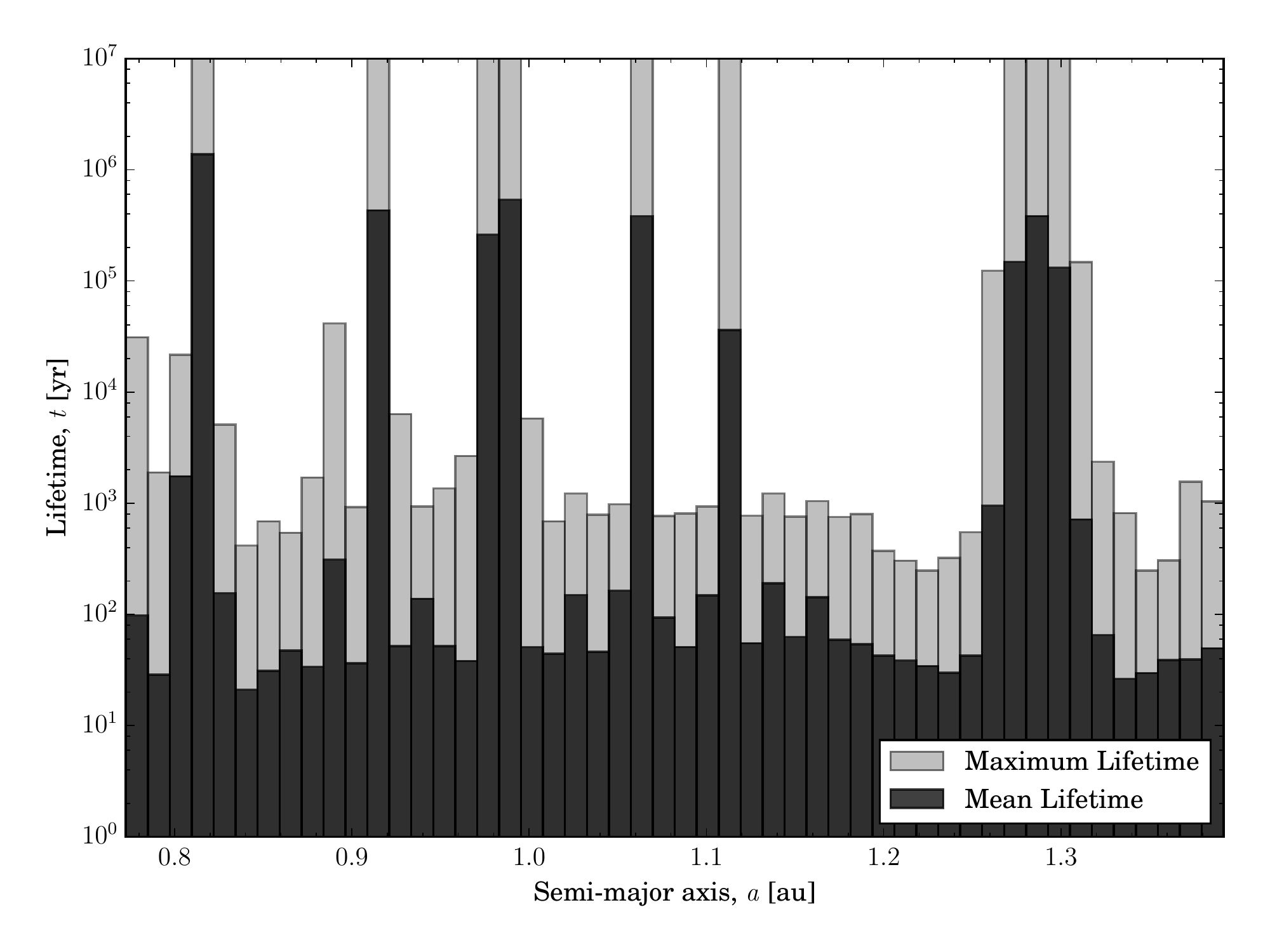}
    \caption{The lifetime histogram for  HD 215497. The lifetimes are shown logarithmically on the $y$-axis. The dark bars represent the average lifetime of all bodies that share the same $a$ values, but different $e$, $\omega$ and $M$ values, whilst the grey bars represent the maximum lifetime of all bodies.}
	\label{fig:hd_215497_plot}
\end{figure}

Systems in our HZ stability simulations which featured heavily perturbed HZs, but managed to retain some stable TPs, are of particular interest as they may represent systems where HZ stability can only be achieved as a result of the stabilising influence of mean-motion resonances. 
We performed a final set of simulations with a $1\ \textrm{M}_{\oplus}$ body in the HZ of those 14 systems.  We use ($a,e$) stability maps and lifetime histograms to examine the nature of the stable islands in HZ. An example of such plots are shown in Figures~\ref{fig:hd_215497_map} and \ref{fig:hd_215497_plot} for the system HD~215497.

Our massive body simulations consist of a suite of $28,560$ simulations for each system with a $1\ \textrm{M}_{\oplus}$ body placed on a unique orbit in each simulation. This includes $35$ simulations ($5\ \omega \times7 \ M$) at each ($a,e$) position, which are binned and coloured by the mean lifetime in the ($a,e$) stability maps.  This is an established technique for studying system stability  \citep[e.g.][]{Horner2010a,Marshall2010,Horner2012a}. On the stability maps, we overlay the lines of several dominant MMRs that correspond with the orbits of the known planets in each system, with the innermost planet at the bottom, and each line above corresponding to the next planet out. Similarly, the lifetime histograms show all simulation outcomes for each system over the 51 semi-major axis bins (see Table~\ref{tab:tp_params}). For each $a$ bin, a total of $560$ simulations were run. We plot both the lifetime of the longest surviving $1\ \textrm{M}_{\oplus}$ body and the mean lifetime of the $560$ simulations. 

\begin{figure}	
	\begin{subfigure}{\linewidth}
    		\centering
    		\includegraphics[width=\linewidth]{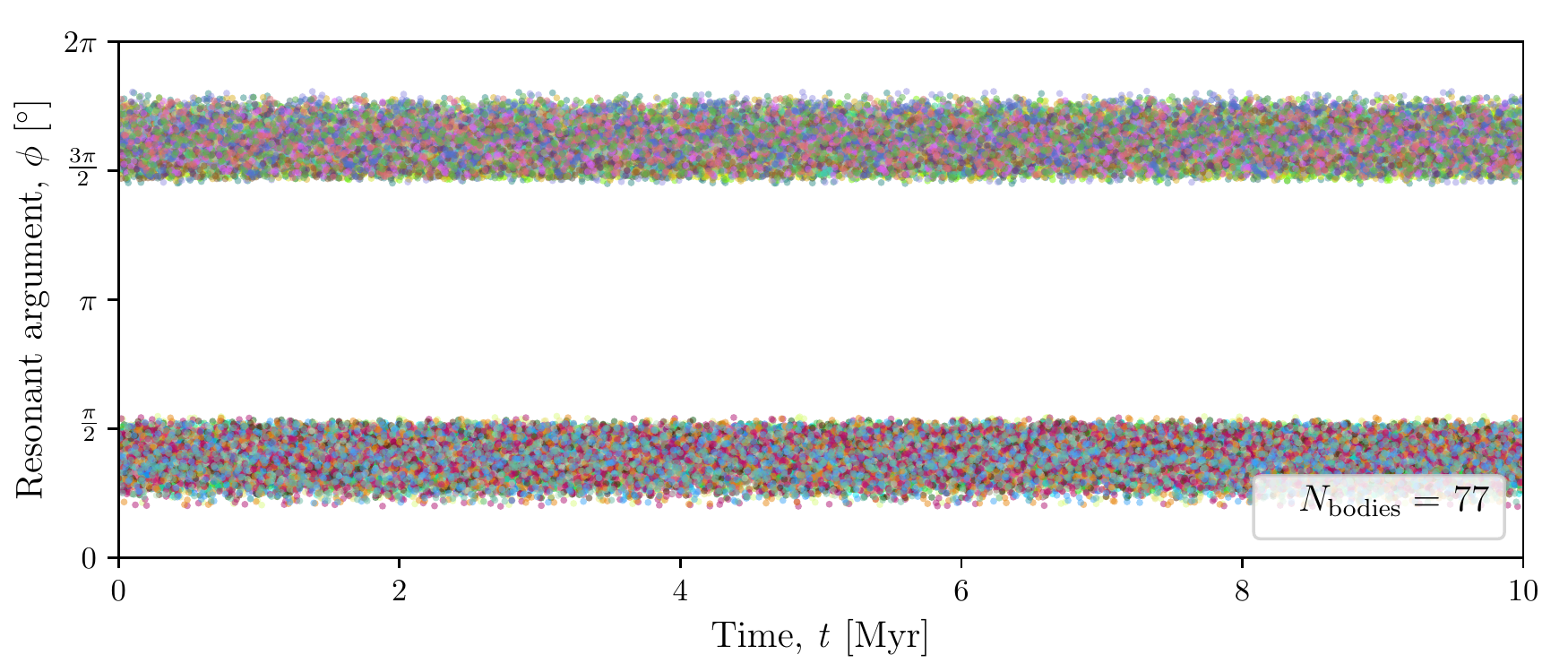}
        \caption{2:1}\label{fig:hd_215497_2_1}
	\end{subfigure}
	
    \begin{subfigure}{\linewidth}
    		\centering
    		\includegraphics[width=\linewidth]{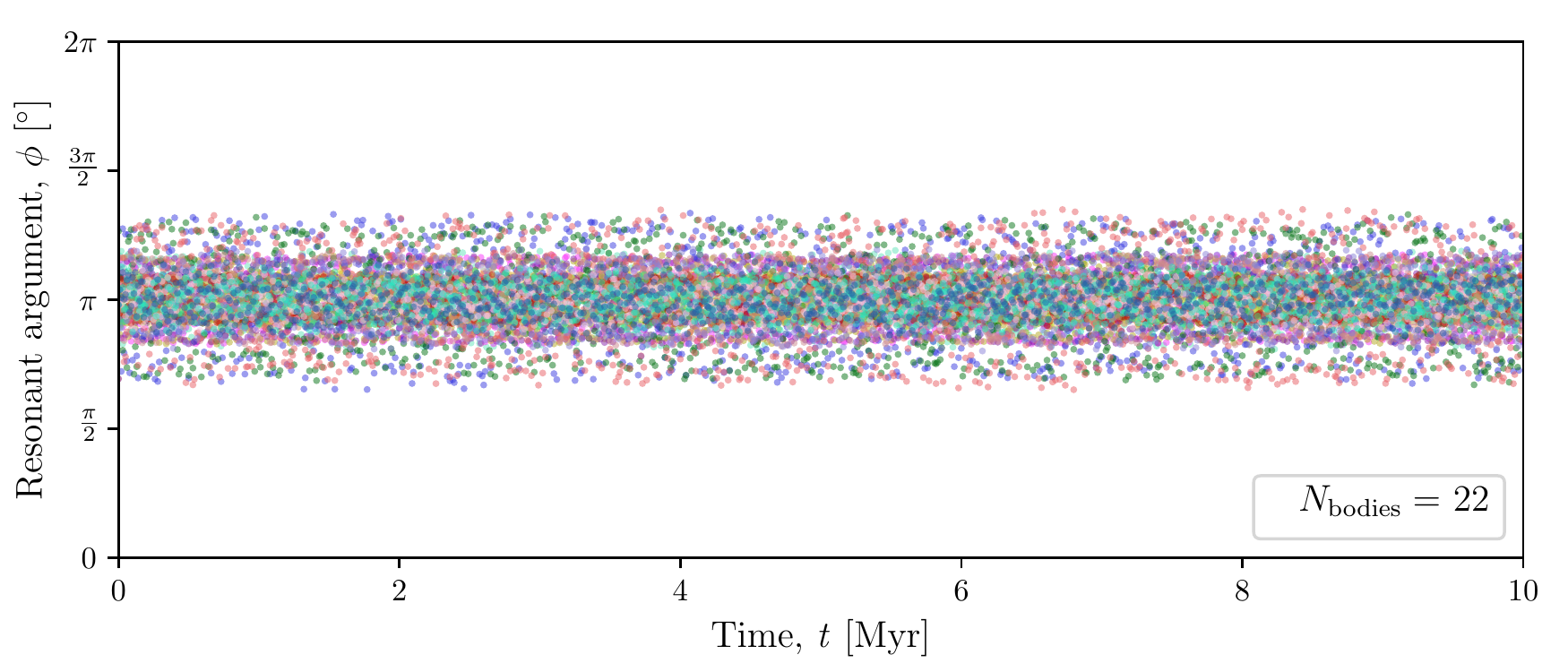}
        \caption{5:3}\label{fig:hd_215497_5_3}
	\end{subfigure}
	
	\begin{subfigure}{\linewidth}
    		\centering
    		\includegraphics[width=\linewidth]{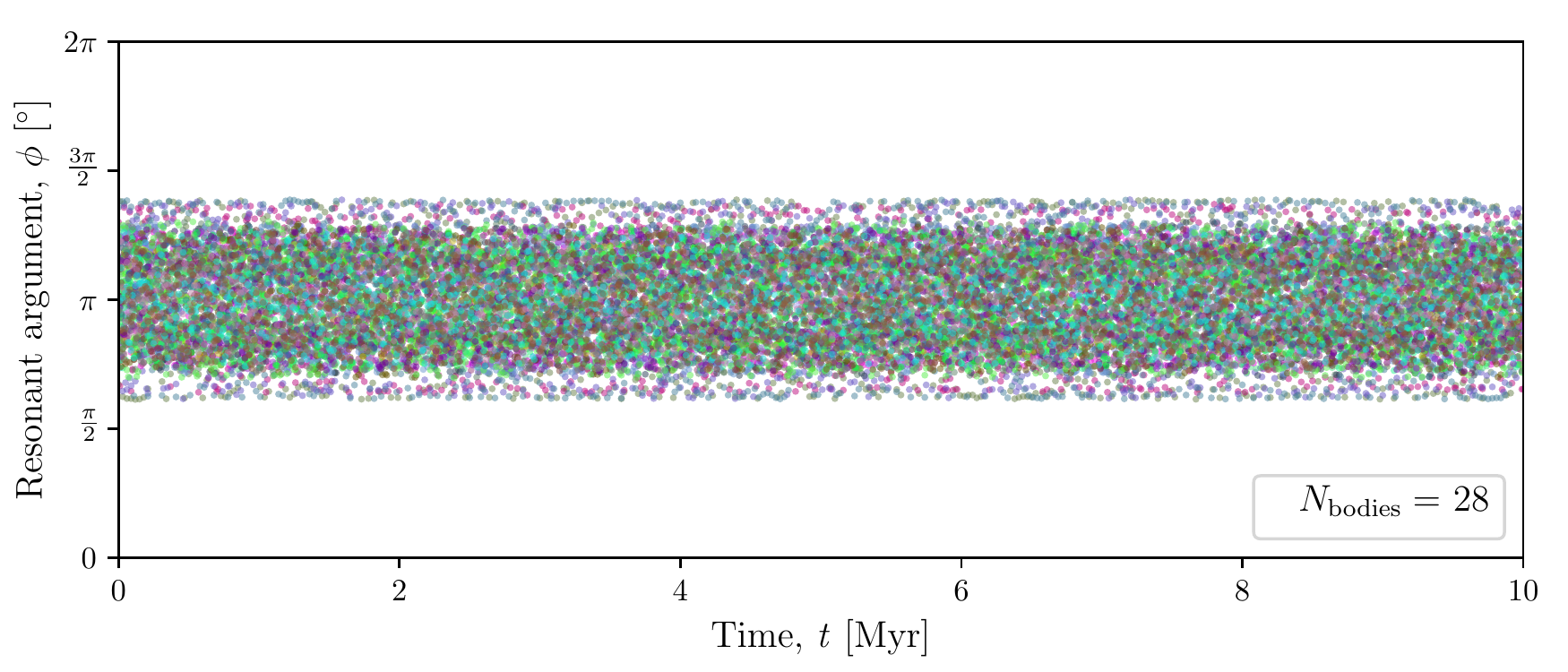}
        \caption{3:2}\label{fig:hd_215497_3_2}
	\end{subfigure}
	
	\begin{subfigure}{\linewidth}
    		\centering
    		\includegraphics[width=\linewidth]{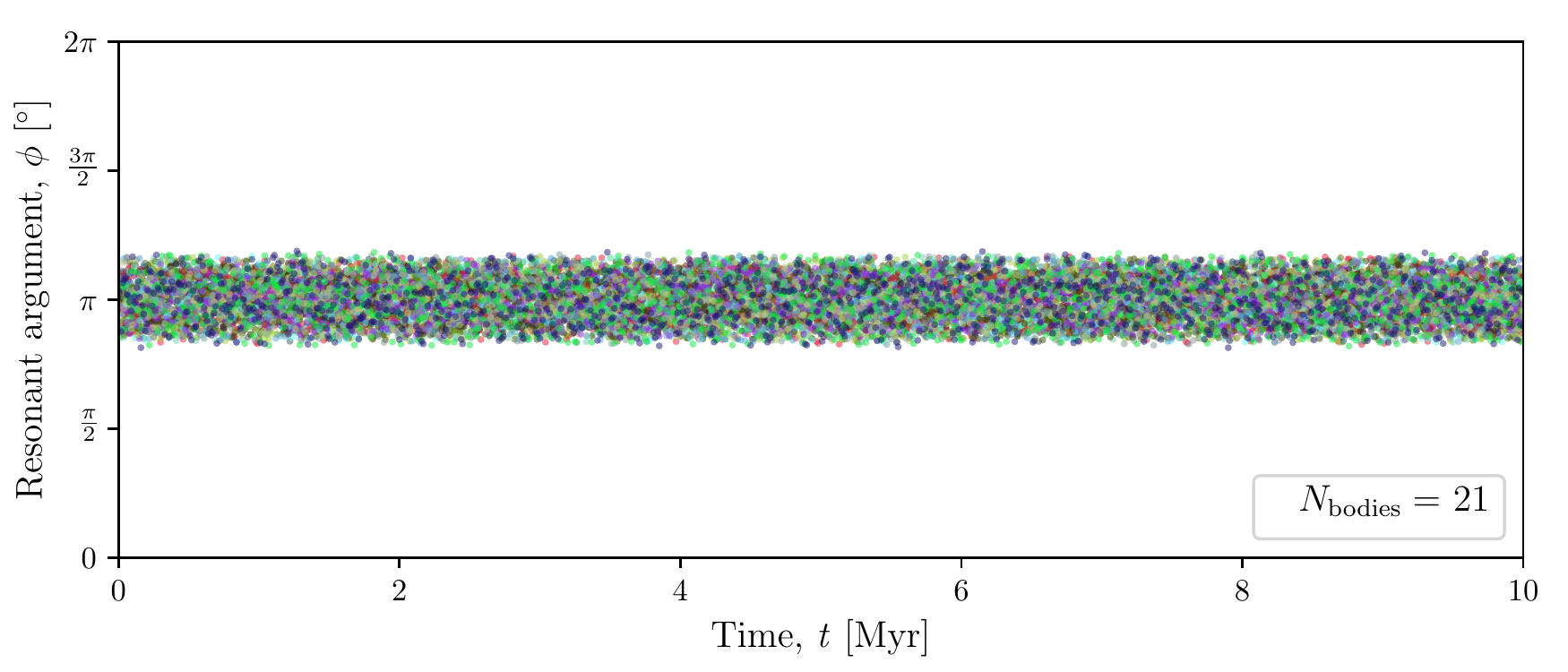}
        \caption{4:3}\label{fig:hd_215497_4_3}
	\end{subfigure}
	
	\begin{subfigure}{\linewidth}
    		\centering
    		\includegraphics[width=\linewidth]{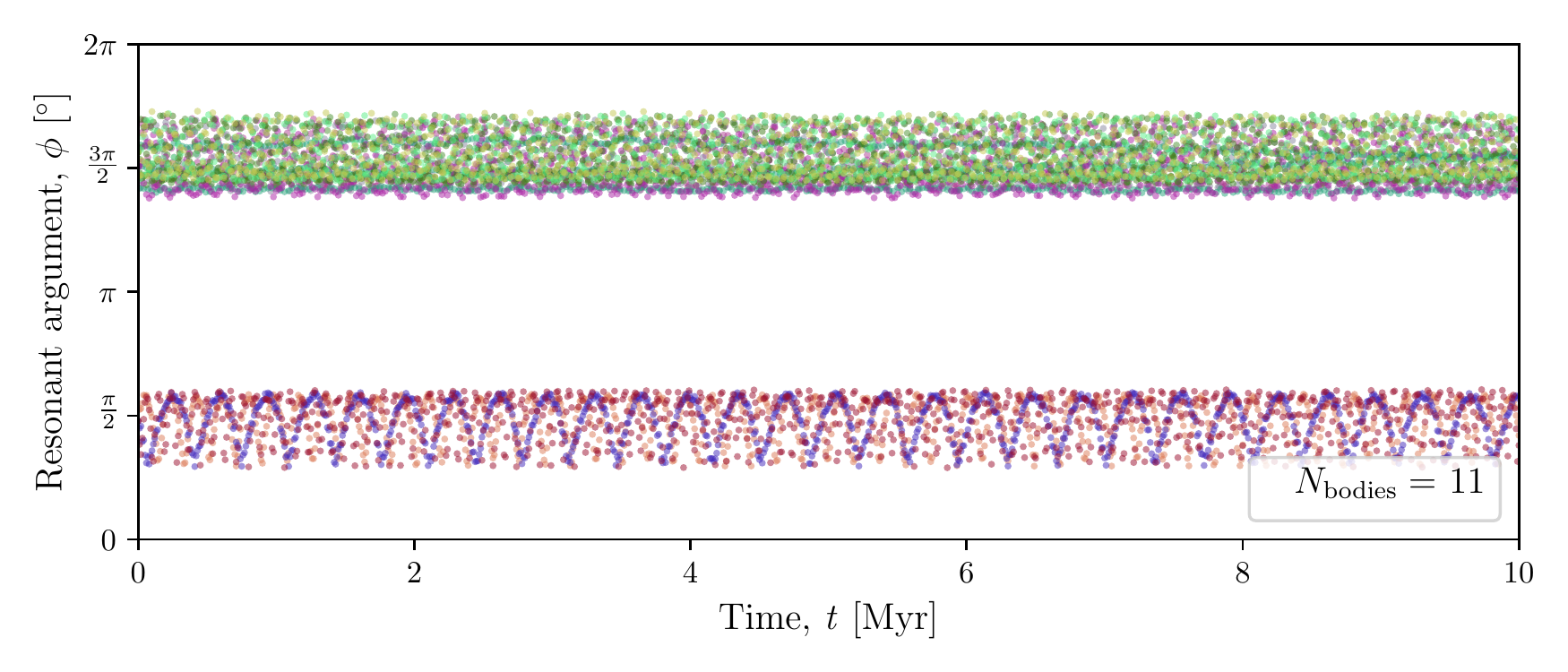}
        \caption{1:1}\label{fig:hd_215497_1_1}
	\end{subfigure}
\caption{The resonant angle, $\phi=(p+q)\lambda'-p\lambda-q\omega'$, against time for all the stable bodies of the $(p+q):p$ MMR in the HD 215497 system. The number of bodies shown in each plot is indicated in the legend, with each $1\ \textrm{M}_{\oplus}$ body stable in its own simulation stacked for these plots. The bound resonant angles demonstrate libration.}\label{fig:hd_215497}
\end{figure}

The stability map and lifetime histogram of HD 215497 demonstrate a beautiful example of MMR stabilisation. We can see that, at each island of stability, there is a corresponding line denoting a MMR with the second planet. However, whilst alignment of a body's semi-major axis with another planet's MMR can be a strong indicator of resonant stabilisation \citep{Agnew2017}, it is critical to determine whether the body is librating within the resonance. This can be achieved by plotting the variation in the resonant argument over the simulation duration. The values of the resonant angle for the $1\ \textrm{M}_{\oplus}$ body was computed dominant MMRs for the HD~215497 system, and the results are shown in Figure~\ref{fig:hd_215497}.

\begin{figure*}
\centering
	\begin{subfigure}{0.31\textwidth}
  		\centering
  		\includegraphics[width=\textwidth]{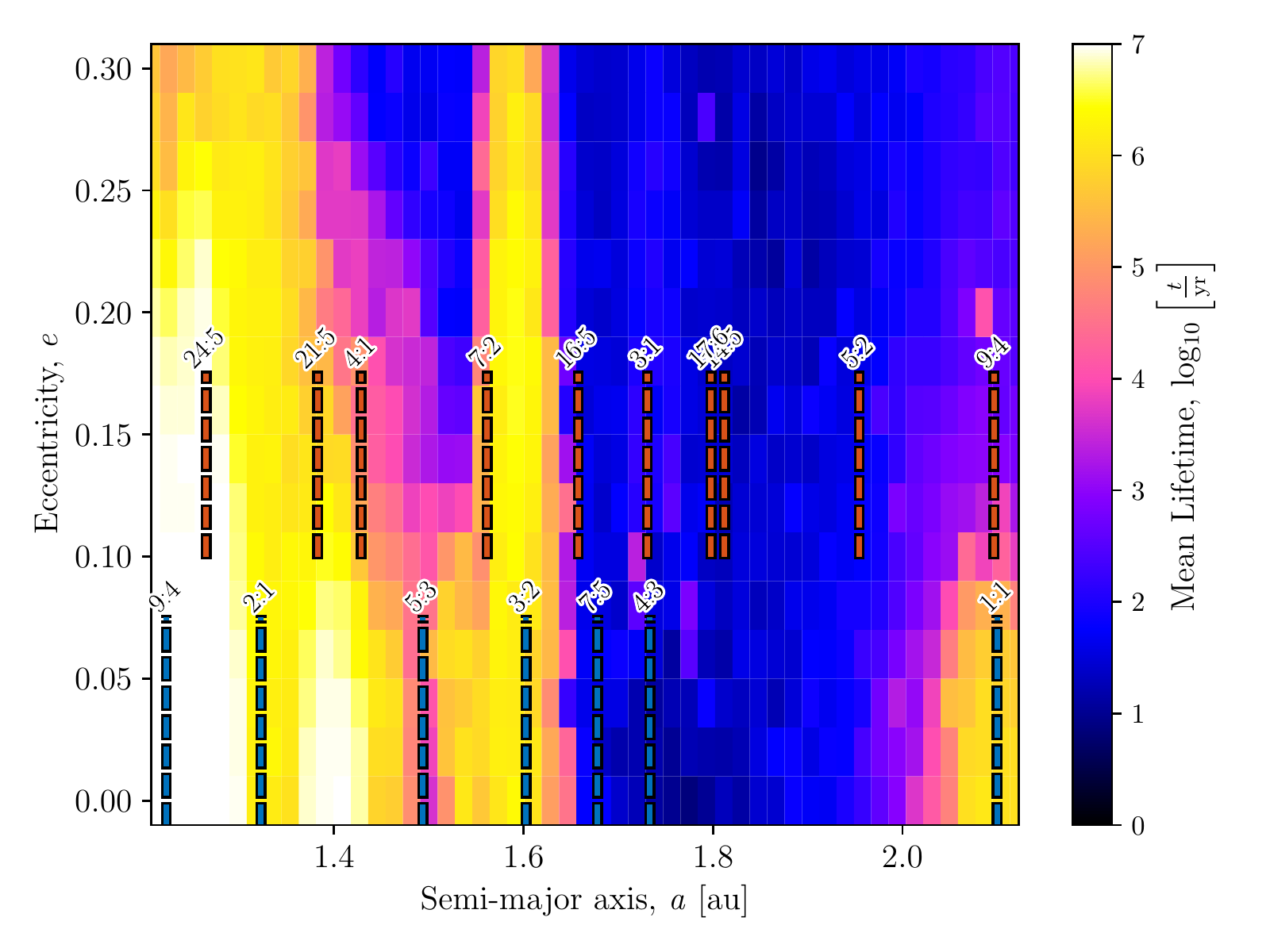}
  		\caption{47 UMa}\label{fig:47uma}
	\end{subfigure}
	\begin{subfigure}{0.31\textwidth}
  		\centering
  		\includegraphics[width=\textwidth]{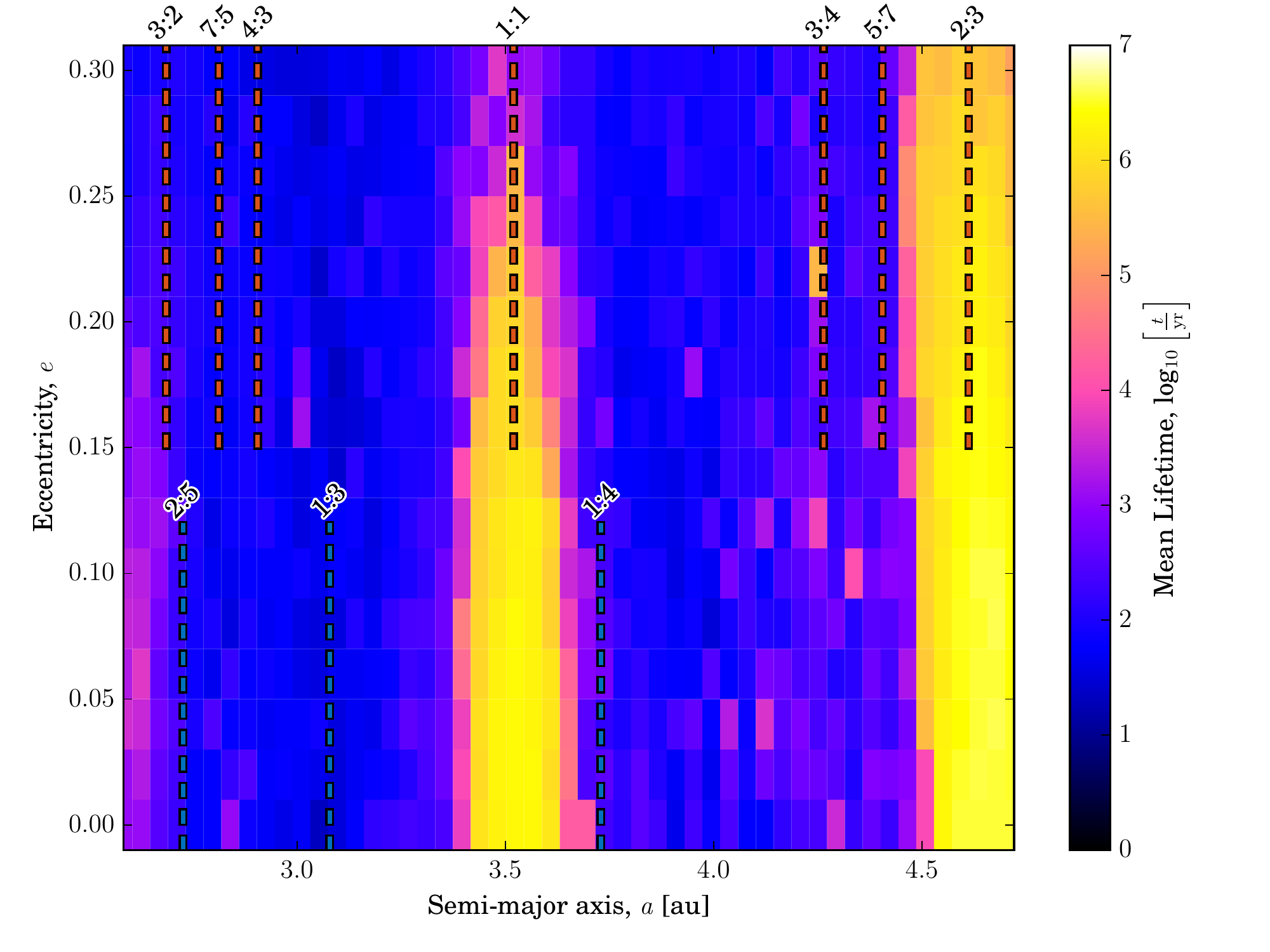}
  		\caption{HD 1605}
	\end{subfigure}
	\begin{subfigure}{0.31\textwidth}
  		\centering
  		\includegraphics[width=\textwidth]{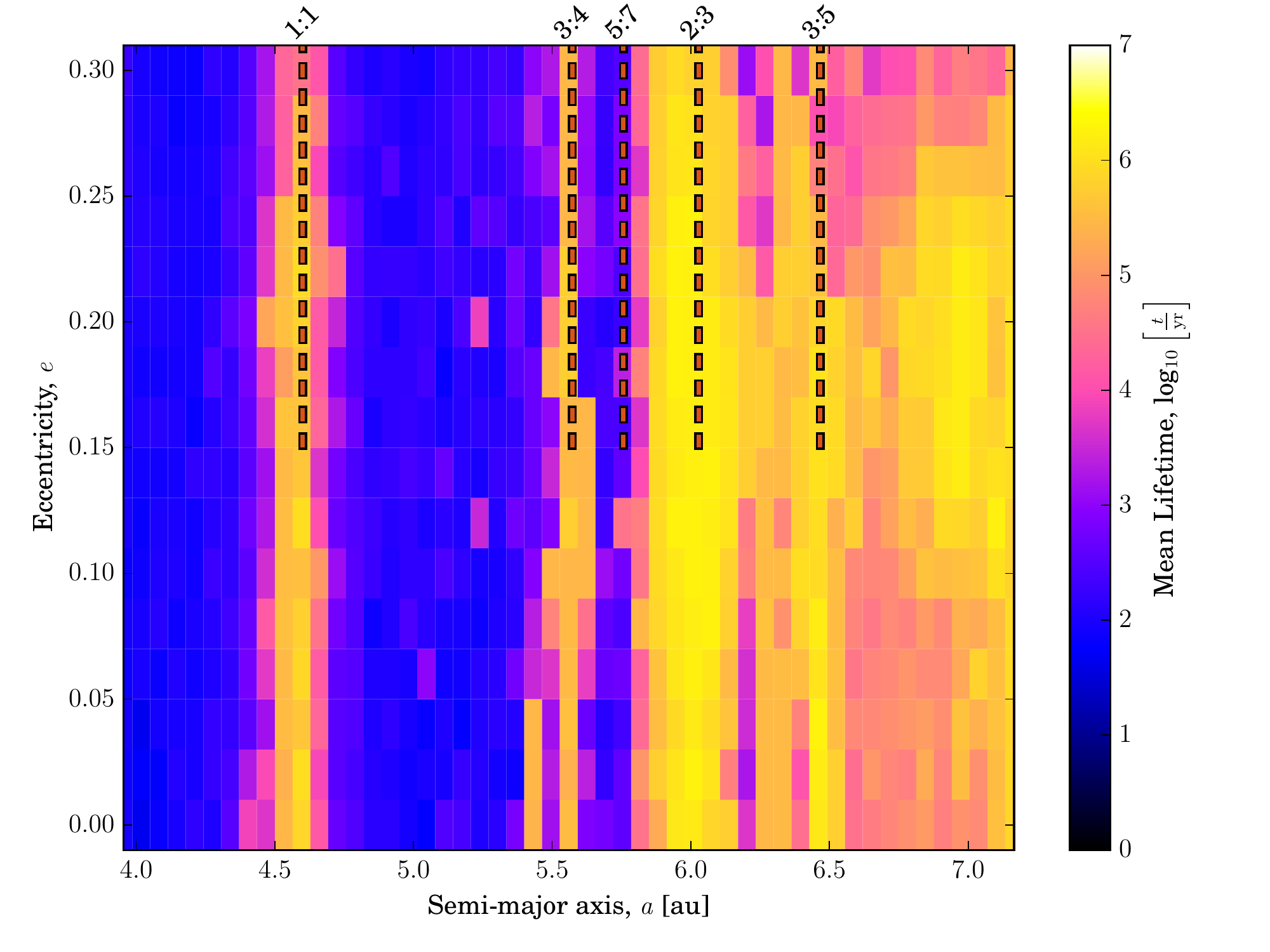}
  		\caption{HD 4732}
	\end{subfigure}
	
	\begin{subfigure}{0.31\textwidth}
  		\centering
  		\includegraphics[width=\textwidth]{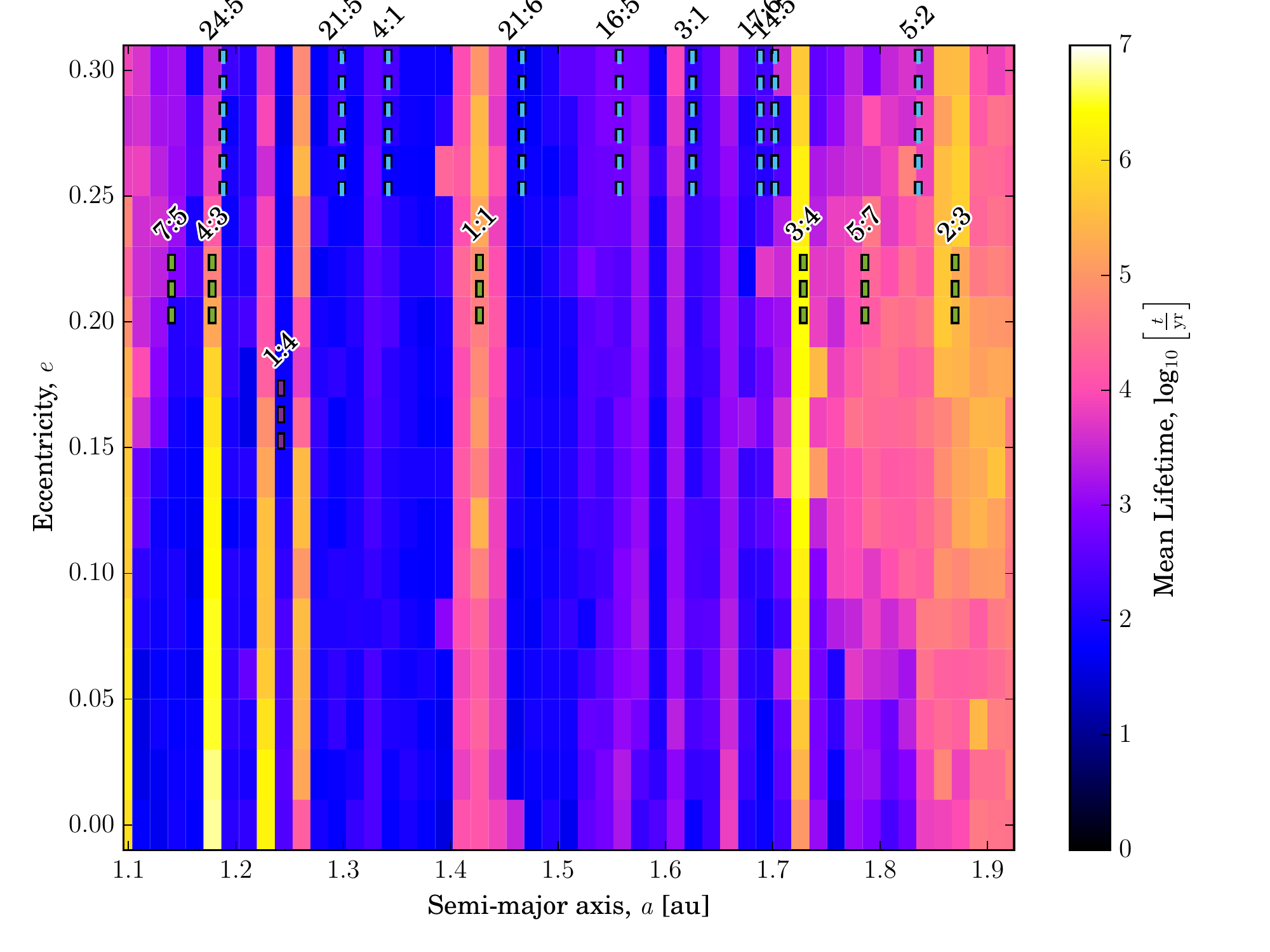}
  		\caption{HD 10180}
	\end{subfigure}
	\begin{subfigure}{0.31\textwidth}
  		\centering
  		\includegraphics[width=\textwidth]{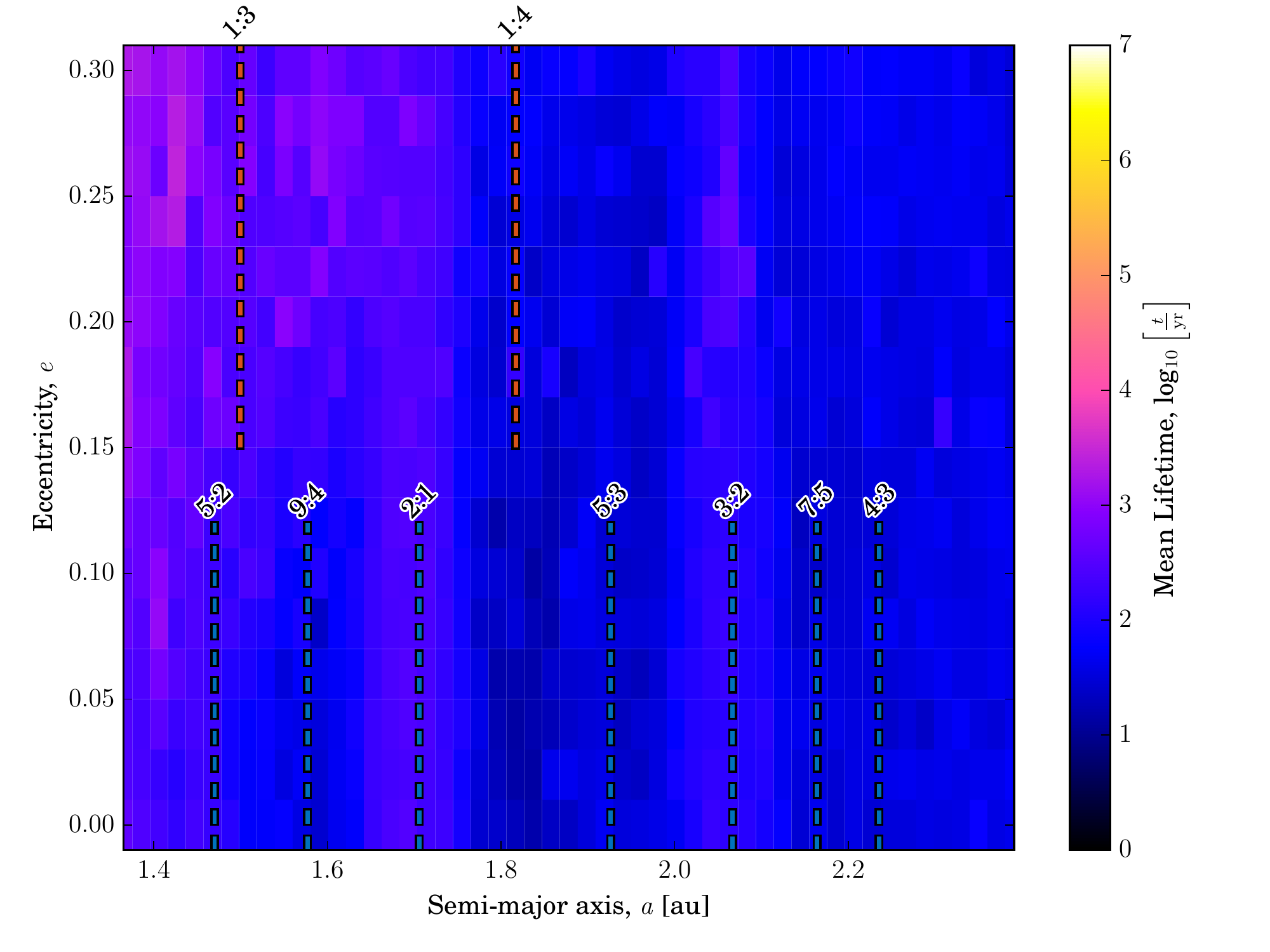}
  		\caption{HD 11506}
	\end{subfigure}
	\begin{subfigure}{0.31\textwidth}
  		\centering
  		\includegraphics[width=\textwidth]{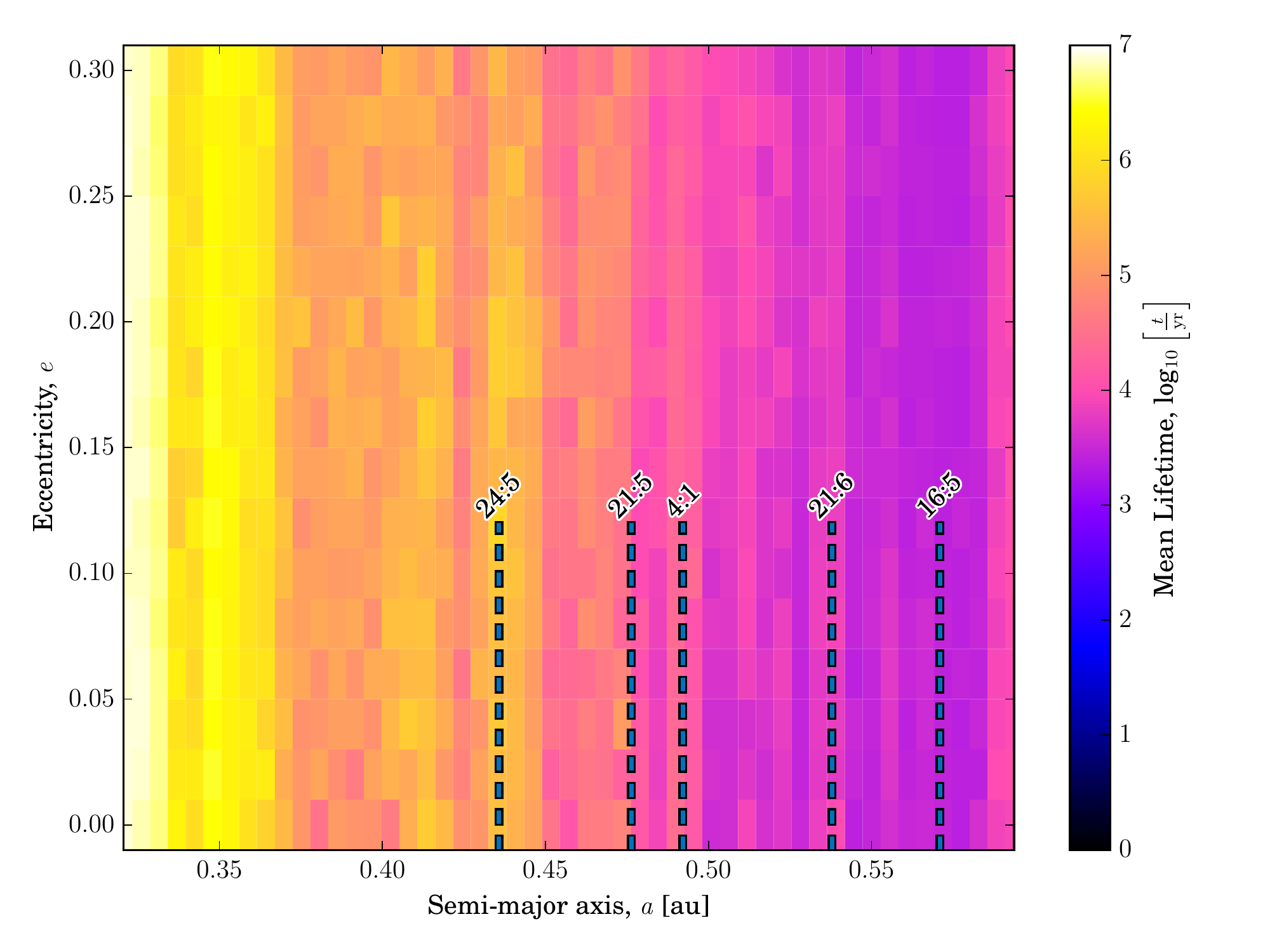}
  		\caption{HD 113538}
	\end{subfigure}
	
	\begin{subfigure}{0.31\textwidth}
  		\centering
  		\includegraphics[width=\textwidth]{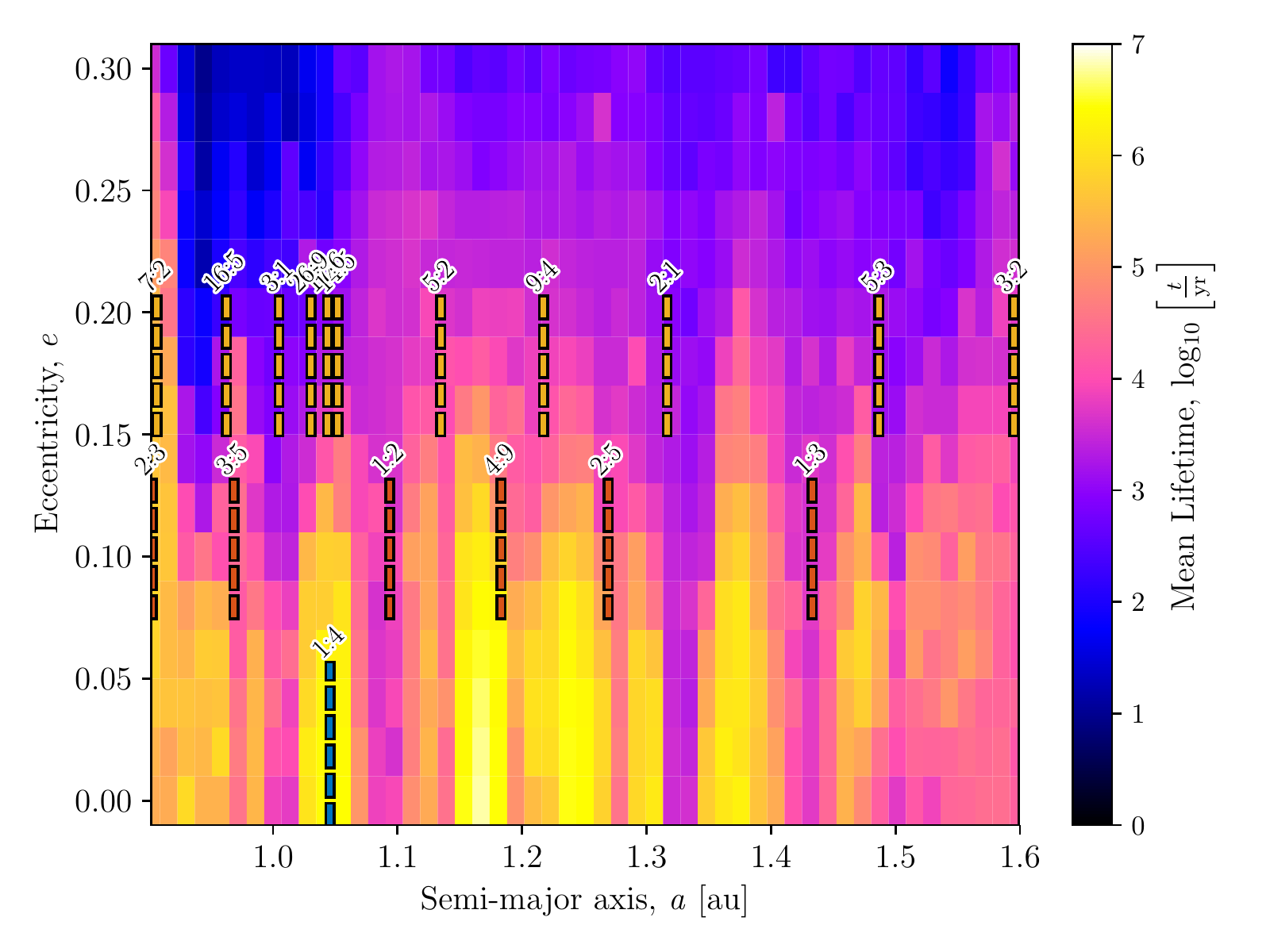}
  		\caption{HD 141399}\label{fig:hd141399}
	\end{subfigure}
	\begin{subfigure}{0.31\textwidth}
  		\centering
  		\includegraphics[width=\textwidth]{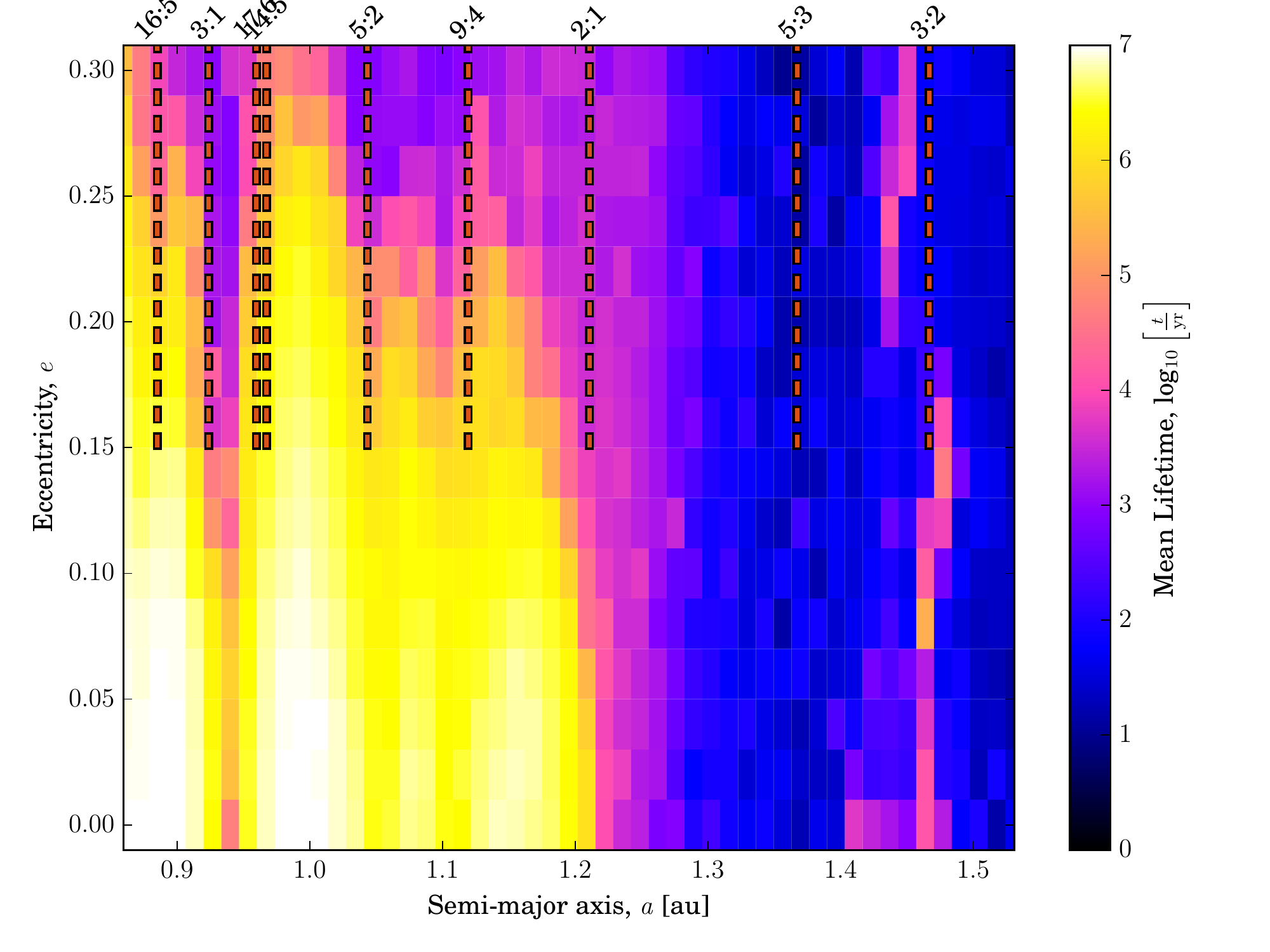}
  		\caption{HD 147018}
	\end{subfigure}
	\begin{subfigure}{0.31\textwidth}
  		\centering
  		\includegraphics[width=\textwidth]{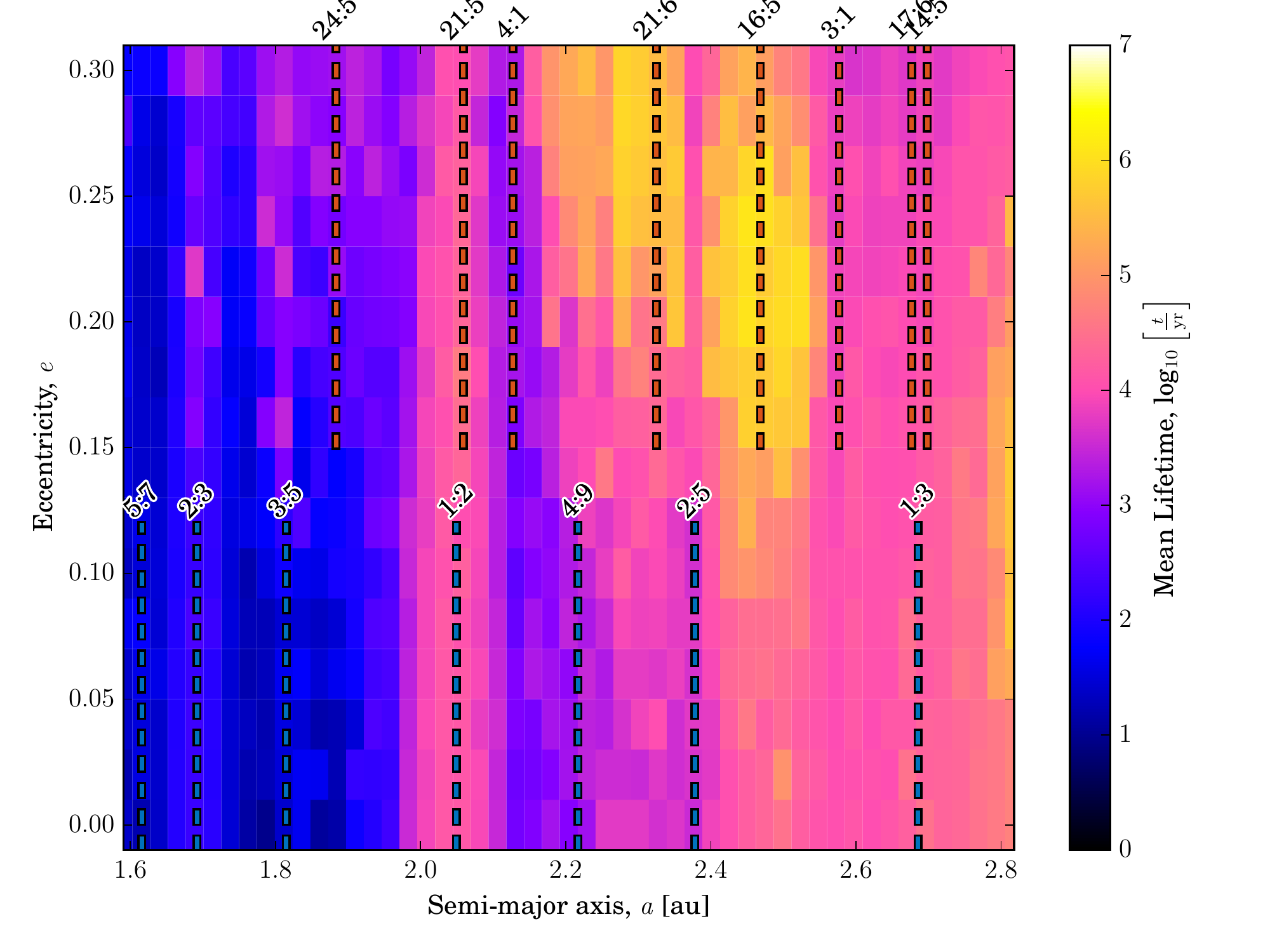}
  		\caption{HD 154857}
	\end{subfigure}
	
	\begin{subfigure}{0.31\textwidth}
  		\centering
  		\includegraphics[width=\textwidth]{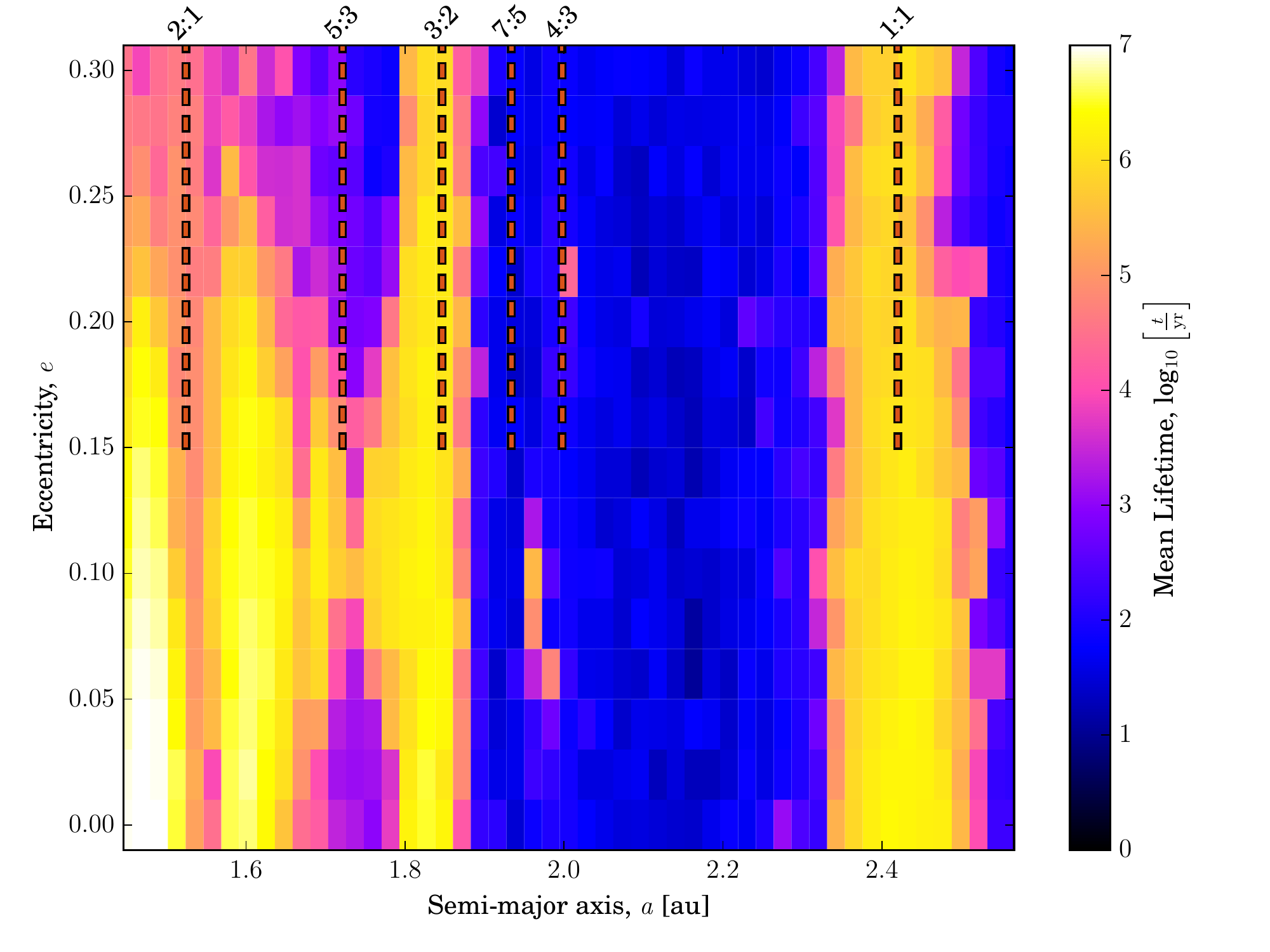}
  		\caption{HD 163607}
	\end{subfigure}
	\begin{subfigure}{0.31\textwidth}
  		\centering
  		\includegraphics[width=\textwidth]{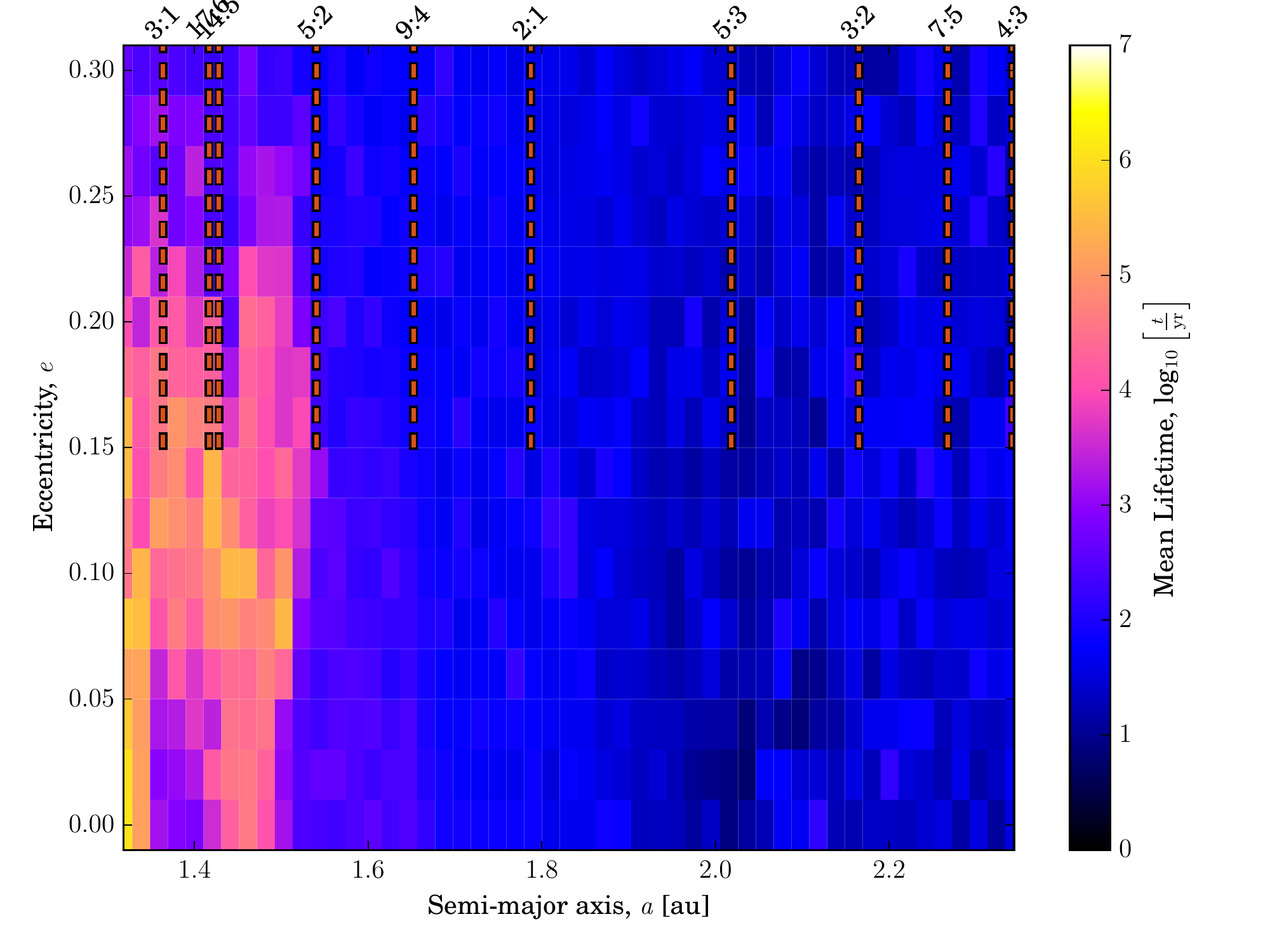}
  		\caption{HD 168443}
	\end{subfigure}
	\begin{subfigure}{0.31\textwidth}
  		\centering
  		\includegraphics[width=\textwidth]{images/HD_215497_map.pdf}
  		\caption{HD 215497}
	\end{subfigure}
	
	\begin{subfigure}{0.31\textwidth}
  		\centering
  		\includegraphics[width=\textwidth]{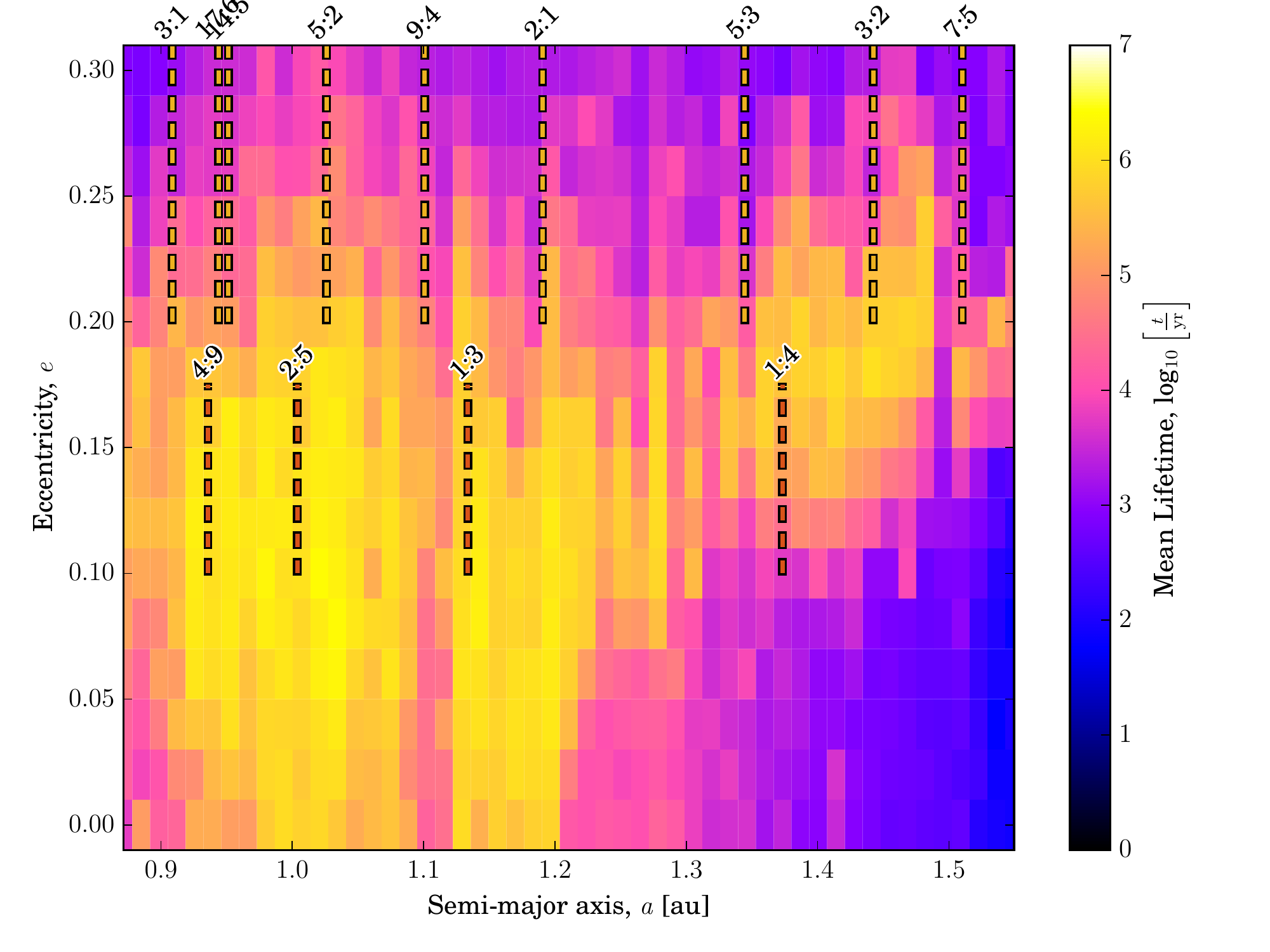}
  		\caption{HIP 14810}
	\end{subfigure}
	\begin{subfigure}{0.31\textwidth}
  		\centering
  		\includegraphics[width=\textwidth]{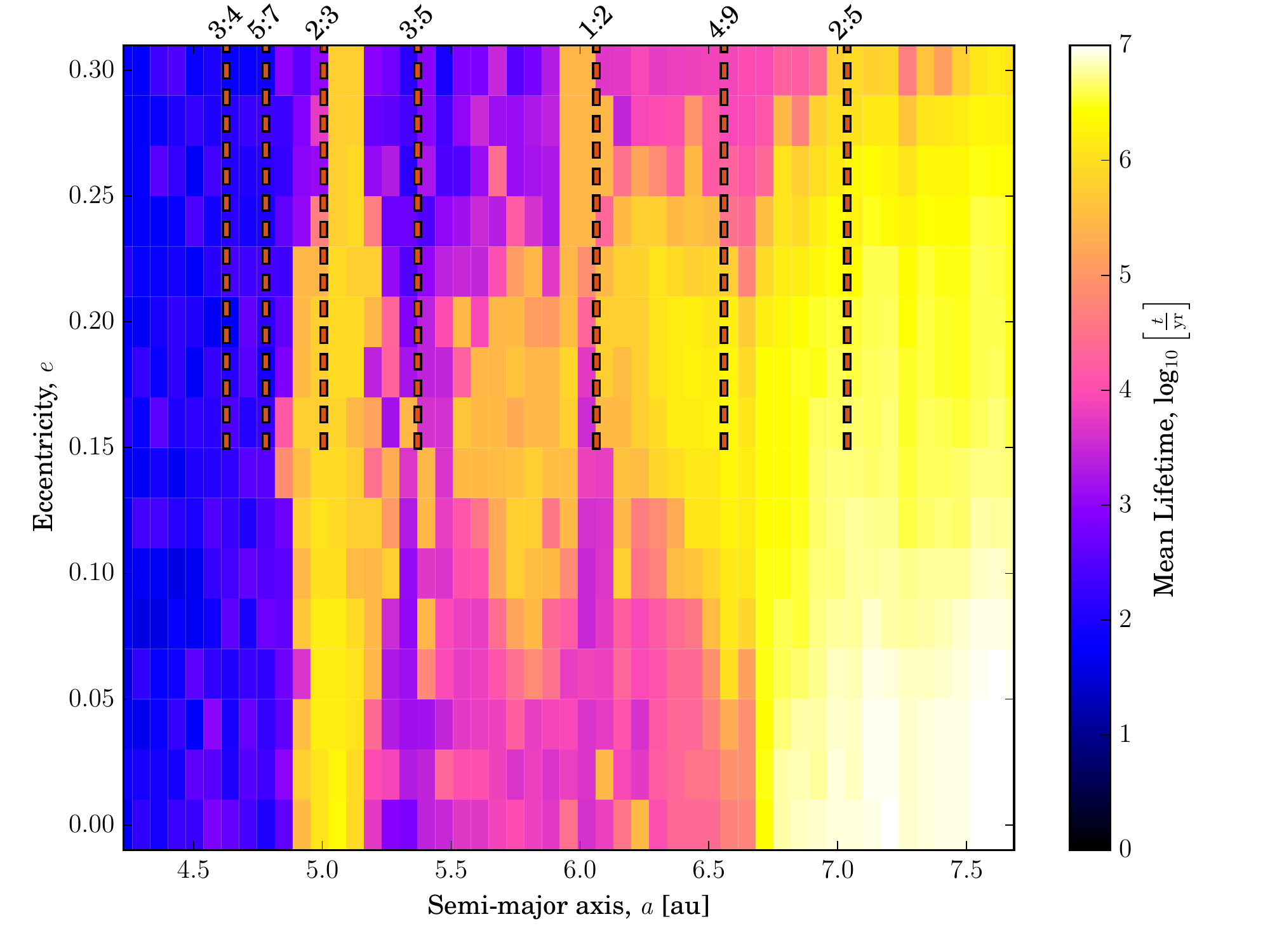}
  		\caption{HIP 67851}
	\end{subfigure}
	\caption{The $1\ \textrm{M}_{\oplus}$ stability maps for the 14 multiple systems with less than $25\%$ survivors in the HZ stability tests.}
	\label{fig:red_maps}
\end{figure*}
Figure~\ref{fig:red_maps} shows the stability maps of all 14 systems of interest (those with less than $25\%$ survivors in the HZ stability simulations). HD~11506 highlights the importance of these massive body simulations, as this system has surviving massless TPs in its HZ, but no surviving massive bodies when mutual gravitational interactions are considered by using a $1\ \textrm{M}_{\oplus}$ body. The rest of the simulations show varying levels of stability, some like HD~10180 and HD~215497 showing particularly narrow MMR stabilised bands, while others such as HD~147018 and HD~113538 show wider unperturbed regions.

We notice is that, in general, the MMR stabilised bands are due to interactions with only one planet. While several systems have more than one planet close enough to the HZ that several of their MMRs can be found in the HZ, it is typically the case that the planet closest to the HZ dominates over the others when it comes to resonant stabilisation. The 1:1 resonance proves particularly strong in stabilising bodies around the L4 and L5 Lagrangian points (in other words as Trojan companions to the massive planet), but as noted by \cite{Agnew2018}, planets sharing an orbit in this manner represents a degenerate scenario for the radial velocity (RV) signal. This is perhaps not surprising, given that lower-order resonances are typically stronger than their higher order cousins, and that both Jupiter and Neptune host significant Trojan populations within the Solar system. However, we note that this rule of thumb would most likely break down when the planet nearer to the HZ is sufficiently less massive than those further from it. 

\subsubsection{Multiple Planet Interactions}
Our sample yields a wide variety of outcomes, from systems that have larger, unperturbed, stable regions within the HZ, to others with small ``islands'' of stability. For several of these stable islands, as demonstrated with HD~215497, the stabilisation is the result of mean motion resonant interactions with a single planet. However, it is also possible for complicated dynamical behaviours such as three-body resonances and resonant chains to create stable systems \citep[e.g.][]{Gallardo2014,Gallardo2016,Mills2016,Luger2017,Delisle2017}.
Figure~\ref{fig:red_maps} demonstrates several systems where complex dynamical behaviour is observed. Two such systems are 47~UMa (Fig~\ref{fig:47uma}) and HD~141399 (Fig~\ref{fig:hd141399}). 

\begin{figure}	
	\begin{subfigure}{\linewidth}
    		\centering
    		\includegraphics[width=\linewidth]{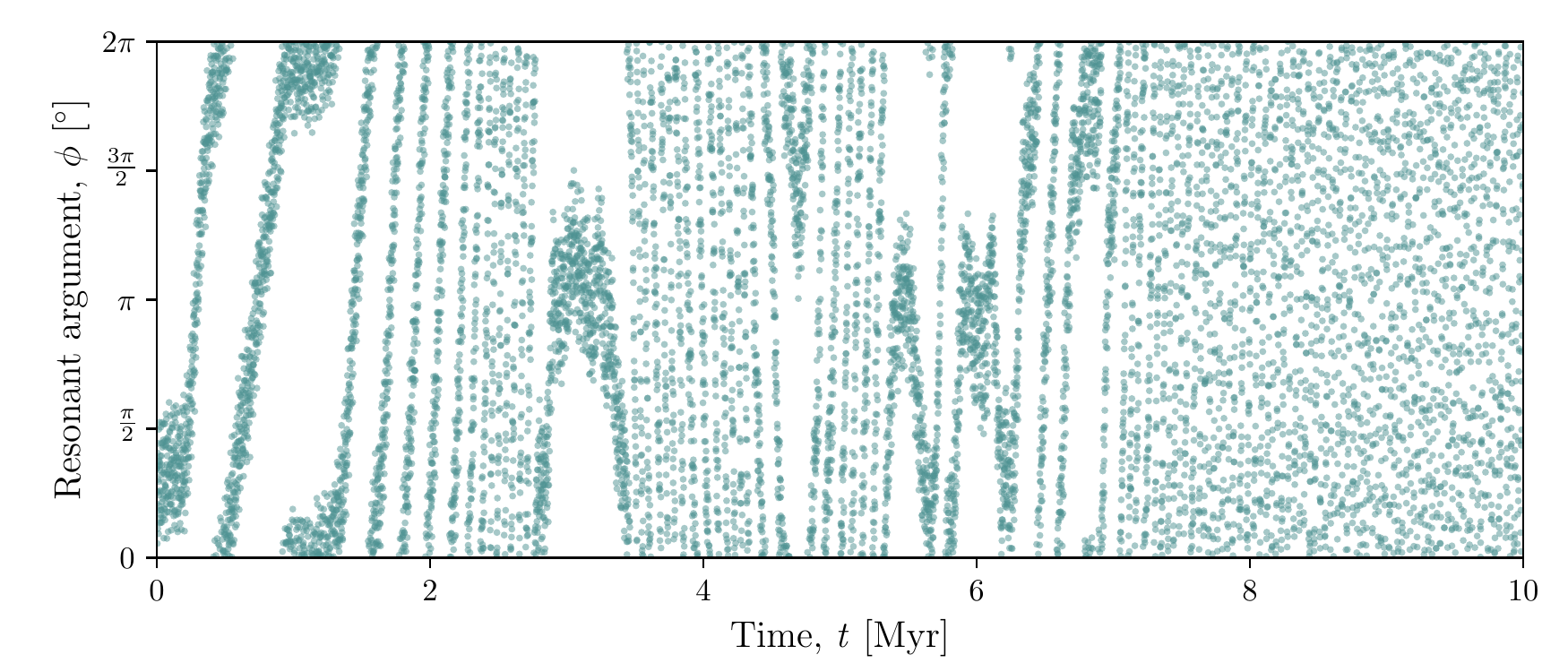}
        \caption{}\label{fig:sec_1}
	\end{subfigure}
	
	\begin{subfigure}{\linewidth}
    		\centering
    		\includegraphics[width=\linewidth]{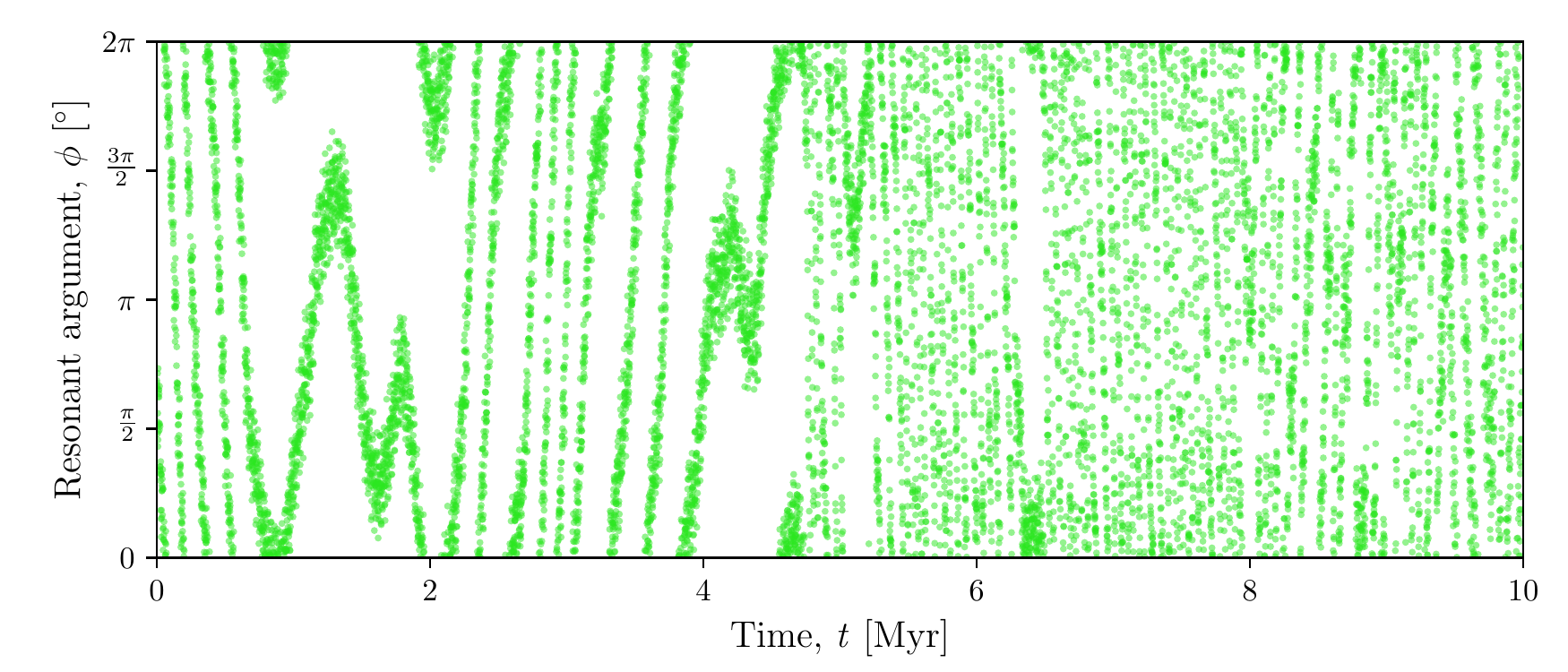}
        \caption{}\label{fig:sec_2}
	\end{subfigure}
	
	\begin{subfigure}{\linewidth}
    		\centering
    		\includegraphics[width=\linewidth]{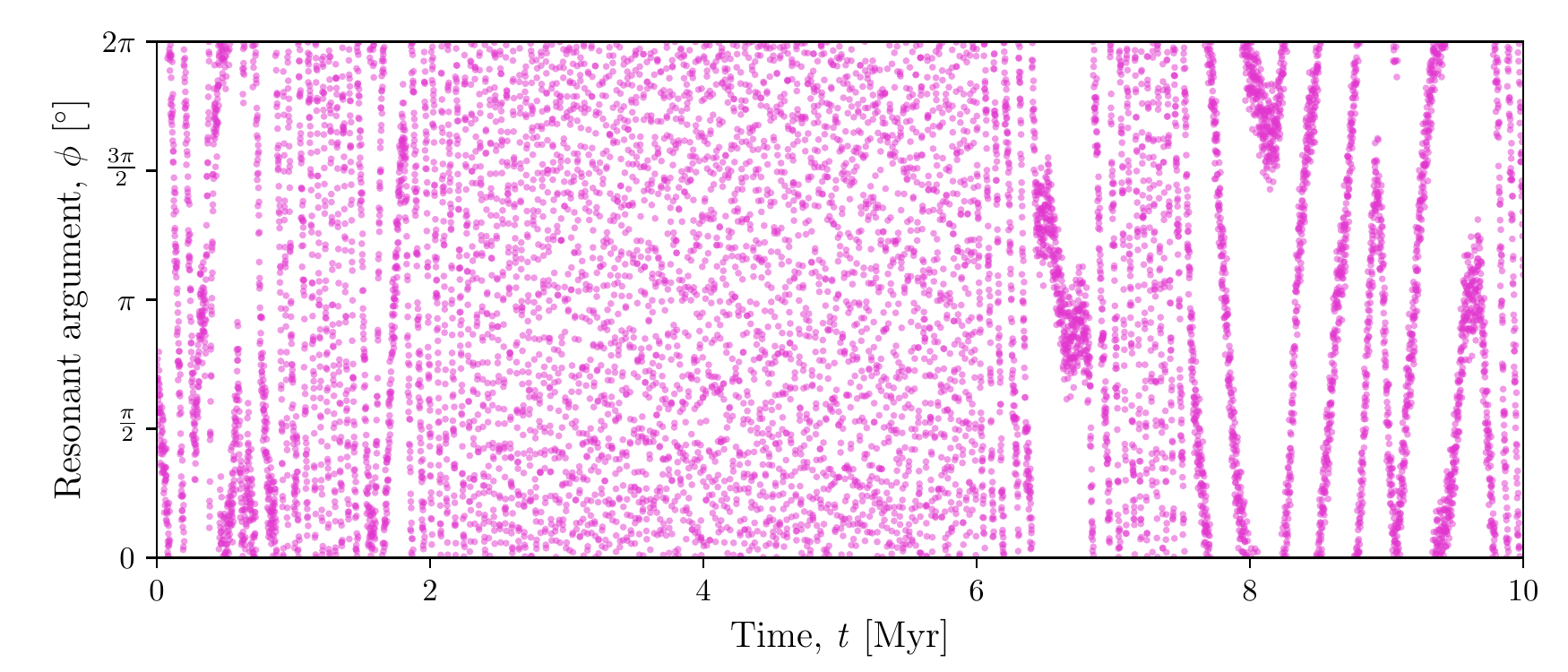}
        \caption{}\label{fig:sec_3}
	\end{subfigure}
\caption{Examples of different body's resonant angles at the 3:2 MMR with 47~UMa~b, i.e. $\phi=3\lambda'-2\lambda-\omega'$. These plots demonstrate the resonant angle structure is perturbed on secular timescales in 47~UMa. In some cases, the bodies experience periods of bound librations for up to $\sim1$~Myr (e.g. between $1$~Myr and $2$~Myr in Fig.~\ref{fig:sec_2}), while in other cases the bodies experience periods of circulation (e.g. between $0$~Myr and $2.5$~Myr in Fig.~\ref{fig:sec_1}). The body also alternates between bound MMR behaviour and circulation and completely unbound, chaotic behaviour.}\label{fig:sec}
\end{figure}

47~UMa has been shown previously to possess stable orbits in the HZ that are the result of stabilising resonances \citep{Laughlin2002,Ji2005}. Our simulations show overlap of some of the lower order MMRs, for example the 3:2 MMR of 47~UMa~b (innermost planet; lower blue-line in Fig.~\ref{fig:47uma}) with the 7:2 MMR of 47~UMa~c (second planet; middle red-lines in Fig.~\ref{fig:47uma}). Plotting the resonant argument of some of these stable $1\ \textrm{M}_{\oplus}$ bodies, Figure~\ref{fig:sec} shows that they do not solely librate about the 3:2 MMR of 47~UMa~b. Instead, there is some very clear structure that shows both circulation, as well as periods of libration (or bound resonant angles) that eventually ``drift'' over much longer timescales to a non-librating (or unbound resonant angle) state. The timescales of $10^{4}-10^{6}$~yr suggests that secular interactions are driving the bodies out of, or into, the MMR. Figure~\ref{fig:sec_1} demonstrates both circulation, with the resonant angle circulating for the first $2$~Myr, and short-term resonant behaviour, with the body librating between $\sim3-3.75$~Myr. Figure~\ref{fig:sec_2} shows a body moving between states of circulation (e.g. the unbound resonant angle regions between $\sim0-1$~Myr) to transient states of resonant behaviour (e.g. the bound resonant angle regions between $\sim1-2$~Myr). Similarly, Figure~\ref{fig:sec_3} shows a long period of non-resonant behaviour (between $\sim2-6$~Myr) followed by resonant behaviour (between $\sim6.25-7$~Myr). These periods of circulation and transient libration with low order resonances can occur for periods of $\sim1$ Myr. 

\begin{figure}	
	\begin{subfigure}{\linewidth}
    		\centering
    		\includegraphics[width=\linewidth]{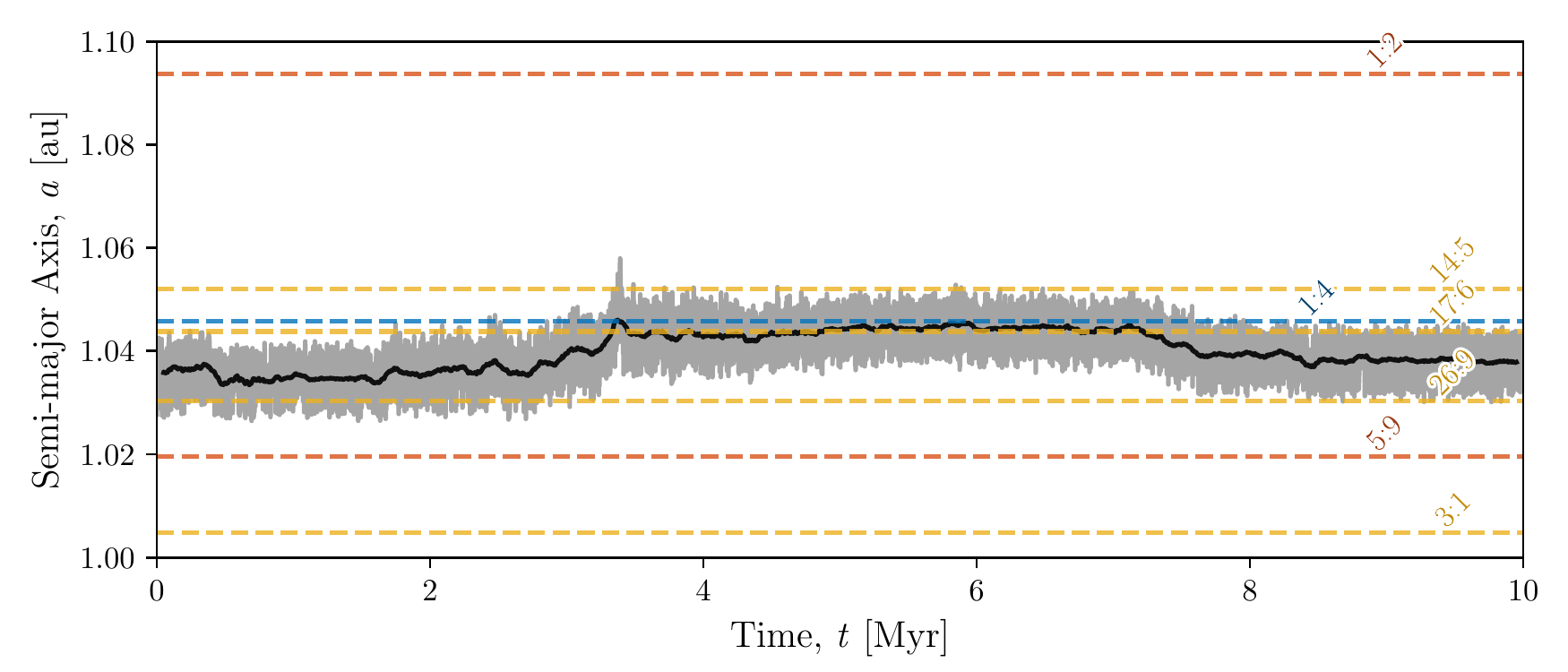}
        \caption{}\label{fig:sticking_1}
	\end{subfigure}
	
	\begin{subfigure}{\linewidth}
    		\centering
    		\includegraphics[width=\linewidth]{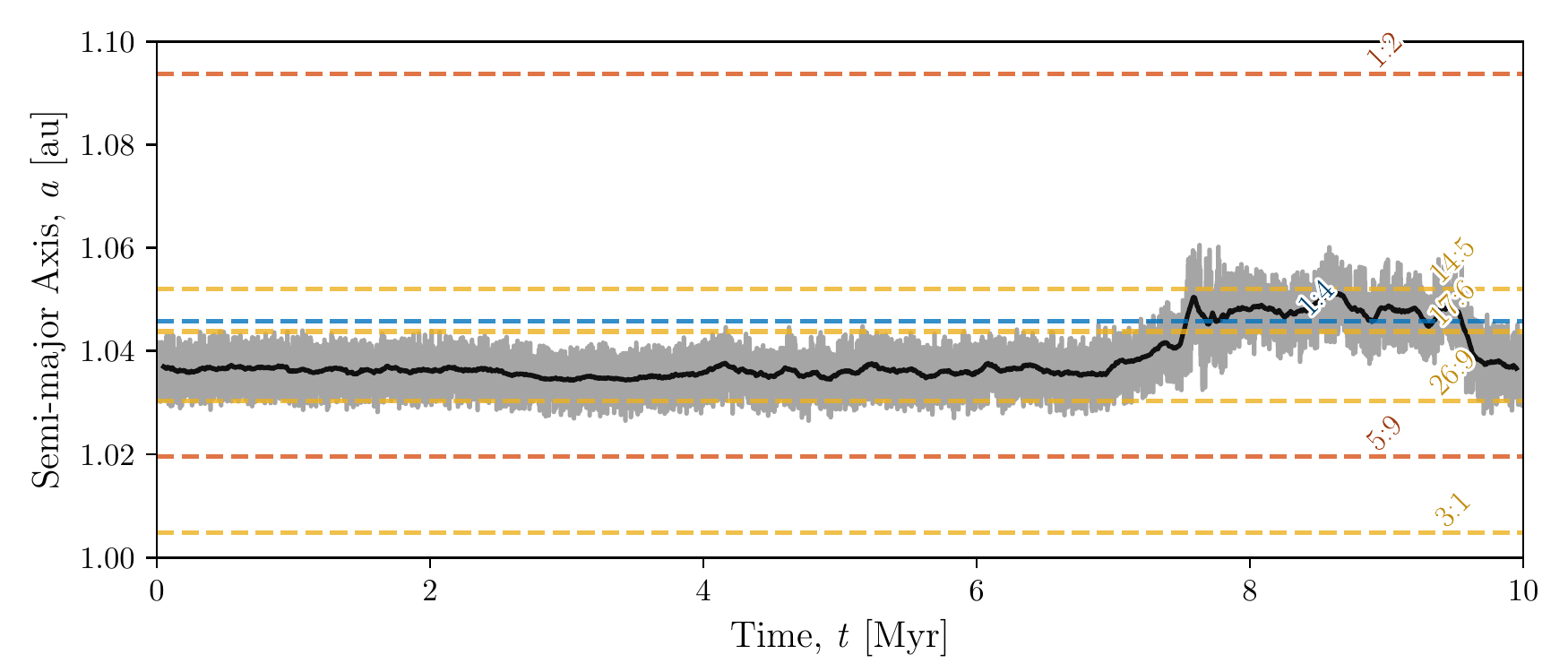}
        \caption{}\label{fig:sticking_2}
	\end{subfigure}
\caption{Two examples of resonant hopping between nearby MMRs around $1.0$~au in HD~141399. The grey line shows the semi-major axis of the a stable $1\ \textrm{M}_{\oplus}$ body, while the black line shows a moving average to more clearly show what semi-major axis value the body is stuck at. The coloured dashed lines indicate the MMRs of the different planets in HD~141399, from the innermost (blue) planet to the outermost (yellow) planet.}\label{fig:sticking}
\end{figure}

Focussing on HD~141399, rather than bodies falling into a transient resonant state with a low order MMR that gradually moves to an unbound resonant angle state, the bodies engage in so-called ``resonance-hopping'', essentially `jumping' between two or more higher order resonances as is evident, and observable, in our own Solar system \citep[e.g.][]{Lykawka2007,Bailey2009,Wood2017}. In the case of HD~141399, the 1:4 MMR of HD~141399~b (innermost planet; lower blue lines in Fig.~\ref{fig:hd141399}) lines up with the tightly packed 14:5, 17:6 and 26:9 MMRs of HD~141399~d (third planet; upper yellow lines in Fig.~\ref{fig:hd141399} This resonant hopping is shown in Figure~\ref{fig:sticking} where it can be seen that the $1\ \textrm{M}_{\oplus}$ bodies jump from one semi-major axis that aligns with an MMR ratio to another (and back again) as the high order resonances are not strong enough to dominate over one another. Further more, in this scenario, HD~141399~b is less massive than HD~141399~d by a factor of 3, and so this also provides a means for the weaker, higher order MMRs to prevent the stronger, lower order MMR from dominating. This ultimately results in the resonant angle of the bodies not librating at any particular MMR.

\subsection{Searching for Exo-Earths in Multiple Jupiter Systems}
The primary goal of our work is to produce a list of candidate systems that could potentially contain Earth-mass planets in their HZ and be detectable with current or near-future instruments. The results presented in Sections~\ref{subsec:tp_results} and \ref{subsec:mass_results} show where a potential exo-Earth could exist on a dynamically stable orbit for a total of 29 multiple planet systems. Of these, 15 feature almost entirely stable HZs, whilst the other 14 show gravitationally stirred HZs that contain regions of stability. Here, we estimate the strength of the radial velocity signal such an exo-Earth would induce on its host star.

We use the equation for the radial velocity semi-amplitude, $K$, where
\begin{align} 
\label{eqn:doppler_shift}
	K &= \left( \frac{2\pi G}{T_{\oplus}} \right)^{\frac{1}{3}} \frac{\textrm{M}_{\oplus}\sin{I}}{\left( M_\star + \textrm{M}_{\oplus} \right)^{\frac{2}{3}}} \frac{1}{\sqrt{1-e_{\oplus}^2}} \, .
\end{align} 
Here, $G$ is the gravitational constant, $M_\star$ is the mass of the host star, $I$ is the inclination of the planet's orbit with respect to our line of sight, and $T_{\oplus}$, $e_{\oplus}$ and $\textrm{M}_{\oplus}$ are the period, mass and eccentricity of the hypothetical exo-Earth. For our calculations, we retain our previously established assumption that the system is co-planar (i.e. $i=0\degr$), and also assume the most optimistic inclination with respect to our line of sight, that is, $I=90\degr$. We note that for shallow orbital inclinations of $5 \degr$ (i.e. $I=85\degr$ or $95\degr$) we would see a decrease in signal strength of $<1\%$.

We produce two candidate lists: one for those that we simulated with a $1 \textrm{M}_{\oplus}$ body (in the 14 perturbed HZ systems), and one for those that we only simulated with massless TPs (in the 15 systems with largely unperturbed HZs).

For the massless TP candidate list, we calculate the value for $K$ that would result from the presence of a $1 \textrm{M}_{\oplus}$, $2 \textrm{M}_{\oplus}$ and $4 \textrm{M}_{\oplus}$ exo-Earth. As these 15 systems have unperturbed or largely unperturbed HZs, it is not unreasonable to compute the predicted radial velocity signals of such exo-Earths as the gravitational strength of the known massive planet (which is of order a Jovian mass) is the critical factor when assessing HZ stability. 

It should be noted that while the HZ boundaries will indeed shift slightly as a function of exo-Earth mass, as mentioned in Section~\ref{subsec:predicting_stable_regions}, these small variations ($<5\%$ change in the width of the HZ between a $1 \textrm{M}_{\oplus}$ and $4 \textrm{M}_{\oplus}$ exo-Earth) relative to the semi-major axis of the exo-Earth are considered to be negligible for the purposes of this predictive exercise.

Exo-Earths are particularly challenging to detect due to their small size and mass. For those that orbit Sun-like stars, these problems are exacerbated by the fact that HZ planets would have orbital periods of approximately $1 \mathrm{yr}$. At present, such planets are essentially undetectable with current radial velocity instruments, which have a limit of around $1 \textrm{ m s}^{-1}$ \citep{Dumusque2012b,Swift2015}. However, the ESPRESSO (\textbf{E}chelle \textbf{SP}ectrograph for \textbf{R}ocky \textbf{E}xoplanet and \textbf{S}table \textbf{S}pectroscopic \textbf{O}bservations) spectrograph \citep{Hernandez2017}, the latest instrument available on the ESO Very Large Telescope to resolve Doppler shifts, has the goal of achieving a resolution as low as $0.1 \textrm{ m s}^{-1}$ \citep{Pepe2014}. At such resolution, it should be possible to detect exo-Earths around Sun-like stars. Looking further into the future, the proposed CODEX (\textbf{CO}smic \textbf{D}ynamics and \textbf{EX}o-earth experiment) spectrograph for the European Extremely Large Telescope is expected to deliver resolutions as low as \~0.01 $\textrm{m s}^{-1}$ \citep{Pasquini2010}. Although such high resolution offers great promise for the search for Earth-like worlds, these discoveries will remain challenging especially when considering the imposed noise due to the activity of the host star that could otherwise result in false positives \citep{Robertson2014}.

\begin{figure}
	\begin{subfigure}{\linewidth}
		\centering
		\includegraphics[width=\linewidth,height=4.5cm]{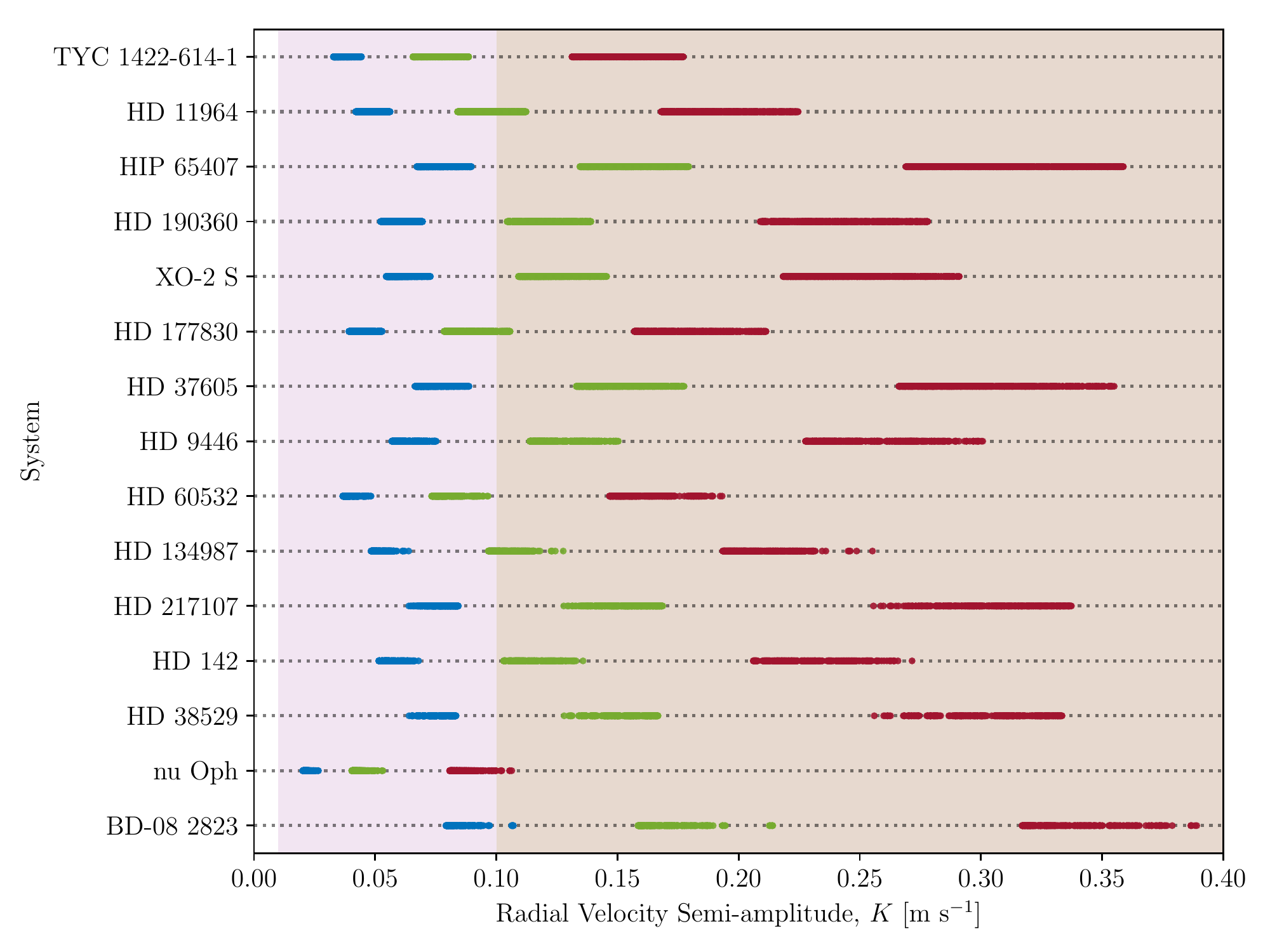}
		\caption{TP predictions of the HZ stability tests}
		\label{fig:massive_preds}
	\end{subfigure}
	\begin{subfigure}{\linewidth}
		\centering
		\includegraphics[width=\linewidth,height=4.5cm]{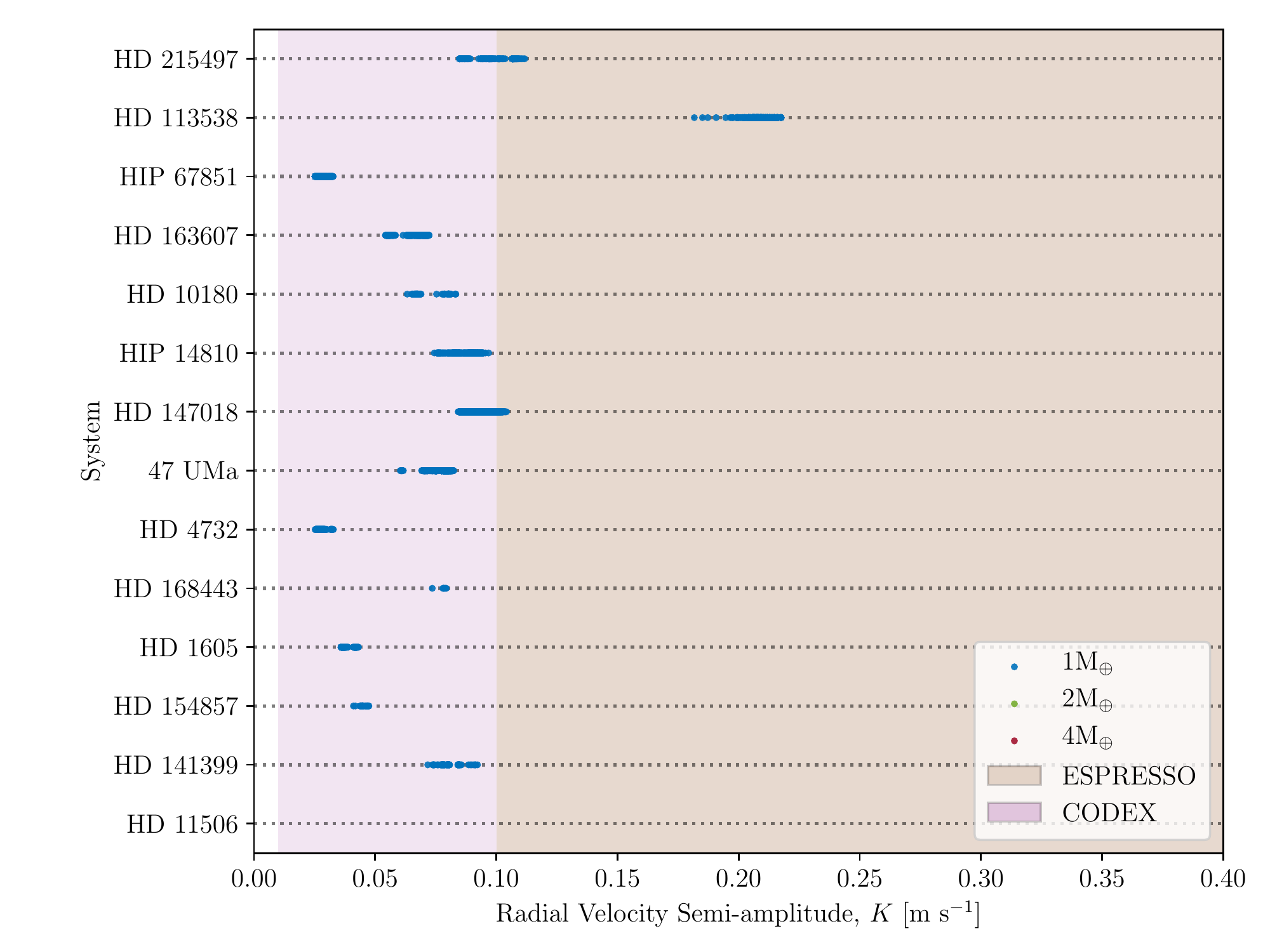}
		\caption{$1\ \textrm{M}_{\oplus}$ predictions of the stirred HZ tests}
		\label{fig:tp_preds}
	\end{subfigure}
	\caption{The semi-amplitude of Doppler wobble induced on all multiple systems that were found to be able to maintain an exo-Earth on stable orbits within their HZ. At a given semi-major axis, the strength of the signal can be computed using equation~\ref{eqn:doppler_shift}. We show the induced radial velocity wobble for a $1\ \textrm{M}_{\oplus}$ (blue), $2\ \textrm{M}_{\oplus}$ (green) and $4\ \textrm{M}_{\oplus}$ (red) exo-Earth. The brown  and pink shaded regions indicate the detection limits of the ESPRESSO ($0.1 \textrm{ m s}^{-1}$) and future CODEX ($0.01 \textrm{ m s}^{-1}$) spectrographs respectively.}
	\label{fig:preds}
\end{figure}

For each body that remains stable for the duration of the simulation, we use equation~\ref{eqn:doppler_shift} to compute the induced radial velocity signal, $K$. This yields several values for $K$ that  depend on the orbital parameters of each surviving body (massless TPs for $> 25\%$ survivor systems or massive bodies for the $< 25\%$ survivor systems). We compare these with the detection limits of the ESPRESSO and CODEX instruments. The induced radial velocity signals of all systems for the three exo-Earth masses of $1 \textrm{M}_{\oplus}$, $2 \textrm{M}_{\oplus}$ and $4 \textrm{M}_{\oplus}$ are shown in Figure~\ref{fig:preds}.

In those figures, the brown and pink shaded regions correspond with the detection limits of ESPRESSO and CODEX respectively. These regions indicate that for a particular exo-Earth mass ($1 \textrm{M}_{\oplus}$, $2 \textrm{M}_{\oplus}$ or $4 \textrm{M}_{\oplus}$), if all points lie within the brown region, then the exo-Earth will be detectable by ESPRESSO \textit{if it exists}. An example is a $4 \textrm{M}_{\oplus}$ exo-Earth (red points) in the HZ of the HD 60532 system. These systems should be a priority for ESPRESSO. Conversely, if all points lie within the pink region, then the exo-Earth would be beyond the detection limit of ESPRESSO, but would be detectable by CODEX \textit{if it exists}. One such example is a $1 \textrm{M}_{\oplus}$ exo-Earth (blue points) in the HZ of the HIP 65407 system. In between these two extremes are systems that straddle both regions. For example, see the case of a $2 \textrm{M}_{\oplus}$ exo-Earth (green points) in the HZ of the HD 11964 system. In such systems, an exo-Earth could exist in a stable orbit within the brown region (i.e. within the detection limit of ESPRESSO) or within the pink region (i.e. beyond the detection limit of ESPRESSO). These systems should be a second priority for ESPRESSO, as a non-detection means the exo-Earth may still exist but is located further from the star such that the induced radial velocity signal is too small to be detected with ESPRESSO.

As the signal is dependent on the host star's mass, some systems will be too challenging to detect a $1 \textrm{M}_{\oplus}$ planet, but detection of a $2 \textrm{M}_{\oplus}$ or $4 \textrm{M}_{\oplus}$ may be possible. The list of systems is summarised in Table~\ref{tab:rv_mins}. Of particular interest is HD~113538 which not only has two gas giants beyond the HZ, just as we see in our own Solar system, but is also the only system that can maintain a $1 \textrm{M}_{\oplus}$ exo-Earth on a stable orbit within its HZ that would also induce a detectable radial velocity signal on its host star.

\begin{table}
\caption{The systems that should be prioritised based on the detectability of an exo-Earth with the ESPRESSO spectrograph. The systems are categorised into three groups by whether an exo-Earth in the HZ that is $1\ \textrm{M}_{\oplus}$, $2\ \textrm{M}_{\oplus}$ or $4\ \textrm{M}_{\oplus}$ is the least massive that may be detected with ESPRESSO.}
   	\label{tab:rv_mins}
   	\centering
	\begin{tabular}{l l}
        \toprule                      
        	{Desired Exo-Earth Mass} & {System} \\    
        \midrule	       
        $1\ \textrm{M}_{\oplus}$	& 	HD 113538 \\
        \midrule					 
        \multirow{10}{*}{$2\ \textrm{M}_{\oplus}$}	& HIP 65407	\\ 
        					& HD 190360	\\  
        					& XO-2 S		\\ 
        					& HD 37605	\\ 
	        				& HD 9446	\\ 
    	    				& HD 217107 	\\
    	    				& HD 142		\\ 
    	    				& HD 38529	\\  
    	    				& BD-08 2823	\\ 
    		    			& HD 215497	\\ 
        \midrule
        \multirow{5}{*}{$4\ \textrm{M}_{\oplus}$} & TYC 1422-614-1 \\ 
        				& HD 11964 \\  
        				& HD 177830 \\ 
        				& HD 60532 \\ 
        				& HD 134987 \\ 
        \toprule
   	\end{tabular}
\end{table}

\section{Conclusions}
\label{sec:summary}
We have investigated the entire population of currently known multiple planet systems that contain at least one Jovian planet in order to determine which systems would be the most promising targets for observations using new instruments designed specifically to search for exo-Earths. 
We have expanded upon the approach developed by \cite{Agnew2017}, and present a more systematic framework to assess the ability for all future discovered single and multiple planet systems to host hidden exo-Earths in their HZs. Whilst our approach is numerical, supplementing it with the AMD stability scheme presented by \cite{Laskar2017} proves to be beneficial in optimising time and computational resources. 

The key findings of our work are as follows:

\begin{itemize}
\item We find 9 systems that do not pass our planetary stability analysis, i.e., the known exoplanets are not stable given their current best-fit orbital parameters. While several of these systems have undergone further numerical investigation to better constrain their orbital parameters, there are still two for which there has not yet been any further analysis: HD~133131~A and HD~160691.
\item The AMD stability criteria presented by \cite{Laskar2017} is a powerful predictor of system stability, as demonstrated by the nearly complete agreement between our planetary stability simulations and the analytical predictions.
\item Massless TP simulations are important in identifying stable regions of the HZ in a computationally efficient manner. In systems where resonant behaviour is responsible for providing stabilisation, TP simulations should not be used to indicate massive body stability due to the absence of mutual gravitational interactions. However, they remain a powerful tool in excluding systems from further investigation as a result TP instability \citep{Agnew2018}. TP simulations are also useful in quickly identifying large regions of HZ stability that are unperturbed or only mildly perturbed by the gravitational effects of existing planets.
\item In general, in systems where low-order MMRs are responsible for stabilising a putative exo-Earth, the planet that is nearer to the HZ will tend to dominate the dynamics and be the sole body responsible for providing stabilisation. Conversely, in systems where higher-order MMRs align with the semi-major axis of the stabilised body but for which the resonant angle does not librate, resonance-hopping between weaker, high-order resonances provides a means of pseudo-stability. In some cases, a low order MMR can stabilise a body but long time-scale secular interactions causes the body to ``wander'' between being trapped in an MMR and being free, i.e. the resonant angle alternates between being bound and unbound. 
\item Of the systems we simulated, there are 28 candidates for which there is the potential for dynamically stable exo-Earths to exist, as yet undetected, in their HZs (see Fig.~\ref{fig:preds}). Of those, 16 of them would be detectable with the ESPRESSO spectrograph \textit{if they exist} (see Table~\ref{tab:rv_mins}). 
\item Of particular interest is HD~113538, which could  host a $1\ \textrm{M}_{\oplus}$ body within its HZ that would be detectable with the ESPRESSO spectrograph, and also has two giant planets located beyond its HZ. Taken in concert, this makes that system a promising potential Solar system analogue \citep{Agnew2018}.
\end{itemize}

In the search for Solar system analogues and a true twin Earth, a focus on a system that resembles our own is a logical starting point. As the Solar system contains several massive planets, we sought to identify candidates that also share this property. Between systems with very stable, unperturbed HZs, and those with stable orbits that result from resonant mechanisms with the known, massive bodies, we have provided a list that can both demonstrably host stable Earth-mass planets in the HZ, but would also be detectable with the new ESPRESSO spectrograph.

\section*{Acknowledgements}

We wish to thank the referee, Rudolf Dvorak, for their helpful comments and suggestions that have improved the paper. MTA was supported by an Australian Postgraduate Award (APA). This work was performed on the gSTAR national facility at Swinburne University of Technology. gSTAR is funded by Swinburne and the Australian Government's Education Investment Fund. This research has made use of the Exoplanet Orbit Database, the Exoplanet Data Explorer at exoplanets.org and the NASA Exoplanet Archive, which is operated by the California Institute of Technology, under contract with the National Aeronautics and Space Administration under the Exoplanet Exploration Program.




\bibliographystyle{mnras}
\bibliography{project_3_final.bib} 



\appendix
\section{Inclination Plots}
\label{appendix:inc}
\begin{figure*}
 \centering
 	\begin{subfigure}{0.47\textwidth}
   		\centering
   		\includegraphics[width=\textwidth]{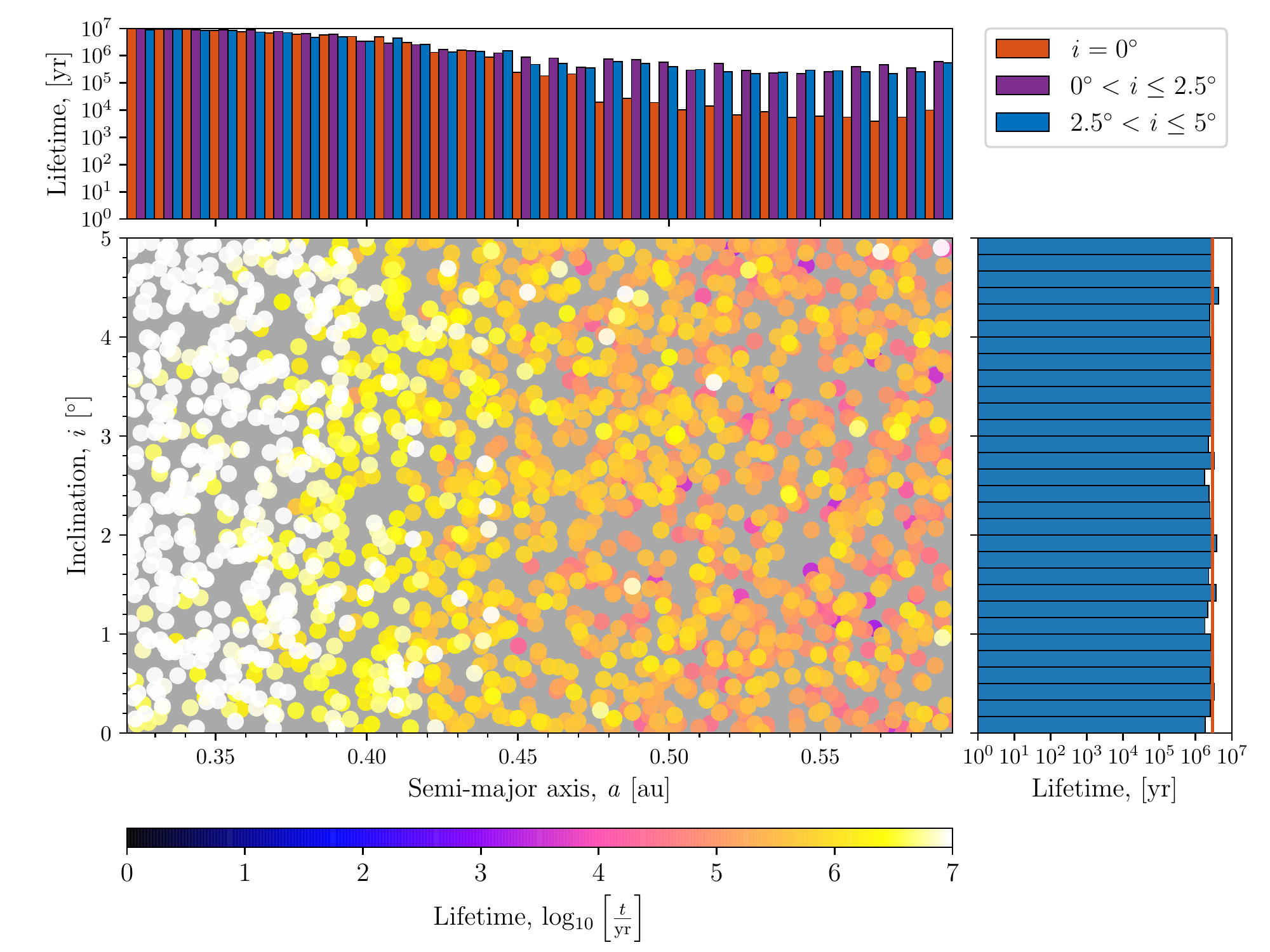}
   		\caption{HD 113538}
 	\end{subfigure}
 	\begin{subfigure}{0.47\textwidth}
   		\centering
   		\includegraphics[width=\textwidth]{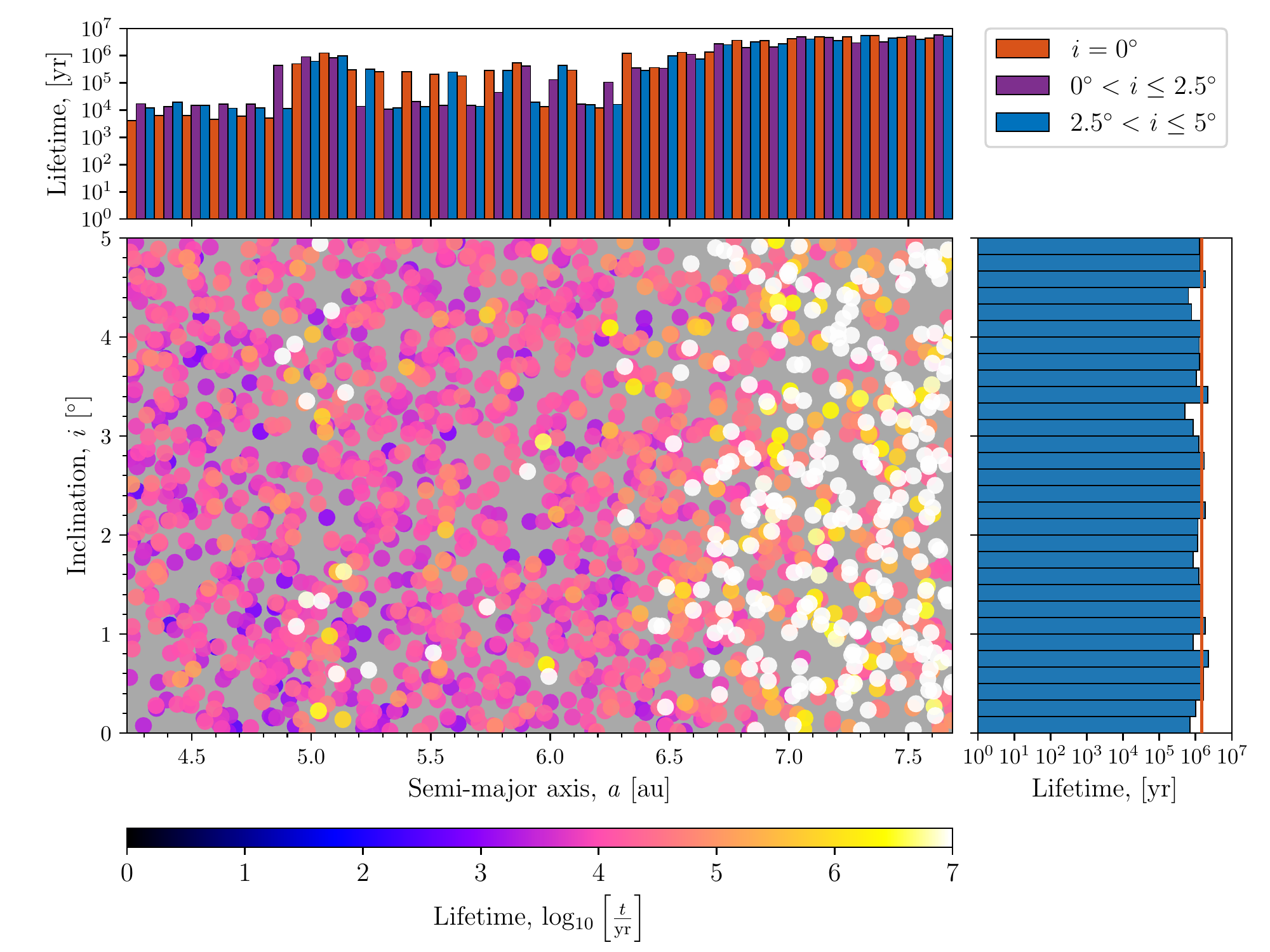}
   		\caption{HIP 67851}
 	\end{subfigure}
    
 	\begin{subfigure}{0.47\textwidth}
   		\centering
   		\includegraphics[width=\textwidth]{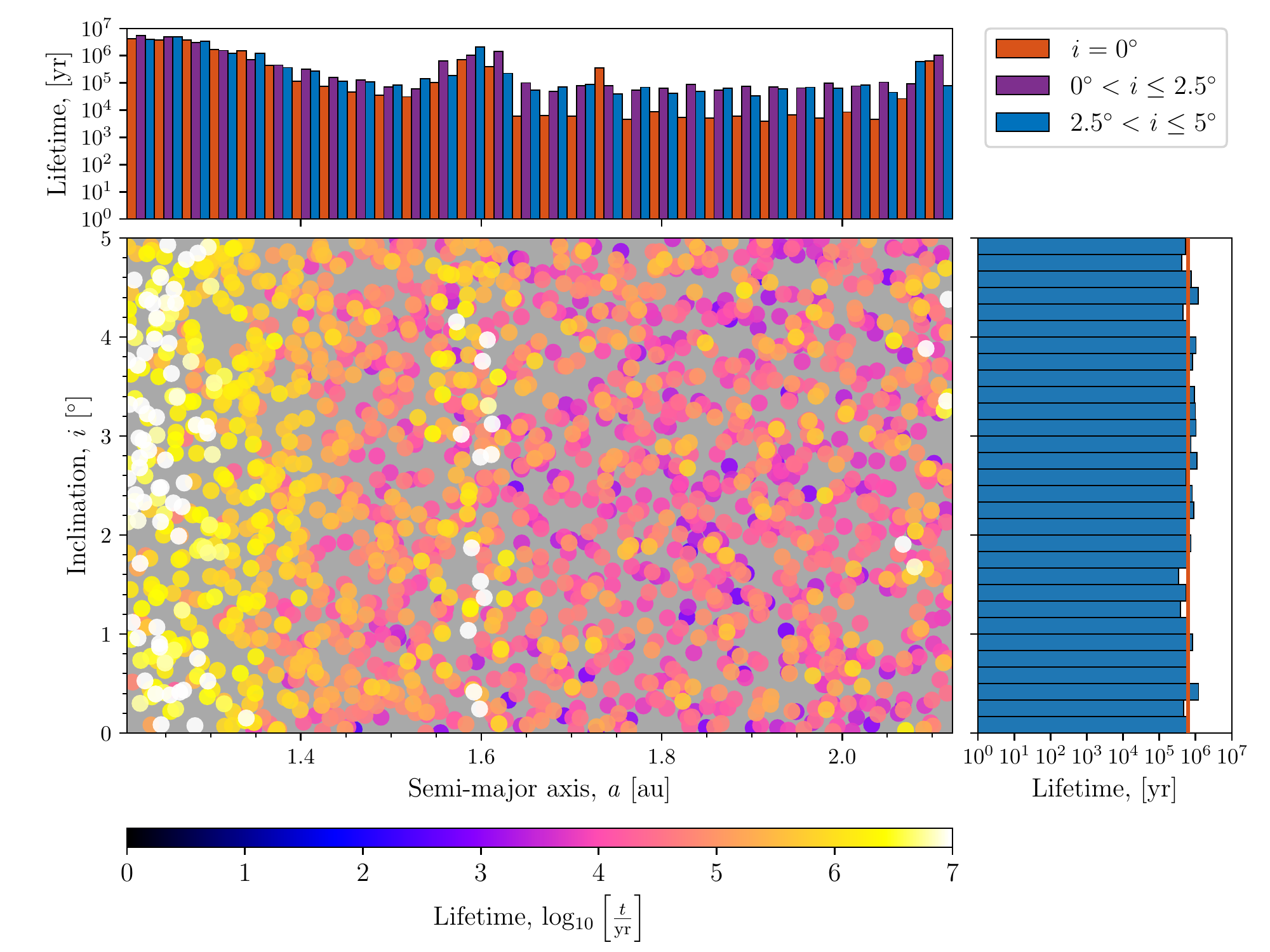}
   		\caption{47 UMa}
 	\end{subfigure}
 	\begin{subfigure}{0.47\textwidth}
   		\centering
   		\includegraphics[width=\textwidth]{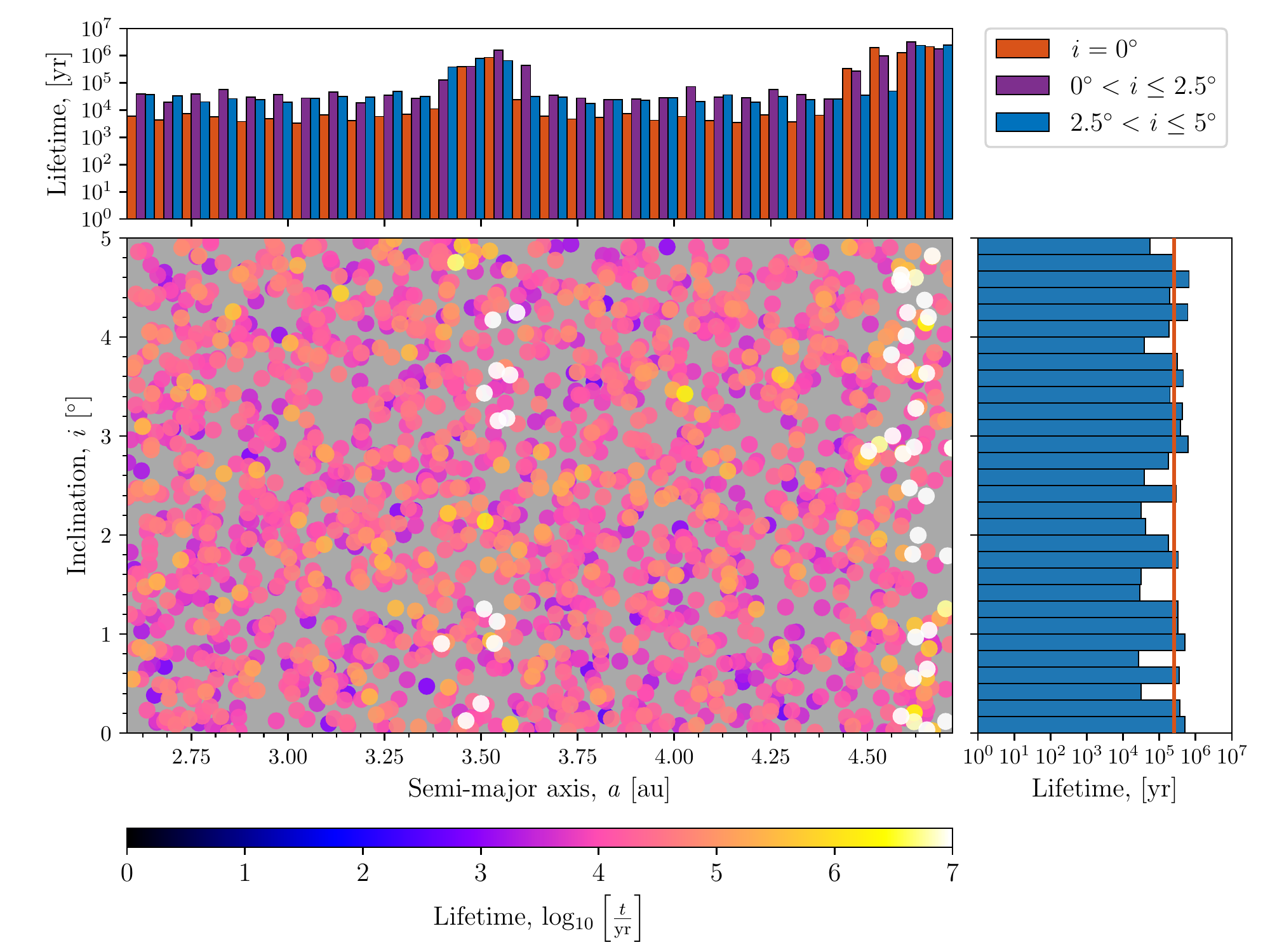}
   		\caption{HD 1605}
 	\end{subfigure}
    
 	\begin{subfigure}{0.47\textwidth}
   		\centering
   		\includegraphics[width=\textwidth]{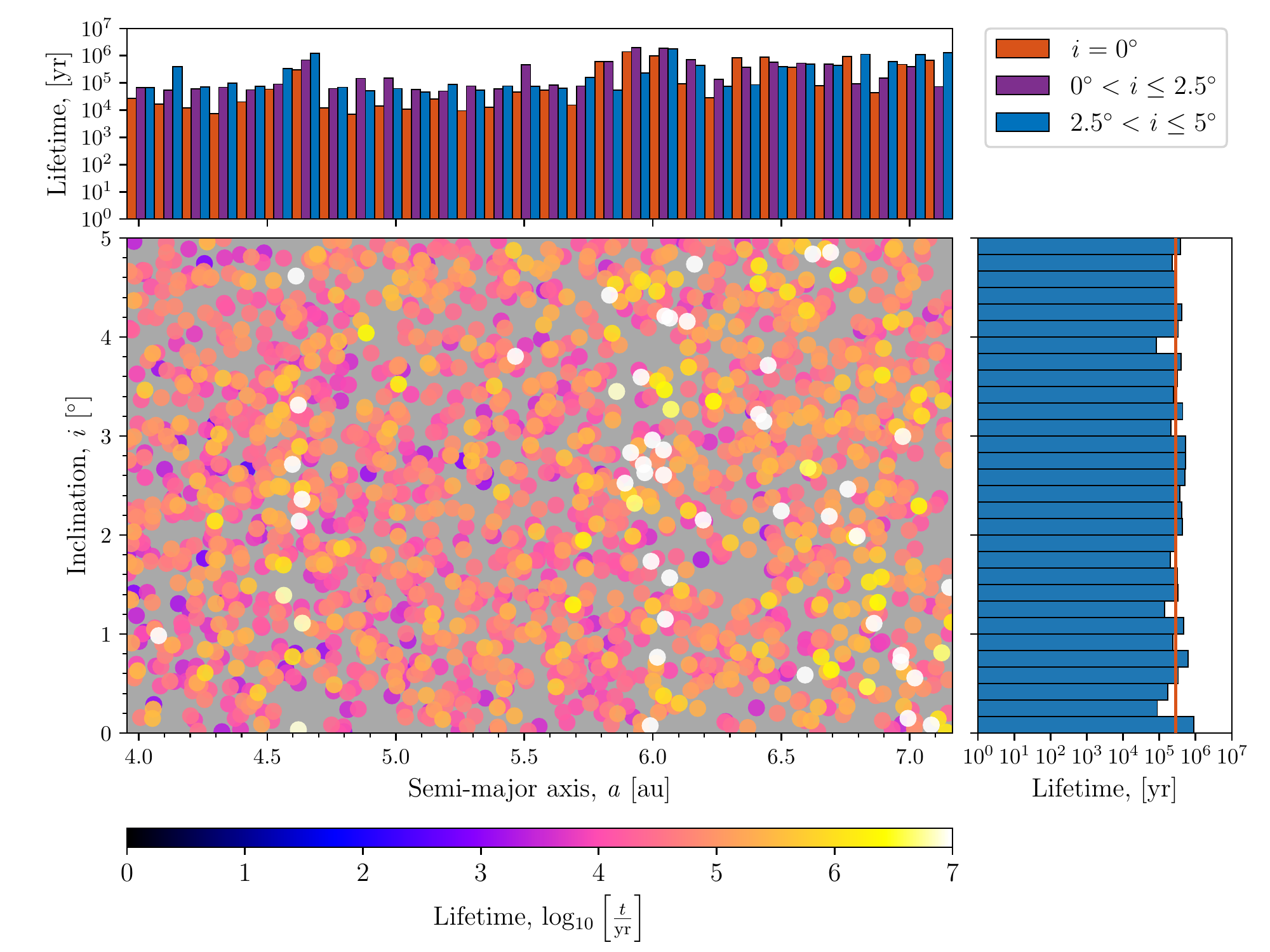}
   		\caption{HD 4732}
 	\end{subfigure}
 	\begin{subfigure}{0.47\textwidth}
   		\centering
   		\includegraphics[width=\textwidth]{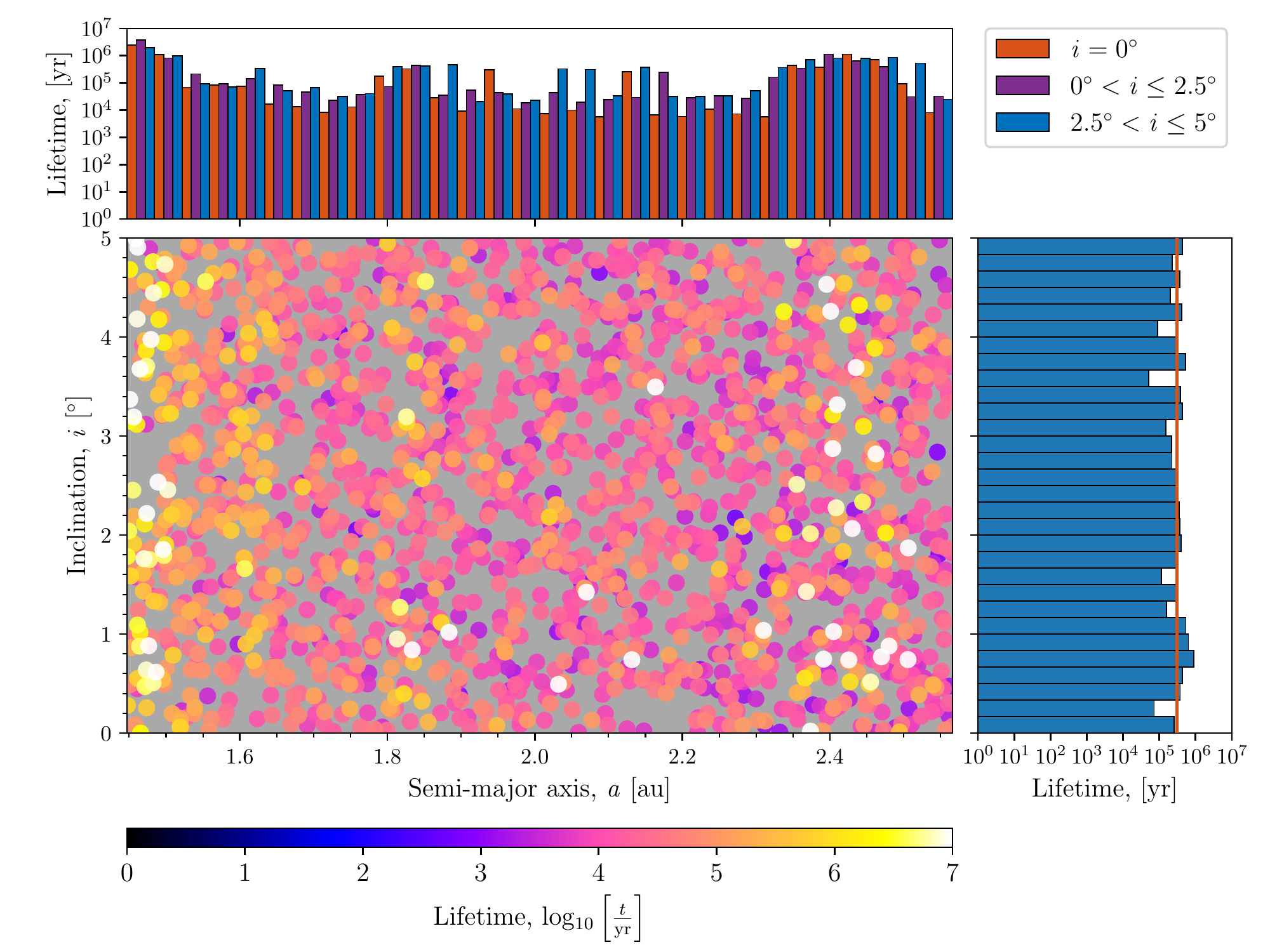}
   		\caption{HD 163607}
 	\end{subfigure}
\end{figure*}
\begin{figure*}
    \ContinuedFloat
 	\begin{subfigure}{0.47\textwidth}
   		\centering
   		\includegraphics[width=\textwidth]{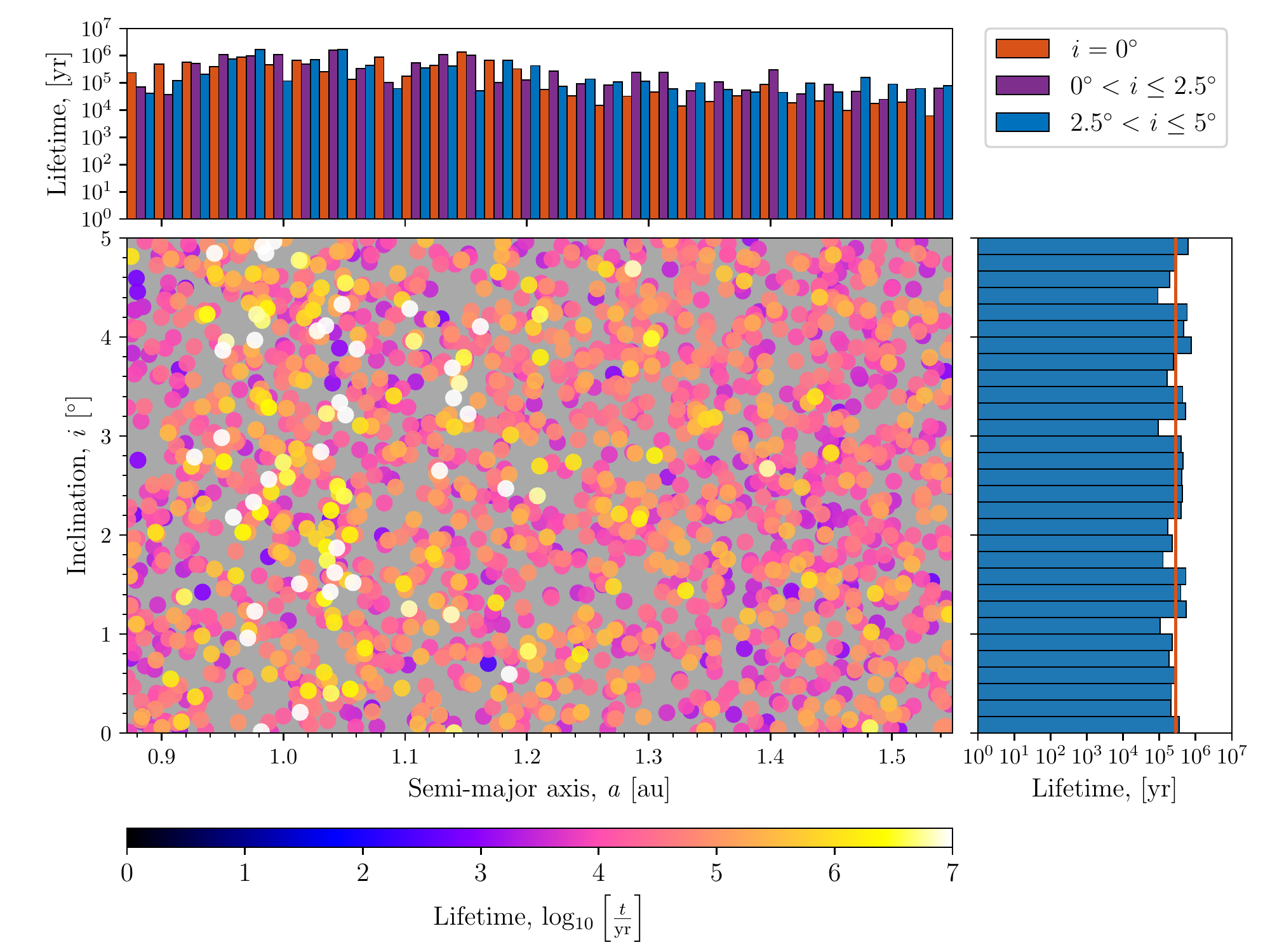}
   		\caption{HIP 14810}
 	\end{subfigure}
 	\begin{subfigure}{0.47\textwidth}
   		\centering
   		\includegraphics[width=\textwidth]{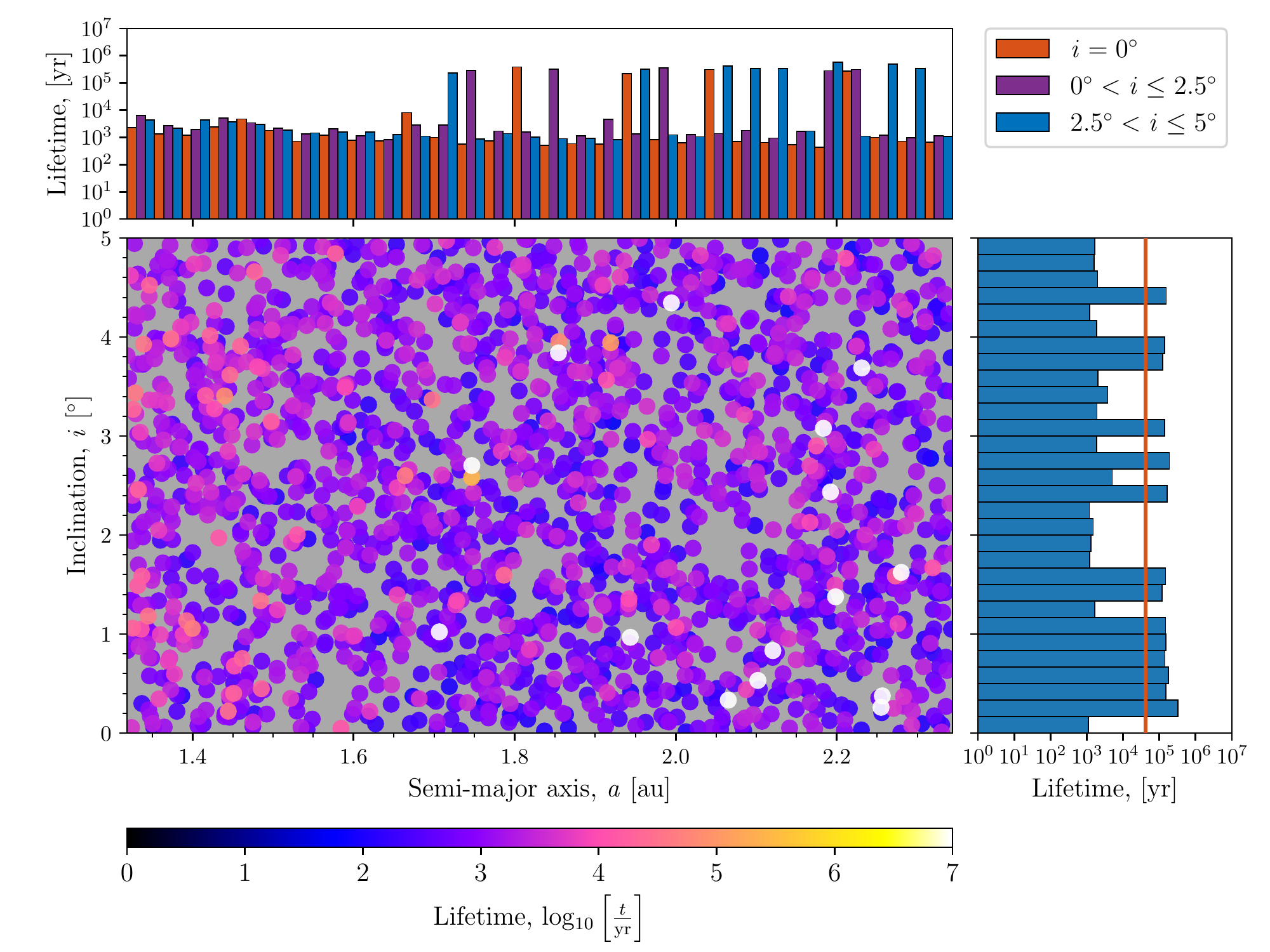}
   		\caption{HD 168443}
 	\end{subfigure}
    
 	\begin{subfigure}{0.47\textwidth}
   		\centering
   		\includegraphics[width=\textwidth]{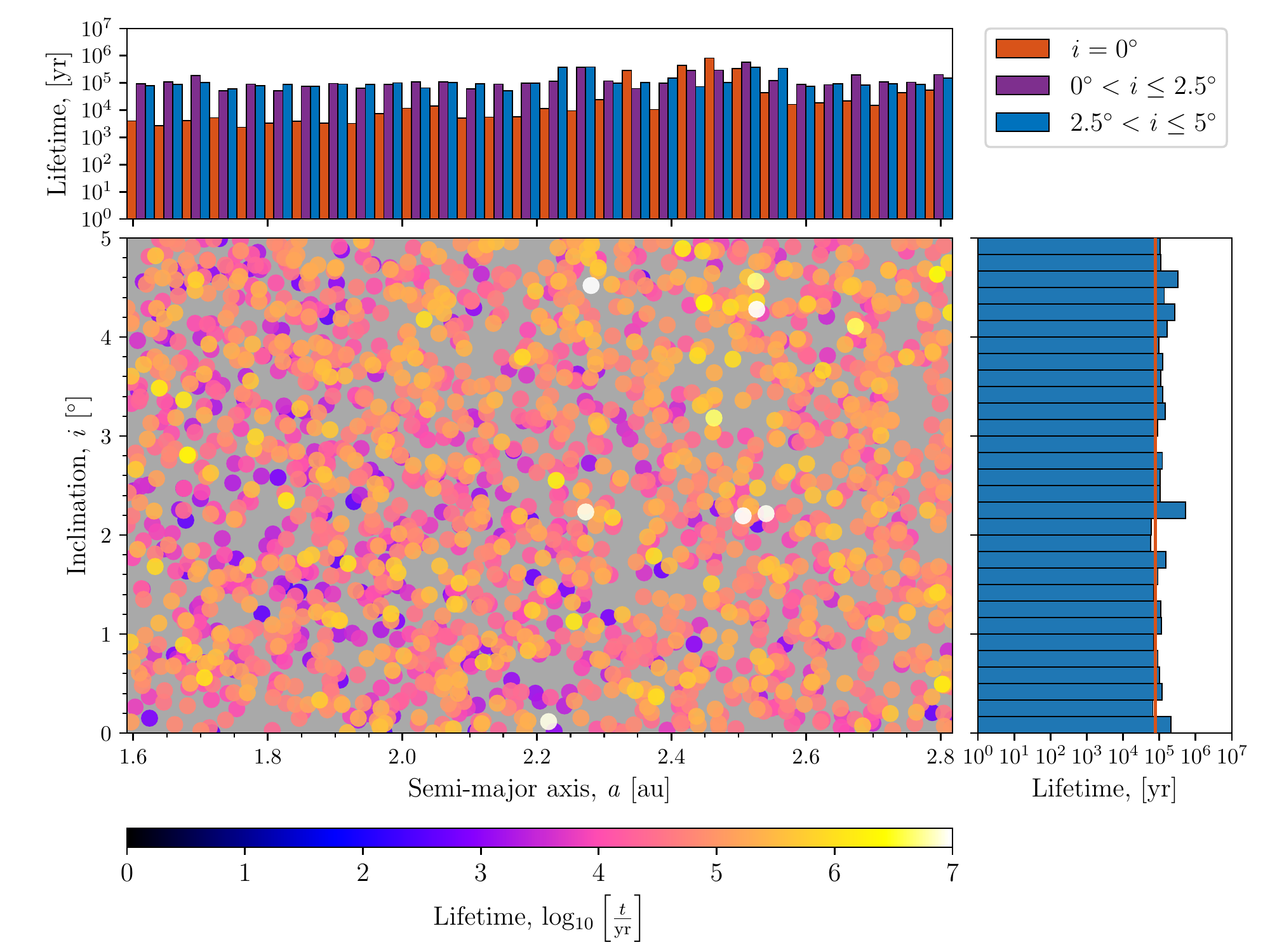}
   		\caption{HD 154857}
 	\end{subfigure}
 	\begin{subfigure}{0.47\textwidth}
   		\centering
   		\includegraphics[width=\textwidth]{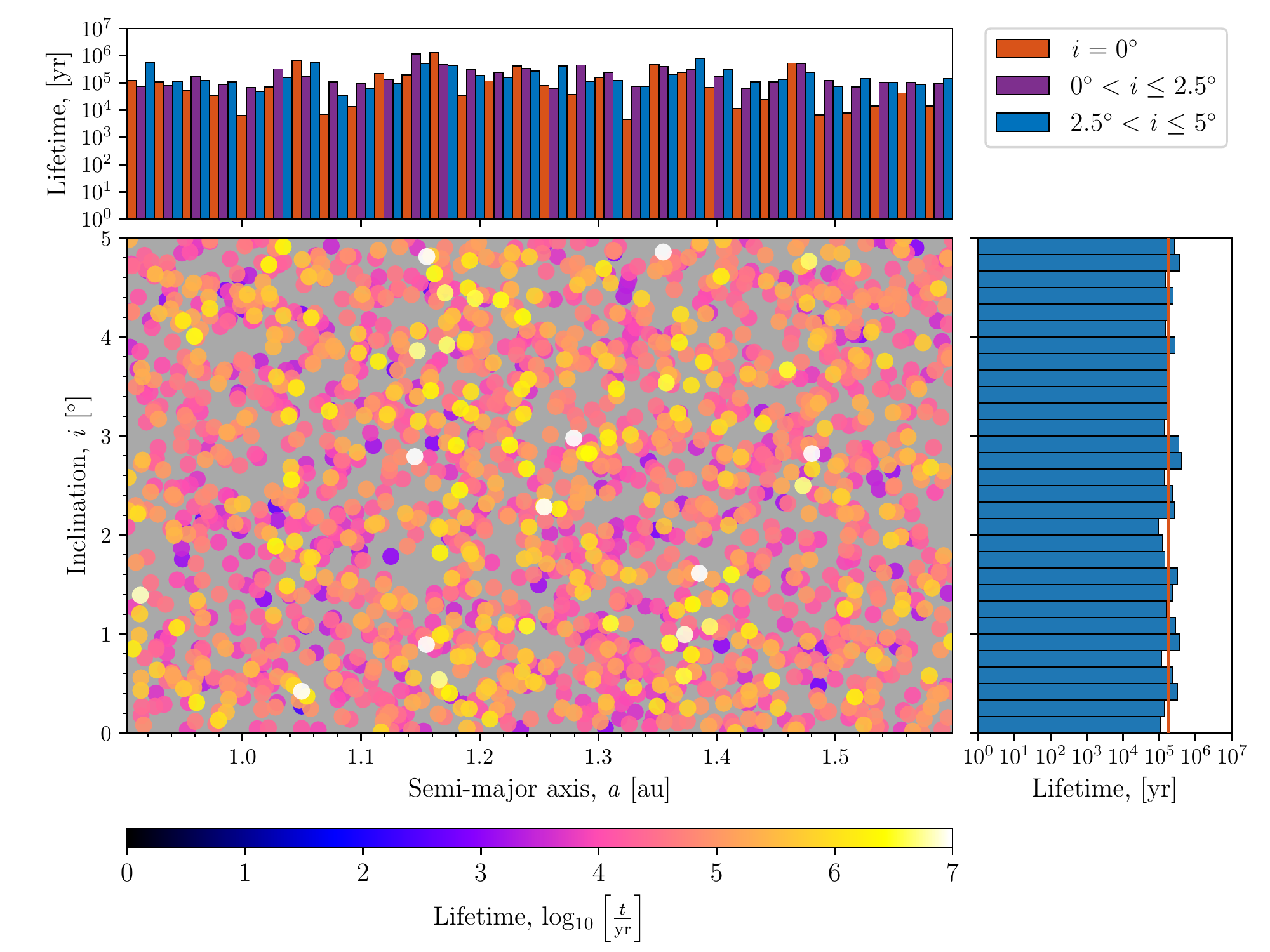}
   		\caption{HD 141399}
 	\end{subfigure}
    
 	\begin{subfigure}{0.47\textwidth}
   		\centering
   		\includegraphics[width=\textwidth]{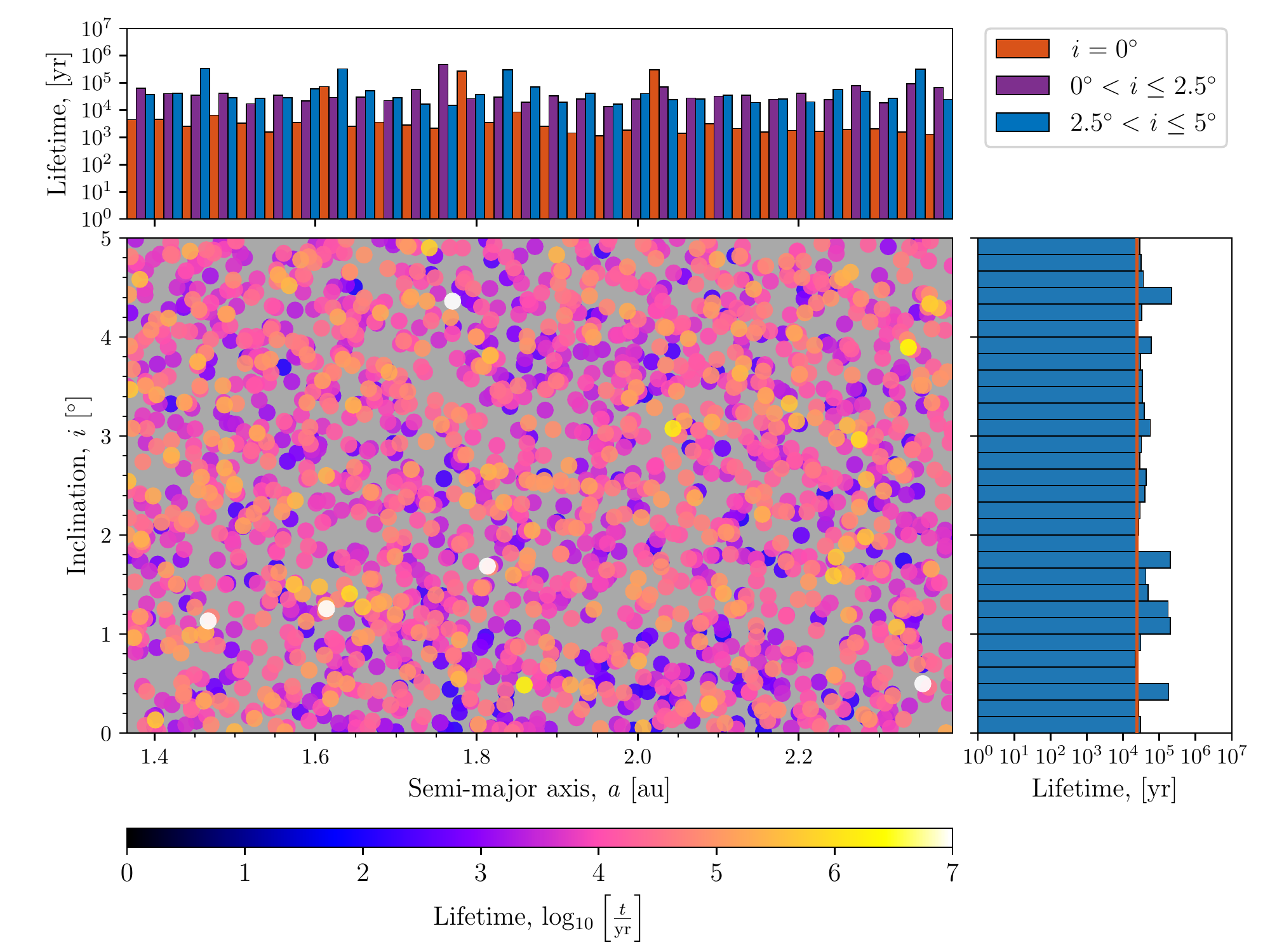}
   		\caption{HD 1506}
 	\end{subfigure}
 	\begin{subfigure}{0.47\textwidth}
   		\centering
   		\includegraphics[width=\textwidth]{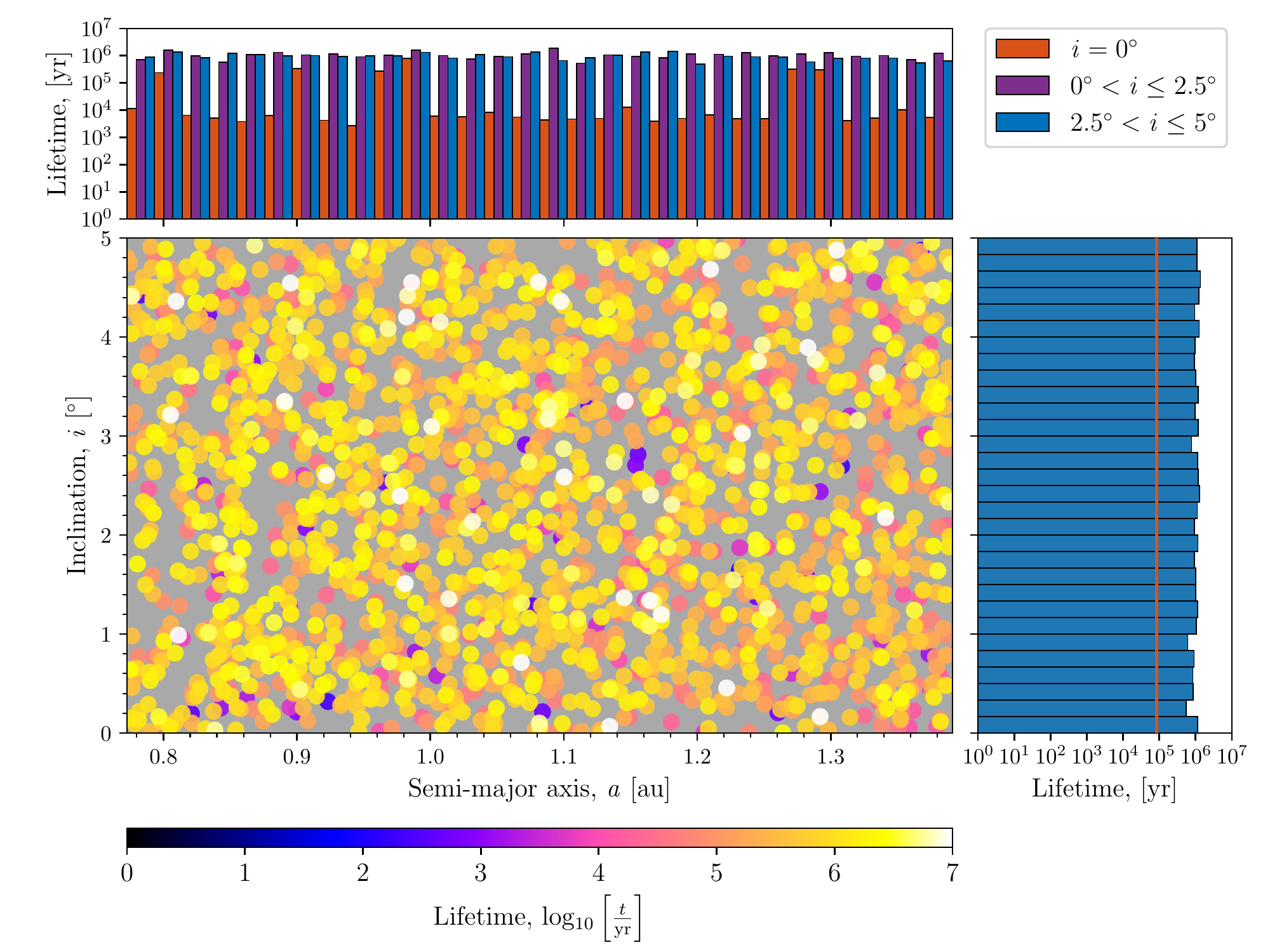}
   		\caption{HD 215497}\label{fig:inc_hd215497}
 	\end{subfigure}
\end{figure*}
\begin{figure*}
    \ContinuedFloat
    \begin{subfigure}{0.47\textwidth}
   		\centering
   		\includegraphics[width=\textwidth]{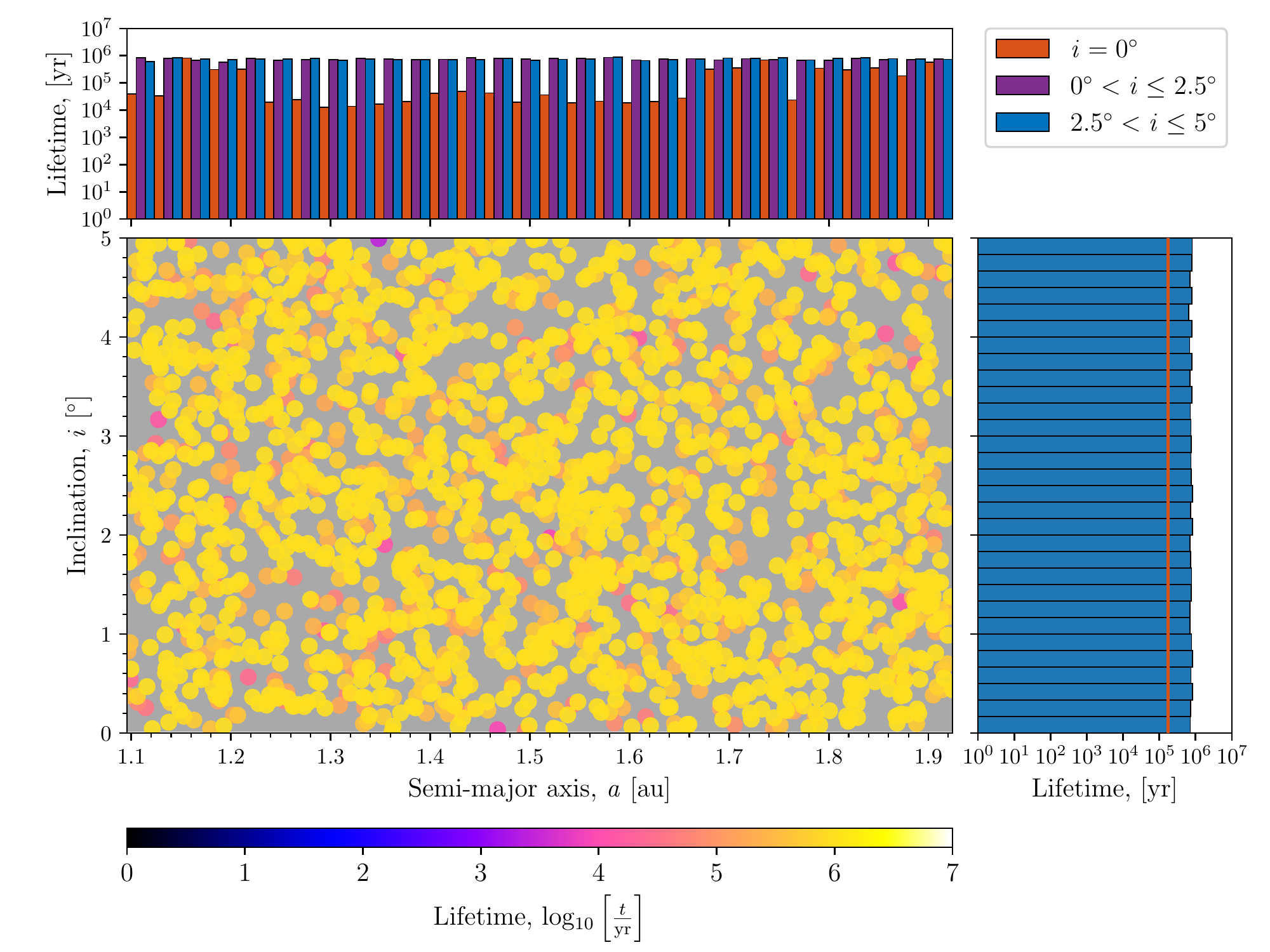}
   		\caption{HD 10180 (First $1$~Myr)}\label{fig:inc_hd10180}
 	\end{subfigure}
\caption{The $a$-$i$ stability maps for all systems from section~\ref{subsubsec:incl} for which we explored the effects of TP inclination on the  stability of the HZ. The colour scale for the TP lifetimes is logarithmic. The top histogram shows the binned mean lifetimes for co-planar TPs (orange), $0\degr<i\leq2.5\degr$ TPs (purple), and $2.5\degr<i\leq5\degr$ TPs (blue). The bins are only $1/3$ of their actual width for readability. The histogram on the right shows the binned mean lifetimes for the $0\degr<i\leq5\degr$ TPs, with the mean lifetime of all co-planar TPs overlaid in orange.}\label{fig:inc_all}
\end{figure*}
    
\newpage
\onecolumn
\section{Orbital Parameters}
\label{appendix:orb} 
\xentrystretch{-0.14}
\begin{center}
\topcaption{The orbital parameters for the exoplanetary systems simulated as they were presented in the NASA Exoplanet Archive (exoplanetarchive.ipac.caltech.edu) as of 27 April 2018.} \label{tab:xtab}
\tablefirsthead{%
\hline 
Star		&	Planet	&	$m\sin{i}$	&	$a$	&	$e$	&	$i$	&	$\Omega$	&	$\omega$	&	 $t_0$	\\ 
		&		&	(M$_{\mathrm{Jup}}$)	&	(au)	&		&	($\degr$)	&	($\degr$)	&	($\degr$)	&	 (days)	\\ \hline}
\tablehead{\multicolumn{9}{l}%
            {{\captionsize \tablename\ \thetable{} --
              continued from previous page}} \\ \hline
Star		&	Planet	&	$m\sin{i}$	&	$a$	&	$e$	&	$i$	&	$\Omega$	&	$\omega$	&	 $t_0$	\\ 
		&		&	(M$_{\mathrm{Jup}}$)	&	(au)	&		&	($\degr$)	&	($\degr$)	&	($\degr$)	&	 (days)	\\ \hline}
\tabletail{\hline \multicolumn{9}{l}{{Continued on next page}} \\ \hline}
\tablelasttail{\hline}
\begin{xtabular*}{\textwidth}{l@{\extracolsep{\fill}}lllllllll}
    HD 113538	&	b	&	0.36	&	1.24	&	0.14	&	0.0	&	0.0	&	74	&	 2455500.0	\\
				&	c	&	0.93	&	2.44	&	0.2	&	0.0	&	0.0	&	280	&	 2456741.0	\\
	\hline
	HD 219134	&	b	&	0.012	&	0.038474	&	0.0	&	0.0	&	0.0	&	0.0	&	 2449999.5	\\
				&	c	&	0.011	&	0.064816	&	0.0	&	0.0	&	0.0	&	0.0	&	 2449998.5	\\
				&	d	&	0.067	&	0.23508	&	0.0	&	0.0	&	0.0	&	0.0	&	 2449964.0	\\
				&	f	&	0.028	&	0.14574	&	0.0	&	0.0	&	0.0	&	0.0	&	 2449983.0	\\
				&	g	&	0.034	&	0.3753	&	0.0	&	0.0	&	0.0	&	0.0	&	 2449972.0	\\
				&	h	&	0.34	&	3.11	&	0.06	&	0.0	&	0.0	&	215	&	 2448725.0	\\
	\hline
	BD-06 1339	&	b	&	0.027	&	0.0428	&	0.0	&	0.0	&	0.0	&	0.0	&	 2455220.5	\\
				&	c	&	0.17	&	0.435	&	0.31	&	0.0	&	0.0	&	41	&	 2455265.2	\\
	\hline
	BD-08 2823	&	b	&	0.045	&	0.056	&	0.15	&	0.0	&	0.0	&	30	&	 2454637.7	\\
				&	c	&	0.33	&	0.68	&	0.19	&	0.0	&	0.0	&	-233	&	 2454193.0	\\
	\hline
	HAT-P-17	&	b	&	0.534	&	0.0882	&	0.342	&	89.2	&	0	&	201	&	 2454803.25	\\
				&	c	&	3.4	&	5.6	&	0.39	&	0.0	&	0.0	&	181.5	&	 2454885.0	\\
	\hline
	Pr0211		&	b	&	1.88	&	0.03176	&	0.011	&	0.0	&	0.0	&	17	&	 2456678.8	\\
				&	c	&	7.79	&	5.5	&	0.71	&	0.0	&	0.0	&	111	&	 2456736.0	\\
	\hline
	HD 181433	&	b	&	0.024	&	0.08	&	0.396	&	0.0	&	0.0	&	202	&	 2454542.0	\\
				&	c	&	0.64	&	1.76	&	0.28	&	0.0	&	0.0	&	21.4	&	 2453235.0	\\
				&	d	&	0.54	&	3	&	0.48	&	0.0	&	0.0	&	-30	&	 2452154.0	\\
	\hline
	HD 215497	&	b	&	0.02	&	0.047	&	0.16	&	0.0	&	0.0	&	96	&	 2454858.95	\\
				&	c	&	0.33	&	1.282	&	0.49	&	0.0	&	0.0	&	45	&	 2455003.48	\\
	\hline
	HD 37605	&	b	&	2.802	&	0.2831	&	0.6767	&	0.0	&	0.0	&	220.86	&	 2453378.241	\\
				&	c	&	3.366	&	3.814	&	0.013	&	0.0	&	0.0	&	221	&	 2454838.0	\\
	\hline
	HD 11964	&	b	&	0.622	&	3.16	&	0.041	&	0.0	&	0.0	&	0	&	 2454170.0	\\
				&	c	&	0.0788	&	0.229	&	0.3	&	0.0	&	0.0	&	102	&	 2454370.0	\\
	\hline
	HD 147018	&	b	&	2.12	&	0.2388	&	0.4686	&	0.0	&	0.0	&	-24.03	&	 2454459.49	\\
				&	c	&	6.56	&	1.922	&	0.133	&	0.0	&	0.0	&	-133.1	&	 2455301.0	\\
	\hline
	HIP 65407	&	b	&	0.428	&	0.177	&	0.14	&	0.0	&	0.0	&	50	&	 2456990.8	\\
				&	c	&	0.784	&	0.316	&	0.12	&	0.0	&	0.0	&	-19	&	 2457047.0	\\
	\hline
	XO-2 S		&	b	&	0.259	&	0.1344	&	0.18	&	0.0	&	0.0	&	311.9	&	 2456413.11	\\
				&	c	&	1.37	&	0.4756	&	0.1528	&	0.0	&	0.0	&	264.5	&	 2456408.1	\\
	\hline
	HIP 14810	&	b	&	3.88	&	0.0692	&	0.1427	&	0.0	&	0.0	&	159.32	&	 2453694.598	\\
				&	c	&	1.28	&	0.545	&	0.164	&	0.0	&	0.0	&	329	&	 2454672.24	\\
				&	d	&	0.57	&	1.89	&	0.173	&	0.0	&	0.0	&	286	&	 2454317.198	\\
	\hline
	HD 108874	&	b	&	1.29	&	1.038	&	0.082	&	0.0	&	0.0	&	232	&	 2454069.0	\\
				&	c	&	0.99	&	2.659	&	0.239	&	0.0	&	0.0	&	27	&	 2452839.0	\\
	\hline
	HD 159868	&	b	&	2.1	&	2.25	&	0.01	&	0.0	&	0.0	&	350	&	 2453435.0	\\
				&	c	&	0.73	&	1	&	0.15	&	0.0	&	0.0	&	290	&	 2453239.0	\\
	\hline
	HD 141399	&	b	&	0.451	&	0.415	&	0.04	&	0.0	&	0.0	&	-90	&	 2456998.0	\\
				&	c	&	1.33	&	0.689	&	0.048	&	0.0	&	0.0	&	-140	&	 2456838.0	\\
				&	d	&	1.18	&	2.09	&	0.074	&	0.0	&	0.0	&	-140	&	 2456923.0	\\
				&	e	&	0.66	&	5	&	0.26	&	0.0	&	0.0	&	-10	&	 2458900.0	\\
	\hline
	HD 217107	&	b	&	1.39	&	0.0748	&	0.1267	&	0.0	&	0.0	&	24.4	&	 2454396.0	\\
				&	c	&	2.6	&	5.32	&	0.517	&	0.0	&	0.0	&	198.6	&	 2451106.0	\\
	\hline
	HD 47186	&	b	&	0.07167	&	0.05	&	0.038	&	0.0	&	0.0	&	59	&	 2454566.95	\\
				&	c	&	0.35061	&	2.395	&	0.249	&	0.0	&	0.0	&	26	&	 2452010.0	\\
	\hline
	HD 38529	&	b	&	0.839	&	0.131	&	0.257	&	0.0	&	0.0	&	92.5	&	 2454012.64	\\
				&	c	&	13.38	&	3.712	&	0.341	&	0.0	&	0.0	&	17.8	&	 2452256.4	\\
	\hline
	HD 4203		&	b	&	1.82	&	1.1735	&	0.52	&	0.0	&	0.0	&	328.03	&	 2451911.52	\\
				&	c	&	2.17	&	6.95	&	0.24	&	0.0	&	0.0	&	224	&	 2456000.0	\\
	\hline
	HD 9446	&	b	&	0.7	&	0.189	&	0.2	&	0.0	&	0.0	&	215	&	 2454854.4	\\
				&	c	&	1.82	&	0.654	&	0.06	&	0.0	&	0.0	&	100	&	 2454510.0	\\
	\hline
	HD 133131 A	&	b	&	1.42	&	1.44	&	0.33	&	0.0	&	0.0	&	16	&	 2452327.0	\\
				&	c	&	0.42	&	4.49	&	0.49	&	0.0	&	0.0	&	100	&	 2452327.0	\\
	\hline
	HD 160691	&	b	&	1.08	&	1.497	&	0.128	&	0.0	&	0.0	&	22	&	 2452365.6	\\
				&	c	&	1.814	&	5.235	&	0.0985	&	0.0	&	0.0	&	57.6	&	 2452955.2	\\
				&	d	&	0.03321	&	0.09094	&	0.172	&	0.0	&	0.0	&	212.7	&	 2452991.1	\\
				&	e	&	0.5219	&	0.921	&	0.0666	&	0.0	&	0.0	&	189.6	&	 2452708.7	\\
	\hline
	HD 187123	&	b	&	0.523	&	0.0426	&	0.0103	&	0.0	&	0.0	&	25	&	 2454343.12	\\
				&	c	&	1.99	&	4.89	&	0.252	&	0.0	&	0.0	&	243	&	 2453580.04	\\
	\hline
	HD 183263	&	b	&	3.67	&	1.51	&	0.3567	&	0.0	&	0.0	&	233.5	&	 2452111.7	\\
				&	c	&	3.57	&	4.35	&	0.239	&	0.0	&	0.0	&	345	&	 2451971.0	\\
	\hline
	HD 190360	&	b	&	1.56	&	4.01	&	0.313	&	0.0	&	0.0	&	12.9	&	 2453542.0	\\
				&	c	&	0.06	&	0.1304	&	0.237	&	0.0	&	0.0	&	5	&	 2454390.0	\\
	\hline
	HD 74156	&	b	&	1.8	&	0.292	&	0.627	&	0.0	&	0.0	&	176.5	&	 2453788.59	\\
				&	c	&	8.06	&	3.85	&	0.432	&	0.0	&	0.0	&	258.6	&	 2453415.0	\\
	\hline
	HD 169830	&	b	&	2.88	&	0.81	&	0.31	&	0.0	&	0.0	&	148	&	 2451923.0	\\
				&	c	&	4.04	&	3.6	&	0.33	&	0.0	&	0.0	&	252	&	 2452516.0	\\
	\hline
	HD 10180	&	c	&	0.0416	&	0.06412	&	0.073	&	0.0	&	0.0	&	328	&	 2454001.445	\\
				&	d	&	0.0378	&	0.12859	&	0.131	&	0.0	&	0.0	&	325	&	 2454022.119	\\
				&	e	&	0.0805	&	0.2699	&	0.051	&	0.0	&	0.0	&	147	&	 2454006.26	\\
				&	f	&	0.0722	&	0.4929	&	0.119	&	0.0	&	0.0	&	327	&	 2454024.67	\\
				&	g	&	0.0732	&	1.427	&	0.263	&	0.0	&	0.0	&	327	&	 2454002.8	\\
				&	h	&	0.2066	&	3.381	&	0.095	&	0.0	&	0.0	&	142	&	 2453433.4	\\
	\hline
	HD 134987	&	b	&	1.59	&	0.81	&	0.233	&	0.0	&	0.0	&	352.7	&	 2450071.0	\\
				&	c	&	0.82	&	5.8	&	0.12	&	0.0	&	0.0	&	195	&	 2461100.0	\\
	\hline
	47 UMa		&	b	&	2.53	&	2.1	&	0.032	&	0.0	&	0.0	&	334	&	 2451917.0	\\
				&	c	&	0.54	&	3.6	&	0.098	&	0.0	&	0.0	&	295	&	 2452441.0	\\
				&	d	&	1.64	&	11.6	&	0.16	&	0.0	&	0.0	&	110	&	 2451736.0	\\
	\hline
	HD 168443	&	b	&	7.659	&	0.2931	&	0.52883	&	0.0	&	0.0	&	172.923	&	 2455626.199	\\
				&	c	&	17.193	&	2.8373	&	0.2113	&	0.0	&	0.0	&	64.87	&	 2455521.3	\\
	\hline
	HD 11506	&	b	&	4.21	&	2.708	&	0.37	&	0.0	&	0.0	&	218.9	&	 2456637.2	\\
				&	c	&	0.36	&	0.721	&	0.24	&	0.0	&	0.0	&	272	&	 2454127.0	\\
	\hline
	HD 163607	&	b	&	0.77	&	0.36	&	0.73	&	0.0	&	0.0	&	78.7	&	 2454185.0	\\
				&	c	&	2.29	&	2.42	&	0.12	&	0.0	&	0.0	&	265	&	 2455085.0	\\
	\hline
	HD 142		&	b	&	1.25	&	1.02	&	0.17	&	0.0	&	0.0	&	327	&	 2452683.0	\\
				&	c	&	5.3	&	6.8	&	0.21	&	0.0	&	0.0	&	250	&	 2455954.0	\\
	\hline
	HD 154857	&	b	&	2.24	&	1.291	&	0.46	&	0.0	&	0.0	&	57	&	 2453572.5	\\
				&	c	&	2.58	&	5.36	&	0.06	&	0.0	&	0.0	&	352	&	 2455219.0	\\
	\hline
	HD 219828	&	b	&	0.06607	&	0.045	&	0.059	&	0.0	&	0.0	&	225	&	 2455998.78	\\
				&	c	&	15.1	&	5.96	&	0.8115	&	0.0	&	0.0	&	145.77	&	 2454180.7	\\
	\hline
	HD 67087	&	b	&	3.06	&	1.08	&	0.17	&	0.0	&	0.0	&	285	&	 2450154.8	\\
				&	c	&	4.85	&	3.86	&	0.76	&	0.0	&	0.0	&	256	&	 2450322.5	\\
	\hline
	HD 177830	&	b	&	1.49	&	1.2218	&	0.009	&	0.0	&	0.0	&	85	&	 2450154.0	\\
				&	c	&	0.15	&	0.5137	&	0.3	&	0.0	&	0.0	&	110	&	 2450179.0	\\
	\hline
	HD 1605		&	b	&	0.96	&	1.48	&	0.078	&	0.0	&	0.0	&	26	&	 2453443.3	\\
				&	c	&	3.48	&	3.52	&	0.098	&	0.0	&	0.0	&	241	&	 2454758.3	\\
	\hline
	HD 60532	&	b	&	1.06	&	0.77	&	0.26	&	0.0	&	0.0	&	-3.7	&	 2454594.7	\\
				&	c	&	2.51	&	1.6	&	0.03	&	0.0	&	0.0	&	179.8	&	 2454973.0	\\
	\hline
	HD 5319		&	b	&	1.76	&	1.6697	&	0.02	&	0.0	&	0.0	&	97	&	 2456288.0	\\
				&	c	&	1.15	&	2.071	&	0.15	&	0.0	&	0.0	&	252	&	 2453453.0	\\
	\hline
	HD 200964	&	b	&	1.85	&	1.601	&	0.04	&	0.0	&	0.0	&	288	&	 2454900.0	\\
				&	c	&	0.895	&	1.95	&	0.181	&	0.0	&	0.0	&	182.6	&	 2455000.0	\\
	\hline
	HD 33844	&	b	&	1.96	&	1.6	&	0.15	&	0.0	&	0.0	&	211	&	 2454609.0	\\
				&	c	&	1.75	&	2.24	&	0.13	&	0.0	&	0.0	&	71	&	 2454544.0	\\
	\hline
	24 Sex		&	b	&	1.99	&	1.333	&	0.09	&	0.0	&	0.0	&	9.2	&	 2454762.0	\\
				&	c	&	0.86	&	2.08	&	0.29	&	0.0	&	0.0	&	220.5	&	 2454930.0	\\
	\hline
	HD 4732		&	b	&	2.37	&	1.19	&	0.13	&	0.0	&	0.0	&	85	&	 2454967.0	\\
				&	c	&	2.37	&	4.6	&	0.23	&	0.0	&	0.0	&	118	&	 2456093.0	\\
	\hline
	HIP 67851	&	b	&	1.38	&	0.46	&	0.05	&	0.0	&	0.0	&	138.1	&	 2452997.8	\\
				&	c	&	5.98	&	3.82	&	0.17	&	0.0	&	0.0	&	166.5	&	 2452684.1	\\
	\hline
	TYC 1422-614-1	&	b	&	2.5	&	0.69	&	0.06	&	0.0	&	0.0	&	50	&	 2453236.5	\\
				&	c	&	10	&	1.37	&	0.048	&	0.0	&	0.0	&	130	&	 2453190.5	\\
	\hline
	nu Oph		&	b	&	24	&	1.9	&	0.1256	&	0.0	&	0.0	&	9.6	&	 2452034.2	\\
				&	c	&	27	&	6.1	&	0.165	&	0.0	&	0.0	&	4.6	&	 2453038.0	\\
	\hline
	BD+20 2457	&	b	&	21.42	&	1.45	&	0.15	&	0.0	&	0.0	&	207.64	&	 2454677.03	\\
				&	c	&	12.47	&	2.01	&	0.18	&	0.0	&	0.0	&	126.02	&	 2453866.95	\\
  	\toprule
\end{xtabular*} 
\end{center}


\bsp	
\label{lastpage}
\end{document}